# Design of a Piezoelectric Wind Energy Harvester for Bacterial Disinfection of Drinking Water

**A Thesis**

**by**

**PRAKASH POUDEL**

**S18027**

**For the award of the degree of**

**MASTER OF SCIENCE (BY RESEARCH)**

**SCHOOL OF ENGINEERING**

**INDIAN INSTITUTE OF TECHNOLOGY MANDI**

**Mandi, Himachal Pradesh-175075**

***March 2022***

# DECLARATION

I hereby declare that the Thesis titled **"Design of a Piezoelectric Wind Energy Harvester for Bacterial Disinfection of Drinking Water"** submitted by me, to the Indian Institute of Technology Mandi for the award of the degree of **Master of Science (by research)** is a bonafide record of the research work carried out by me in the **School of Engineering**, Indian Institute of Technology Mandi, under the supervision of **Prof. Rajeev Kumar and Prof. Satish Chandra Jain.** The contents of this thesis, in full or in parts, have not been submitted to any other Institute or University for the award of any degree or diploma. In keeping with the general practice, due acknowledgments have been made wherever the work described is based on findings of other investigators.

IIT Mandi (H.P.)
Date:

Prakash Poudel

# THESIS CERTIFICATE

This is to certify that the Thesis titled **"Design of a Piezoelectric Wind Energy Harvester for Bacterial Disinfection of Drinking Water"** submitted by **Prakash Poudel** to the Indian Institute of Technology Mandi for the award of the degree of **Master of Science (by Research)**, is a bonafide record of the research work done by him under our supervision in the School of Engineering, Indian Institute of Technology Mandi. The contents of this thesis, in full or in parts, have not been submitted to any other Institute or University for the award of any degree or diploma.

Dr. Rajeev Kumar
Professor
School of Engineering
IIT Mandi

Date:

Dr. Satish Chandra Jain
Emeritus Professor
School of Engineering
IIT Mandi

Date:

# ACKNOWLEDGEMENT

I would like to thank from my inner heart to all the people who have been instrumental in making this work possible. First of all, I would like to express the deepest appreciation to my thesis supervisor, Professor Rajeev Kumar, who has the great attitude and the substance of a genius. He continuously and convincingly guided me in my research and academics. He has not only provided me the utmost support and guidance from the beginning of my research work but also provided me freedom and motivation to work on the research problems of my choice. In addition to being my thesis supervisor, he has always been a great teacher to me, imparting the basic concepts required for my work. Without his guidance and persistent help this thesis would not have been possible.

I would also offer my sincerest gratitude to my thesis co-supervisor, Professor Satish Chandra Jain, for his innovative inputs to my research and for providing me guidance and encouragement throughout this work.

I convey my deepest appreciation and thanks to Prof. Rahul Vaish for his inestimable help and support in my research work. He has always provided me his invaluable mentorship and advice in every possible aspect of my academic experience at IIT Mandi. I am extremely thankful for all the encouragement and motivation he has provided to me.

In addition, a sincere thank you to Dr. Saurav Sharma, who has always helped me in my academics and research work. He has provided easy solutions to my problems and assisted me in my experimental work. I am extremely thankful for all the encouragement and motivation he has provided to me.

I also like to acknowledge the members of my thesis committee, Dr. Vishal Singh Chauhan, Dr. Mohammad Talha, and Dr. Manoj Thakur for their valuable appraise of my annual research progress.

The work of this thesis would not have been possible without the direct or indirect contributions from my friends and colleagues at IIT Mandi. I sincerely thank my research colleagues, Dr. Sumeet Sharma, Jitendra Adhikari, Satish Kumar, Kamalpreet Singh, Rishikant Thakur, Gokul Krishna and Diwakar Singh all the other members of the Smart Materials and Structures lab for all their help and support in my research work. Much of my efforts would not have been successful without proper help and support from my adorable friends: Niraj K.C., Lawaj Thapa, Prakash Giri, Sabin Kafle, Jitendra Chaudary, Prakash Neupane, and Puspendra Sukla.

Last but not least, I submit my gratitude to my parents for always encouraging me to achieve my academic goals and my wife, Susma for always being on my side and being the source of all the positivity and optimism in my life.

This thesis includes research findings published in the journals, *energies* and *Global Challenges*. The author has obtained permission for inclusion in this thesis.

# PREAMBLE

Energy harvesting techniques on small scale tend to be promising for supplying continuous energy for wireless sensors as well as portable electronic devices. Different types of aerodynamic instabilities; namely vortex-induced vibrations (VIV), flutter, galloping, and wake galloping occur when a wind flows on the structures. These instabilities can be used to obtain an electrical power output for various applications with appropriate design of the harvester.

In this thesis, a piezoelectric wind energy harvester is proposed for bacterial disinfection of drinking water. A mathematical model of a piezoelectric wind energy harvester is developed to predict the electrical energy from the wind induced vibration based on the galloping phenomenon. Structure of a piezoelectric wind energy harvester is a cantilever piezolaminated elastic beam with a bluff body attached to the free end. Linear time invariant and lumped parameter approach is considered to model the structure. In order to improve the performance of the harvester, bluff body is modified by attaching different attachment in the form of circular-shaped, triangular-shaped, square-shaped, Y-shaped, and curve-shaped. Based on the shape of the bluff body, aerodynamic modeling is carried out to predict the wind force acting on the energy harvester. This modeling is based on the empirical coefficients that define the force coefficients for different shaped bluff bodies. Electrical modeling of the energy harvester is based on the Kirchhoff's current law. Based on the mathematical modeling, a MATLAB Simulink model is developed to carry out the numerical studies. Experimental setup is also developed to validate the numerical studies. It has been found both numerically and experimentally that the harvester with a curve-shaped attachment to the bluff body provides the best electrical output.

Bacterial disinfection of drinking water is performed using the electrical output of the piezoelectric wind energy harvester. Energy harvester with a curve-shaped attachment to a bluff body is used for this application because it provides enhanced electrical output as compared to other shaped harvesters. The output of the energy harvester is connected to a coaxial copper electrode modified with copper oxide (CuONWs). The simulation analysis is performed to observe the effect of nanowires on the center electrode. It has been found that the concept of using centre electrode, with nanowires grown perpendicular to the surface of the electrode highly enhances the electric field and thus enables bacterial disinfection with low applied voltage. The electrical energy is supplied to the bacterial water and the bacterial sample is studied for different time periods. Finally, the bacterial log inactivation efficiency is calculated to observe the reliability of the disinfection process.

Numerical and experimental studies show that the proposed bacterial disinfection technique is efficient, low cost, by-product free, and has strong potential for water treatment in the storage system.

# TABLE OF CONTENTS

# List of Figures

# List of Tables

# CHAPTER 1: INTRODUCTION

This chapter presents a brief concept of energy harvesting using piezoelectricity. Later, the various phenomena of wind induced vibrations, and the applications of wind energy harvesting are presented. A detailed literature review on wind energy harvesting is presented afterward. The objectives of the thesis are defined thereafter based on the literature review. Lastly, the outline of the thesis is presented.

## 1.1 Background

Energy harvesting is defined as capturing of energy from external sources, including wind, solar, vibrations, electromagnetic, etc. Basically, the ambient energy present in the nature is converted into useful electrical energy. The sources and uses of energy harvesting in both macro and micro scale is shown in Figure 1.1.

| *Type of Energy harvesting* | *Energy Source* | *Application* | *Ultimate Goal* |
|---|---|---|---|
| **Macro scale** | Renewable sources like solar, wind, ocean tides, etc. | Energy management solutions in a large scale | Reduce oil dependency |
| **Micro scale** | Small scale sources like mechanical vibrations, human motion, ambient wind, electrostatic energy, etc. | Ultra-low-power solutions | Driving low energy consuming devices |

Figure 1.1. Types of energy harvesting [1].

Energy harvesting techniques on small scale tend to be promising for supplying continuous energy for wireless sensors as well as portable electronic devices [1], [2], [3], [4]. Powering these devices using traditional chemical batteries can have limitations, as these have limited lifespan and have difficulty in replacement in some cases. Also, there occur problems with maintenance as well as environmental pollution issues with electrochemical batteries. Obtaining electrical power from ambient sources would be good choice to power micro electromechanical systems (MEMS). Energy can be extracted from different energy sources, like wind, solar, human movement, and mechanical vibration to device self-powered wireless sensor networks. Various researches have been carried out on energy harvesting techniques and researches are able to calculate appropriate power output for different engineering applications [5], [6]. The overall phenomena of energy harvesting is shown in Figure 1.2. The energy available in various sources is extracted with the help of different transducers available. Once the energy is harvested, there is a need of proper conditioning of the electrical signals obtained. The noises and other fluctuations present on the signal are removed for generating smooth signals, using power conditioning devices. Then, the electrical energy is stored in small rechargeable batteries or storage capacitors for further use. Finally, the electrical energy in a small scale can be used in a self-powering electronic devices, biosensors, wireless sensors, etc. for various applications.

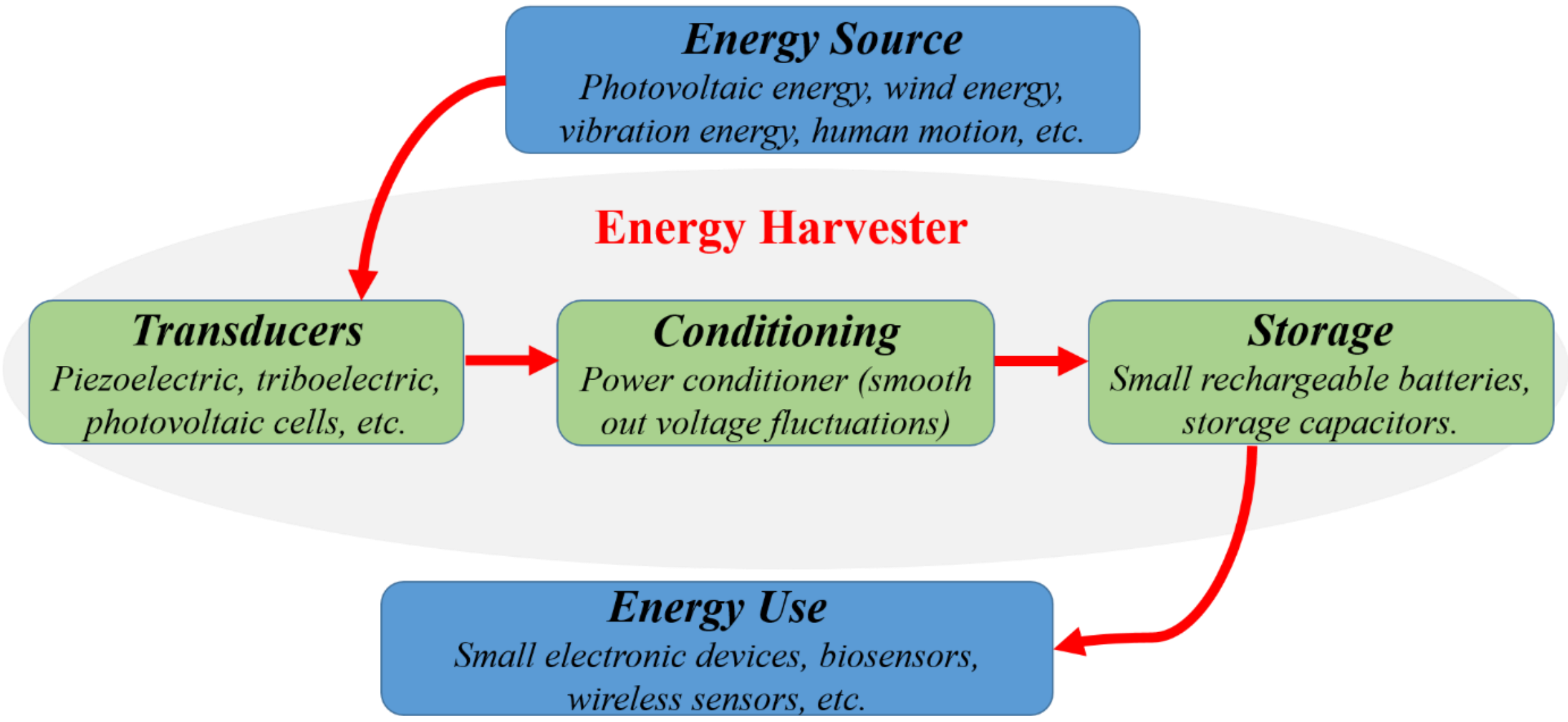

Figure 1.2. Overview on energy harvesting.

Harvesting ambient energy form wind is an important energy harvesting technique in recent years. The use of wind turbines to obtain electrical power in large scale is carried out in many countries where strong wind is available. In contrast, wind turbines are less efficient when employed to harvest energy in low scale. Wind energy can be tapped in small scale using wind energy harvester and it can serve as a power source for various sensors. Wind energy harvester will play an important role to provide sustainable energy for eco-friendly systems. Wind induced vibrations in structures may lead to failures, due to the occurrence of oscillations with large amplitude, thus are not desired in civil and aerospace engineering [7]. However, these vibrations can be usefully converted into electrical energy using special transducers; namely, piezoelectric, electromagnetic, electrostatic transducers. Piezoelectric transducers have been widely used for this purpose, because they have high power density, easy working mechanism and do not require initial voltage [8]. The oscillations in the system offer some mechanical strain energy, which is then transformed into an electric voltage due to direct piezoelectric effect. Piezoelectric transducers can be explained in three different modes of modes of operation: namely $d_{31}$ mode, $d_{33}$ mode, and $d_{15}$ mode. The $d_{31}$ mode of piezoelectric material is widely used because it offers low resonant frequency, and it is more likely to match the vibrations produced due to wind flow over the harvester. A harvester with specially designed bluff body will offer instability in a system, when subjected to a direction of wind flow. It is because there occurs interaction between a solid and a fluid medium, which creates large amplitude vibrations in a bluff body. These vibrations offer some strain on the piezoelectric patch that is attached on the beam of the harvester and finally due to effect of piezoelectricity, there results electric voltage output in small scale.

Different types of aerodynamic instabilities; namely vortex-induced vibrations (VIV), flutter, galloping, and wake galloping occur when a wind flows on the structures. These instabilities can be used to obtain an electrical power output for various applications with appropriate design of the harvester. Instability based on galloping phenomenon is appropriate in energy harvesting technique because galloping-based harvesters can operate in wind range of wind speed and vibrations with large amplitude are possible. Energy harvesting utilizing ambient wind involves coupling with different domains. A piezoelectric wind energy harvester based on galloping phenomenon requires an aeroelectromechanical coupling behaviour. Wind that flows over the harvester offers nonlinear aerodynamic forces that gives rise to mechanical vibrations in the harvester, which involves fluid-structure coupling. The fundamental concept and mechanism of piezoelectricity and the instabilities offered by the wind onto the harvester are elaborated in the following.

## 1.2. Piezoelectricity: Fundamental Concept and Working Mode

Piezoelectricity is defined as a property of certain materials in which the material responds to the applied mechanical stress, with production of electrical charges [9]. The electrical voltage thus obtained is proportional to the applied mechanical stress. This method of obtaining the electrical voltage and power from applied mechanical stress is known as direct piezoelectric effect. When connected to a circuit, the electric dipoles contained in the piezoelectric material create the electric potential on the application of mechanical stress. When the piezoelectric material is without any stress, the positive and negative charges contained on it balance each other. Thus, the material is neutrally charged, and no electric output is obtained in the connected circuit. If a stress is developed with an application of external force, the relative position of the positive and negative charges is altered, that creates change in dipole moment. Potential difference is obtained within the material since the charges do not cancel each other. On the other hand, mechanical strain can be achieved with the application of an electric field, referred as converse piezoelectric effect. When a potential difference is applied on the material, the change in the position of charges occur, due to electrostatic attraction or repulsion created. There are certain materials that exhibit piezoelectric behaviour which are categorized as: ceramics, crystals and polymers. Among them, piezoelectric ceramics possess large piezoelectric coefficient compared to others. Piezoelectric materials can also be classified as: naturally available and man-made. Naturally occurring materials include: berlinite, quartz, cane sugar, topaz, Rochelle and tourmaline. Man-made piezoelectric materials are available in different forms, including lead zirconate titanate (PZT), polyvinylidene fluoride (PDVF), barium titanate ($BaTiO_3$), macro fiber composites (MFC), and so on. These piezoelectric materials are commercially available and are used for various sensing and actuation purposes.

Poling is a treatment performed in piezoelectric materials to obtain intrinsic piezoelectricity. Figure 1.3. shows the poling process in piezoelectric materials. For a material without poling treatment, the electric dipoles are randomly oriented. A high electric field is applied in order to pole the piezoelectric material. During poling process, all the electric dipoles are oriented in the direction of applied electric field. After poling, the electric field is removed, and the electric dipoles seem to align roughly in the direction of electric field. This slight misalignment in the electric dipoles after poling is due to the microscope defects present in the lattice of the piezoelectric crystals.

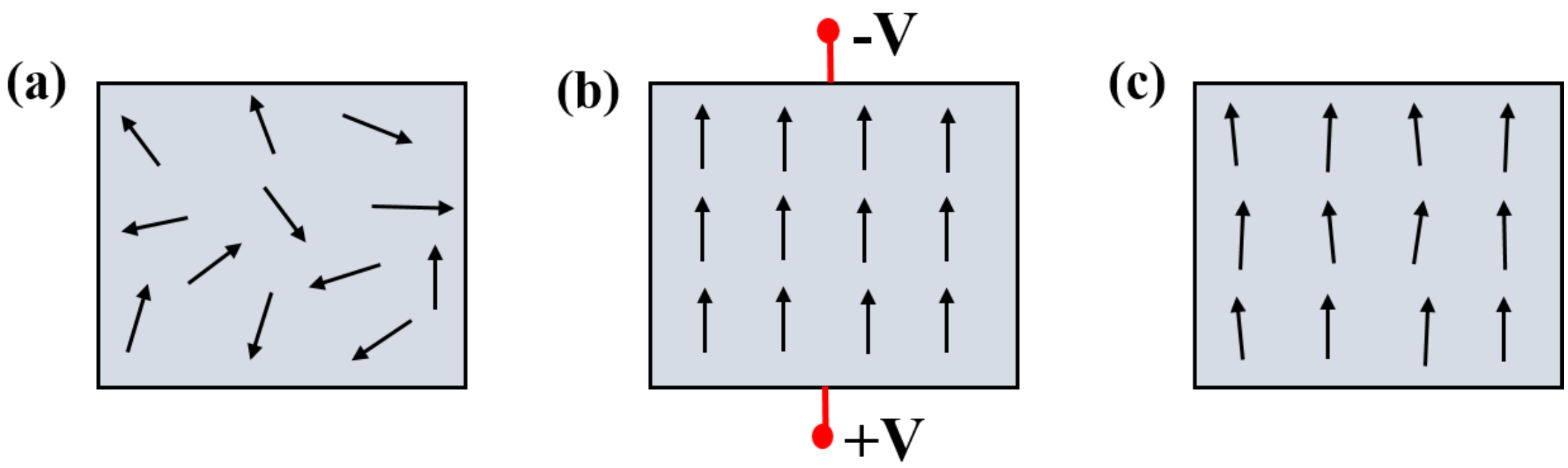

Figure 1.3. Poling procedure: (a) Before poling; (b) During poling; (c) After poling.

There are some piezoelectric constants that greatly affect the piezoelectricity of piezoelectric materials. The relation between the mechanical strain developed on the material with applied electric field is defined as piezoelectric strain constant, $d$. Its unit is defined in (m/V).

$$d = \frac{strain\ developed}{applied\ electric\ field} \quad (1)$$

Another important parameter used to define the power conversion efficiency in piezoelectric material is electromechanical coupling coefficient, $k$. It is defined as,

$$k_{ij}^2 = \frac{W_i^e}{W_j^m} \tag{2}$$

Where, $W_i^e$ is the electrical output obtained in the $i$ axis and $W_j^m$ is the input mechanical energy in the $j$ axis.

The relationship among the field variables of piezoelectricity is established in the form the constitutive equations,

$$S_{ij} = s_{ijkl}^E T_{kl} + d_{kij} E_k \tag{3}$$

$$D_i = d_{ikl} T_{kl} + \varepsilon_{ik}^T E_k \tag{4}$$

Where, $S_{ij}$ is called the mechanical strain, $s_{ijkl}^E$ is the mechanical compliance measured at constant electric field, $d_{ikl}$ is the piezoelectric strain coefficient, $E_k$ is the electric field, $D_i$ is the electric displacement, $T_{kl}$ is the mechanical stress, and $\varepsilon_{ik}^T$ is the dielectric permittivity measured at zero mechanical stress. Considering the symmetry of tensors, the constitutive equations of piezoelectricity in the matrix form are expressed as,

$$\begin{bmatrix} S \\ D \end{bmatrix} = \begin{bmatrix} s^E & d' \\ d & \varepsilon^T \end{bmatrix} \begin{bmatrix} T \\ E \end{bmatrix} \tag{5}$$

Where a superscript ($'$) denotes a transpose. The values of the above-mentioned piezoelectric constants are affected by the poling direction and the direction of applied mechanical stress. Based on the direction of applied mechanical stress and poling direction, the modes of operation of piezoelectric materials are classified into three types: namely, $d_{33}$ or longitudinal mode, $d_{31}$ or transverse mode, and $d_{15}$ or shear mode, as shown in Figure 1.4.

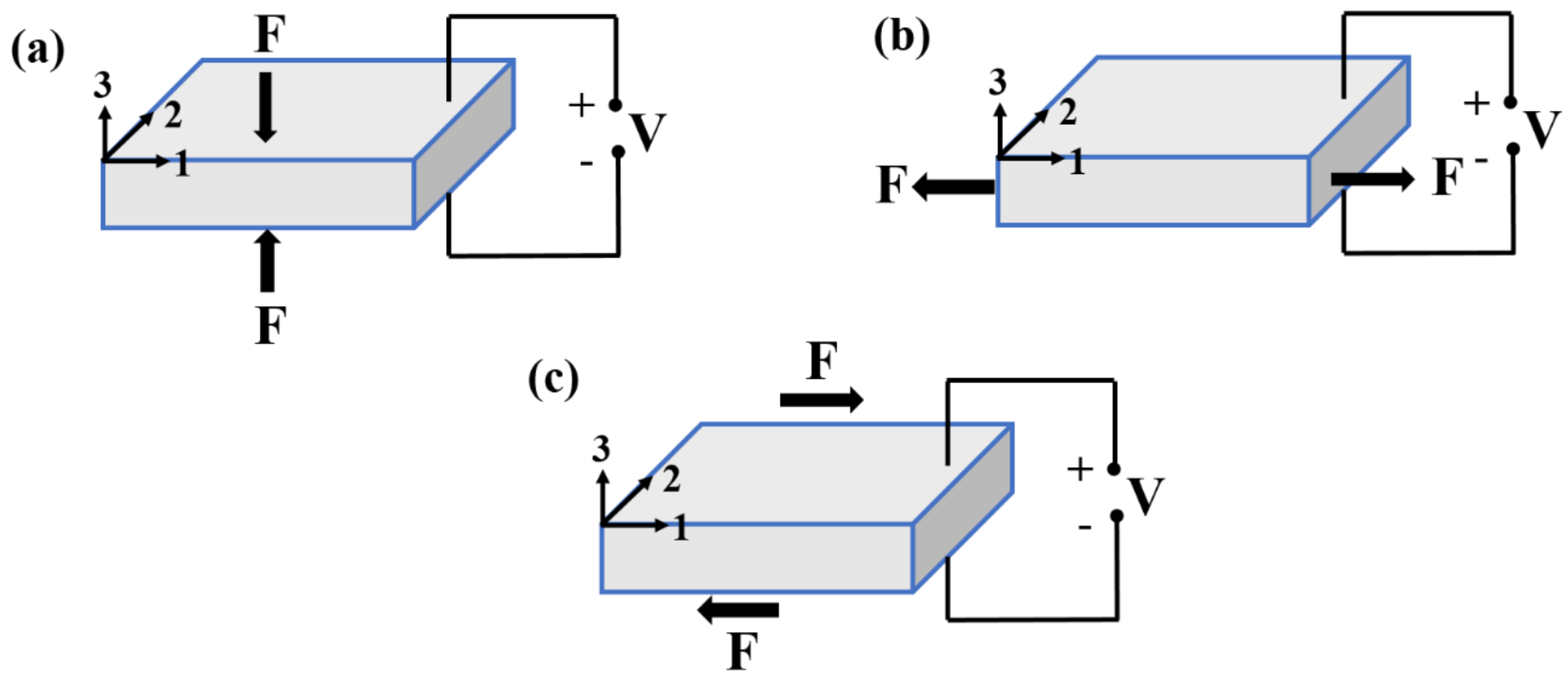

Figure 1.4. Modes of operation of piezoelectric materials: (a) $d_{33}$ mode; (b) $d_{31}$ mode; (c) $d_{15}$ mode.

In longitudinal mode of operation, the direction of the force applied on the material and the poling direction are the same, as shown in Figure 1.4(a). This mode is applicable for impact-based loading. Piezoelectric tiles, piezoelectric speed sensors are the examples for longitudinal mode piezoelectric devices [10]. Transverse mode or $d_{31}$ mode is the most commonly used working mode for piezoelectric sensing, actuation and energy harvesting applications [11]. In this mode, the force is applied perpendicular to the poling direction, as shown in Figure 1.4(b). Piezoelectric sensors and actuators based on cantilever and undergo bending deformation. The $d_{31}$ mode of operation has two significant advantages over $d_{33}$ mode. First, the harvester can be easily tuned at low resonant frequency, thus making it suitable for energy harvesting techniques in natural environment. Secondly, in a harvester in $d_{31}$ mode, higher mechanical strains can be obtained from low input force. In this thesis, we use $d_{31}$ mode of operation for design of small piezoelectric wind energy harvester because of above mentioned advantages. In shear or $d_{15}$ mode, a piezoelectric element is subjected to shear strain rather than axial strain, as applied in other two modes discussed above [12]. It can be seen in Figure 1.4(c) that the electric field is obtained in 3$^{rd}$ direction, with the application of shear strain in 1-3 plane. Shear mode is not commonly employed in piezoelectric applications due to the difficulties associated with the fabrication and poling process.

## 1.3. Bluff Body Aerodynamics

Bluff body can be defined as a body which has surface that is not aligned with the streamlines of the fluid flow. In contrast, streamline body has surface aligned with the streamlines when subjected to fluid flow. Due to its shape, bluff body has a separated flow in the wake region, thus creating asymmetric vortices, as shown in Figure 1.5(a). In bluff bodies, the drag force created is mainly due to the pressure force, whereas in case of streamline bodies the drag force is due to the viscous drag. Bluff body has a great importance in various engineering applications, such as bridges, towers, pipelines and the fluid flow behaviour around the bluff body is required for better design of structures. In a bluff body, the boundary layers are separated, thus forming unsteady vortex flows past the body. Thus, these asymmetric vortices past the bluff body will cause oscillations of the body.

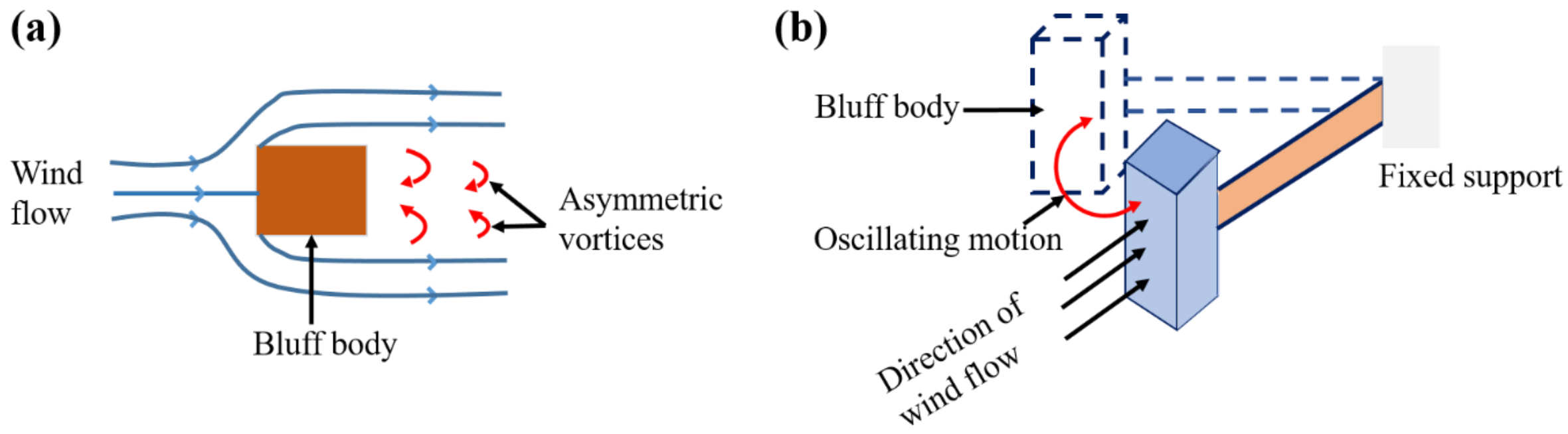

Figure 1.5. Bluff body aerodynamics: (a) Flow around a bluff body; (b) Bluff body motion.

Figure 1.5 (b) shows the motion of a bluff body that is attached to a free end of a cantilever beam. When a cantilever beam with a bluff body attached onto it is subjected to a wind flow, there occur an oscillatory motion of the bluff body with respect to the direction of undisturbed wind flow. Small scale wind energy harvesters often use bluff body attached cantilever beam so that high amplitude oscillation of the body is possible for better electrical output, with the use of piezoelectric transducer.

## 1.4. Wind Induced Vibrations

When the engineering structures are subjected to fluctuating wind, these structures respond dynamically to the wind flow, which lead to vibration. These vibrations are termed as wind induced vibrations, and are often encountered in various engineering structures, such as tall buildings, electrical transmission lines, electrical towers, large suspension bridge cables. The mechanism and behaviour of different types of wind induced vibrations; namely vortex-induced vibration (VIV), galloping, flutter and wake induced vibrations, are discussed in this section.

### 1.4.1. Vortex-induced vibration

One of the most common aeroelastic instability phenomena that occurs in many engineering structures, such as bridges, offshore structures, marine cables, cooling towers, and other many aerodynamic and hydrodynamic structures is vortex-induced vibration (VIV). When a bluff body is subjected to a steady and uniform flow, the separation of fluid flow occurs, forming two shear layers on either side of the body, which creates vortices on the downstream of the bluff body as shown in Figure 1.6 [13], [14], [15]. The formation of this alternating vortices at the wake region is called vortex shedding and it is very important to understand the behaviour of this phenomena both theoretically and experimentally.

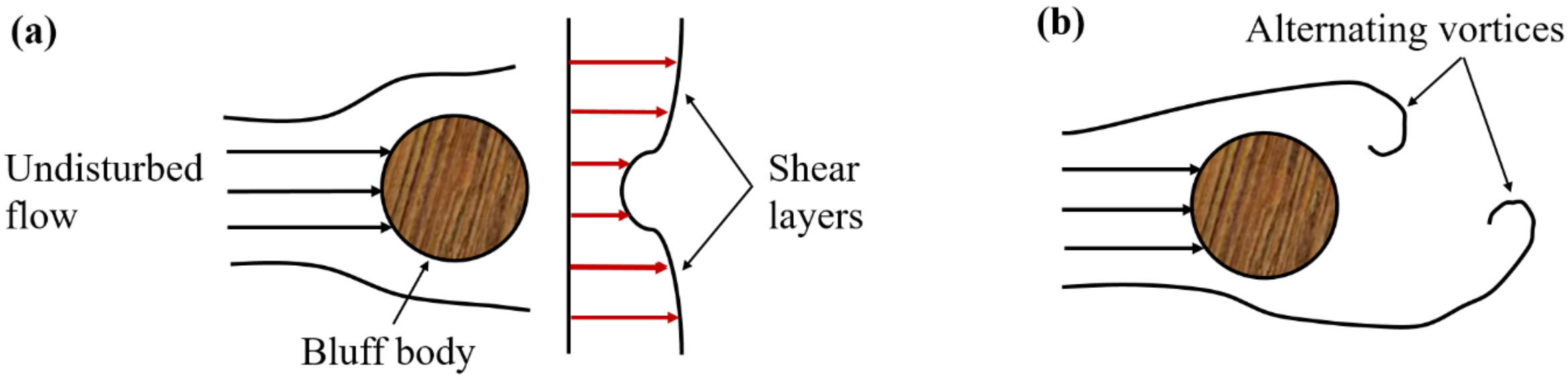

Figure 1.6. Flow past a bluff body: (a) Formation of shear layers; (b) vortex shedding.

The frequency at which the shedding occur is defined as vortex shedding frequency, $w_f$ and is given as,

$$w_f = 2\pi S_t \frac{U}{l} \tag{6}$$

where, $S_t$ is called Strouhal number which depends on the shape of the bluff body. Similarly, $U$ represents the undisturbed wind flow velocity and $l$ is the reference length scale. Vortex-induced vibration is a interaction between the bluff body and the alternating vortices created at the wake of the body. Wind energy harvesters that are based on vortex-induced vibration phenomenon operate on only limited range of wind speed for generating effective power. This is considered as the main constraint for VIV-based wind energy harvesting.

### 1.4.2 Galloping

Another type of an aeroelastic phenomenon that is observed particularly with slender structures, such as bridge cables, transmission lines, etc. is called galloping. Galloping cannot be achieved in bluff body with a symmetric cross-section. Galloping occurs with a bluff body having asymmetry cross-section and its features are high-amplitude (about 20~300 times the diameter of the conductor) and low-frequency (about 0.1-3 Hz). The important parameters that affect the galloping phenomenon are the cross-sectional geometry and the angle of attack of the wind flow [16], [17].

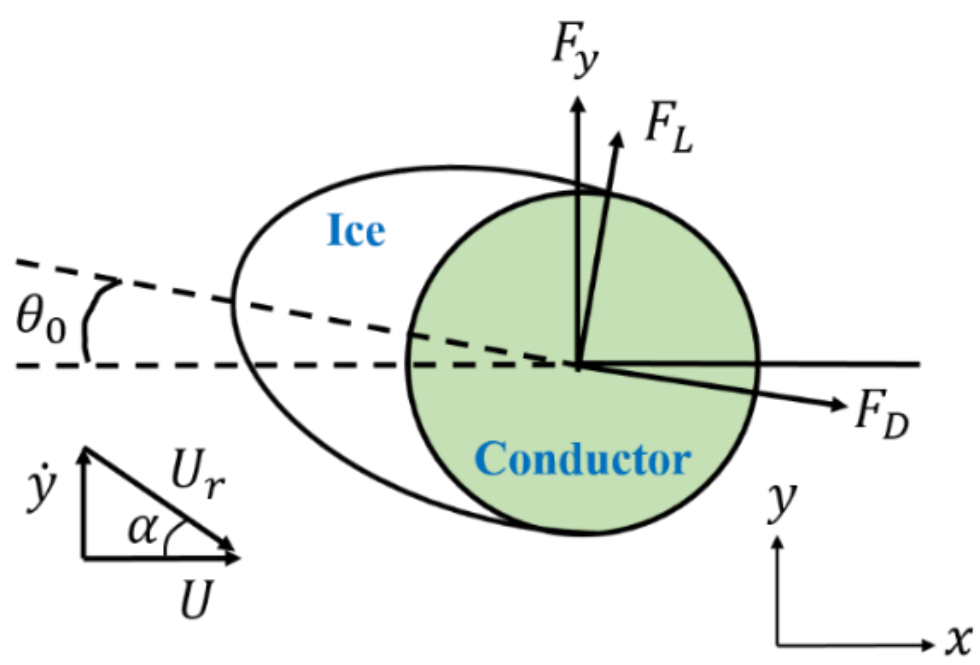

Figure 1.7. Iced conductor showing various forces involved in galloping phenomenon.

Galloping is usually considered as a destructive phenomenon in engineering structures. Wind-induced galloping of buildings, bridges, cables and traffic signs, transmission lines, etc. has been reported extensively. The most significant occurrence of galloping was observed in 1940, which leads to the catastrophic oscillation and collapse of Tacoma Narrow Bridge in the United States. Similarly, the conductor galloping is frequently encountered in transmission lines under the action of wind, due to the deposition of ice on the surface of wire [18], [19]. An iced conductor model, as shown in Figure 1.7, can be used to study the galloping phenomena, under the action of aerodynamic loads. A detail study on conductor galloping was first performed by Den Hartog [20], and various studies are done galloping as well as anti-galloping mechanisms afterwards.

Wind induced galloping is devastating phenomenon for engineering structures, however in small scale energy harvesting applications, it is quite beneficial. The large amplitude oscillations of the bluff body with low frequency are of great importance to extract ambient wind energy for producing significant electrical output. In this thesis, energy harvester is based on galloping phenomenon, therefore detail mathematical analysis of this phenomenon is done in mathematical modeling part, later in Chapter 2.

### 1.4.3. Flutter

Flutter is a type of aeroelastic instability phenomenon, often encountered in a flight vehicle, which occurs due to the interaction of elastic, inertial, and aerodynamic forces. Beyond a critical wind speed value, called as flutter speed, the magnitude of oscillation of the body increases with increase in the wind speed. In case of vortex-induced vibration, the wake instability generates the alternating shed vortices, also known as Karman Vortex Street, which in turn generates an oscillatory lift force coupled with structural oscillation. Thus, in this case, the wake instability and unsteady effects of flow separation dominate the oscillatory aerodynamic forces. However, in flutter, the dynamic instability does not necessarily be associated with a fluctuating wake. In case of flight vehicle, the dynamic instability may arise out of the out-of-plane motion of an aircraft panel in relation to surface fluctuating pressure.

The flutter instabilities can be employed to harvest the flow energy by making specially designed arrangements. There exist several forms of flutter, such as the crossflow flutter [21], axial flow flutter [22], and the coupled torsion and bending flutter [23]. Recent wind energy harvesting techniques based on flutter of an inverted piezoelectric flag [24], leaf shaped flexible piezoelectric materials, such as stalks and polymer film [25], and piezoelectric energy harvester based on flag flutter [26] have been developed. Numerical and experimental studies performed in flutter behaviour evidences the importance of flutter-based energy harvesting techniques.

### 1.4.4. Wake induced vibrations

Aerodynamic instabilities due to wake induced phenomena are often encountered in engineering applications. The arrangement of transmission lines where two or more wires are arranged in parallel for electricity transmission can face risk of failure due to wake induced vibrations. Other examples include twin cables in the offshore structures, arrays of heat-exchanger tubes, and twin slender chimneys, etc. The distance between the centres of the cylinders is denoted by $L$ and $D$ is the diameter of the cylinder, as shown in Figure 1.8.

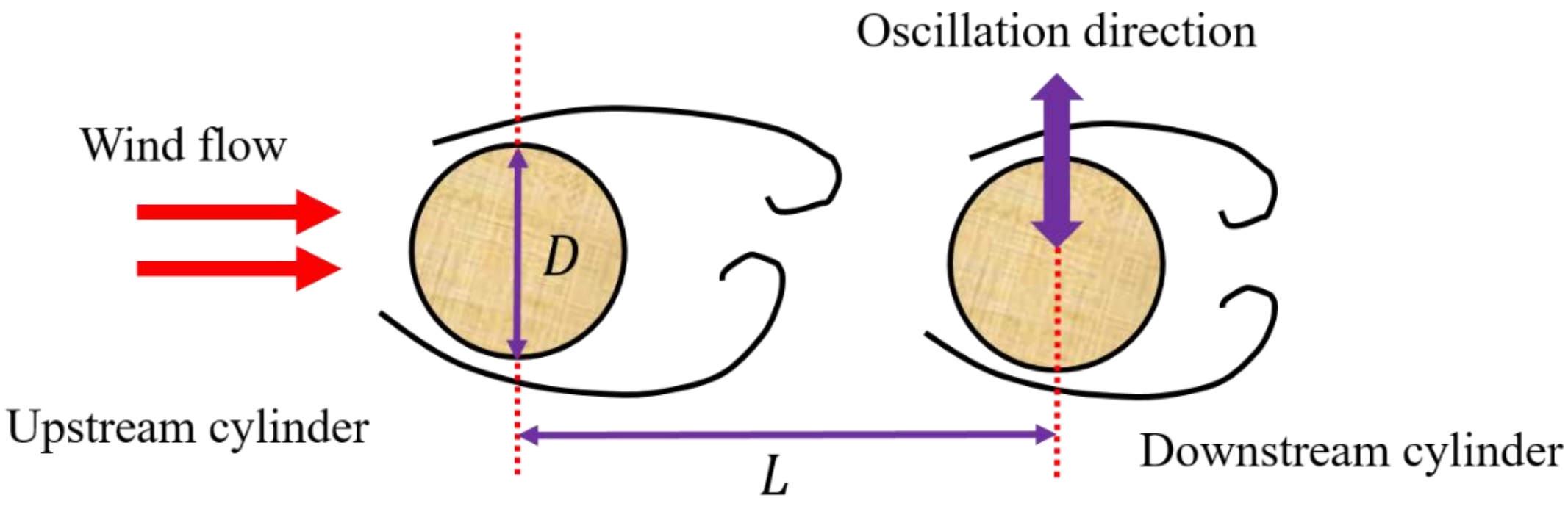

Figure 1.8. Schematic of wake galloping in parallel cylinders.

Depending on the spacing ratio, wake induced vibrations can be categorized as: interference galloping, wake galloping, and wake flutter. Interference galloping occurs when the cylinders are placed very close to each other, such that, $L/D \leq 3$. For wake galloping, the spacing ratio, $L/D$ lies between 1.5 and 6 [27]. Wake-induced flutter occurs to the twin or groups of cylinders with large gaps, $L/D = 10{\sim}20$. Wake induced oscillation is an important source of energy in wind energy harvesting system. Energy harvesting using wake galloping phenomenon [28], and an interference galloping [29] have been performed for obtaining electrical power output in wider range of wind speeds.

## 1.5. Potential Applications of Wind Energy Harvesting

The ambient wind energy can be used in various purposes to extract electrical power in micro scale to power small electronic devices. Different applications of wind energy harvesting is shown schematically in Figure 1.9. Small scale wind energy harvesters are likely to provide power to wireless sensor networks and self-powered electronic devices. Self-powered wireless sensor networks can be implemented in both indoor and outdoor atmospheres. In the indoor environment, the wind energy required to drive the harvester can be achieved from air flow in the HVAC systems [30]. In the outdoor environment, the source of wind energy is natural wind flow. Wind energy harvesters also play an important role in structural health monitoring. Small scale wireless sensor networks that are powered by wind energy are frequently in conditioning monitoring of large buildings, bridges and transmission lines. The stresses in these structures can be calculated using wireless sensors and also the cracks can be detected so that the failure of the structures can be early detected to avoid catastrophic damage. Another important application of wind energy harvesting is in environmental monitoring. Wind energy harvesters can be fixed in outdoor atmosphere such as trees and electric poles to monitor the air quality, meteorological parameters including air humidity, rainfall and to detect forest fires. The electrical power obtained from wind energy harvesters can also be used to disinfect bacteria present in water. One of the important objectives of this research is to apply the electrical power obtained from the harvester in bacterial disinfection of drinking water in water storage tanks.

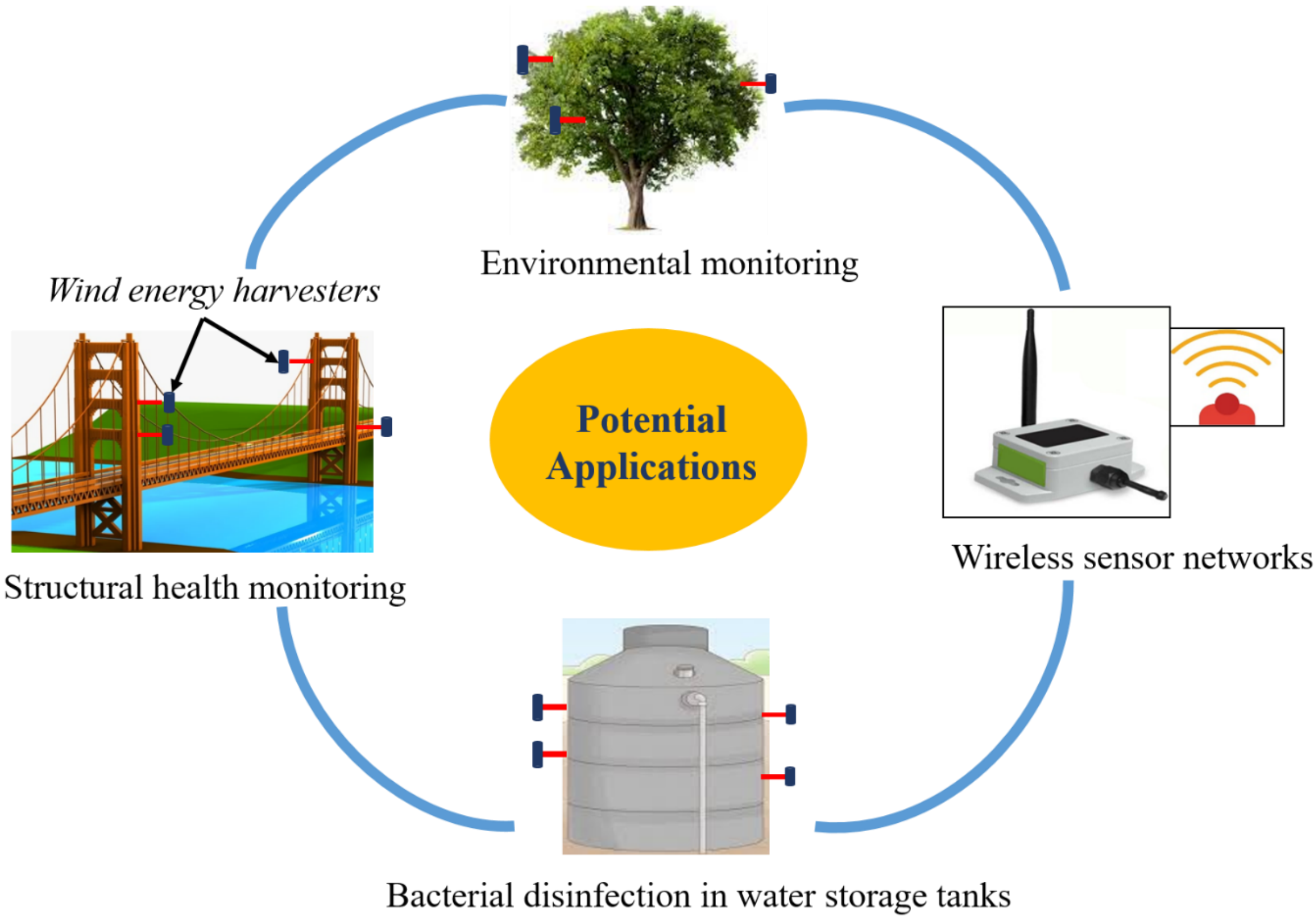

Figure 1.9. Potential applications of wind energy harvesting.

## 1.6. Literature Review

### 1.6.1 Energy harvesting techniques

Energy harvesting has become a very promising technique for supplying power to variety of self-powered micro systems, with a variety of research been carried out in the recent years. The history of energy harvesting started a long time ago, with the development of the windmill and the waterwheel for electrical power production. People have explored different methods of energy conversion and storage from heat and vibrations for many years. The important aim for researchers is to invent efficient methods to power sensors and small electronic devices without batteries.

Energy harvesting in the form of current from natural source was first witnessed in 1826, when Johann Seebeck invented the flow of current in a closed circuit made of two dissimilar metals that are maintained at different temperatures, referred as Seebeck effect [31]. The principle of generating electricity from magnetism, also known as electromagnetic induction, was discovered by Michael Faraday in 1831. Energy harvesting due to photovoltaic effect was first observed in 1839, when Edmund Becquerel was performing experiment with an electrolytic cell composed of two metal electrodes [32]. Energy harvesting in the form of charge was first observed in 1880, when Pierre and Jacques Curie introduced the concept of piezoelectricity. Piezoelectricity can be considered as the phenomenon in which certain crystals would exhibit a surface charge when subjected to mechanical stress [33]. The phenomenon where a material becomes electrically charged, when it is separated from a different material with which it was in contact is referred as triboelectric effect. Energy harvesting using triboelectric nanogenerators only gained importance in 2012, when Wang and co-workers published their research work on triboelectric effect [34]. Pyroelectric energy harvesting technique that is based on pyroelectric effect can produce an AC electrical output from the temperature fluctuations in pyroelectric materials [35].

### 1.6.2. Small scale wind energy harvesting

The energy harvesting devices are constructed in such a way that they convert the wind flow into vibrations. Wind energy harvesters can be categorized as rotational and aeroelastic harvesters, based on the mechanism involved for trapping wind flow vibrations [36]. Zakaria et al. [37] developed a swirl type centimetre scale micro wind turbine with improved power density and efficiency at wider range of wind speeds. This centimetre scale wind turbine model is very efficient in incorporating various losses present in the system. Apart from electromagnetism, the wind energy harvesters based on rotation mechanism also employ the piezoelectric and electrostatic transduction principle. Priya et al. [38] presented a small-scale wind energy harvesting technique based on direct piezoelectric effect with miniaturized rotating traditional windmill. The proposed device consisted of 12 piezoelectric cantilevers attached to a small rotating fan, and the wind flow on the device caused oscillatory stress on the cantilevers due to rotation of the fan. Another rotational piezoelectric wind energy harvester, based on impact-based resonance was proposed by Yang et al [39]. The device consisted of 12 micro-cantilevers attached on the fan and this harvester was able to enhance the output power, with effective excitation of the vibration modes. In case of rotational wind energy harvesters, strong spin of the active part of the harvester is possible. But this mechanism requires a relatively larger space for getting sufficient rotation. In other words, this mechanism is not suitable for portability. Therefore, wind energy harvester for micro-electromechanical system application that is entirely based on rotation mechanism has not been developed so far.

Wind energy harvesters that are based on aeroelastic mechanism can be broadly categorized into vortex induced vibrations and movement induced vibrations-based energy harvesters [40]. VIV based energy harvester operate due to the vortex shedding phenomenon, caused by formation of vortices in the wake of bluff body. An electromagnetic wind energy harvester based on VIV was presented by Zhu et al. [41] and the proposed harvester could produce 470 μW electrical power output at a wind speed of 2.5 m/s. However, the difficulty associated with portability and the requirement of high cut-in wind speed was a major issue. Wind energy harvesters based on VIV phenomenon can also use piezoelectric transduction mechanism to convert the oscillations produced into useful electrical energy [42]. Usually, in this type of wind energy harvesters, a cantilever beam is fixed on one end, and the other end has a cylindrical bluff body attached onto it. Piezoelectric patch is placed in the area close to the fixed end of the cantilever. A nonlinear distributed parameter model of VIV based piezoelectric wind energy harvester consisting of a cantilever beam with a circular cylinder attached onto it was presented by Dai et al. [43]. Parametric analysis performed conclude that the cylinder's mass, length of the piezoelectric patch, and the electrical load resistance can be optimized depending upon the free stream wind velocity for achieving enhanced performance of the harvesting system. A novel self-tuning piezoelectric wind energy harvester based on VIV phenomenon was designed with a slidable bluff body configuration [44]. A self-tuning energy harvesting system presented in this harvester enhanced the overall energy efficiency with enlarged frequency lock-in range. A novel piezoelectric VIV based wind energy harvester using nonlinear magnetic forces was proposed by Zhang et al. [45]. The use of nonlinear magnetic forces was beneficial for designing efficient and broadband energy harvesting systems based on VIV phenomenon, with increase in average power output by 29%.

Aeroelastic harvesters based on movement induced vibrations can be further categorized as: flutter-based, galloping based, and flapping-leaf-based energy harvesters. These types of aeroelastic harvesters are affected directly from the wind flow force. Flutter-based wind energy harvesters generally consist of a cantilever beam with an airfoil attached onto the free end. When the air flow crosses some critical value, known as flutter speed, negative damping occurs resulting the flutter deformation [46]. Galloping based wind energy harvesters usually consist of a prismatic body, with a cross-section than circular, attached to a free end of a cantilever beam [47]. Sirohi et al. [48] developed a wind energy harvester with D-shaped bluff body that operates on galloping phenomenon. An aluminium bimorph cantilever beam was able to generate 1.14 mW of electrical power when operated at a wind speed of 4.69 m/s. Tang et al. [49] presented the model of a galloping piezoelectric wind energy harvester with an equivalent circuit representation, where the mechanical parameters and the piezoelectric coupling in the system were represented by standard electronic components.

The aeroelastic mechanism, in compared to rotation mechanism, is suitable for designing small scale wind energy harvesting devices. The required space for the device is small, since it just needs a narrow beam for oscillation with some bodies attached onto it. In addition, the fabrication for this type of mechanism is simple, without complex parts. However, the major problem associated with aeroelastic wind energy harvesters is that it requires the wind to flow in the direction to which the bluff body is placed. But wind flow in arbitrary directions so it is practically impossible to extract electrical power all the time using aeroelastic energy harvesters.

## 1.7. Research Gaps

After the thorough review of the literature related to piezoelectric wind energy harvesters and their applications, following research gaps have been identified,

- The output of the state-of-art piezoelectric wind energy harvesters is still low, limiting the range of their applications.

- Limited studies have been carried out on the concept of modification of the bluff body using different shaped attachments to enhance the performance of the piezoelectric wind energy harvesters.
- Structural design of piezoelectric wind energy harvester must include fracture-fatigue studies to increase the reliability, stability, and durability of the harvester.
- The applications of previously studied piezoelectric wind energy harvesters were limited to power small electronic devices and wireless sensor networks.
- Improving the design of electrical circuits and managing rectification and storage losses is essential.
- Pyro-piezoelectric wind energy harvesters are yet not attempted.

## 1.8. Objectives of the Thesis

The aim of this thesis work is to develop a piezoelectric wind energy harvester for bacterial disinfection of a drinking water. The research work presented herein is intended to achieve the following objectives,

- To develop a mathematical model of the galloping piezoelectric wind harvester.
- To perform the parametric analysis in order to study the behaviour of different parameters on the harvested power.
- To perform bacterial disinfection of water using the proposed harvester.
- To validate the numerical results with experimental results.

## 1.8. Organization of the Thesis

The thesis is divided into five chapters. The chapter wise breakdown of the thesis is as follows,

**Chapter 1: Introduction**

This chapter discusses the basic concept of piezoelectricity and aerodynamics of a bluff body. Later, a brief discussion of wind induced vibration, and the applications of wind energy harvesting is presented. A detailed literature review on piezoelectric wind energy harvesters is presented afterward. Finally, the objectives of the thesis, which are based on the literature review are defined.

**Chapter 2: Mathematical modeling of a piezoelectric wind energy harvester based on galloping phenomenon**

In this chapter, a lumped parameter model of a galloping piezoelectric wind energy harvester is presented based on some assumptions and principles. Based on the shape of the bluff body, aerodynamic modeling is carried out to predict the wind force acting on the energy harvester. Aerodynamic modeling is based on the empirical coefficients that define the force coefficients for different shaped bluff bodies.

**Chapter 3: Comparison of an output of piezoelectric wind energy harvester with different shaped attachments on the bluff body**

A comparison study is carried out between the harvesters with different shaped modifications made to the bluff body. Simulation of the pressure field and the velocity field variation around the different shaped bluff bodies is performed. A theoretical model of the energy harvester is validated by performing a set of experiments.

**Chapter 4: Bacterial disinfection of drinking water**

In this chapter, the electrical output of the piezoelectric wind energy harvester with curve-shaped attachments is utilized to perform the bacterial disinfection of drinking water. The electrode that supplies the electrical energy to the bacterial water is modified with nanowires in order to enhance the electric field that is required to kill the bacteria. Finally, the bacterial log activation efficiency is calculated to demonstrate the effectiveness of disinfection procedure.

**Chapter 5: Conclusions and future scope**

In this chapter, the key conclusions are discussed that are based on the studies conducted in this thesis work. Finally, some suggestions regarding the future scope and application for the present work are provided.

The outline of the thesis is shown in Figure 1.10.

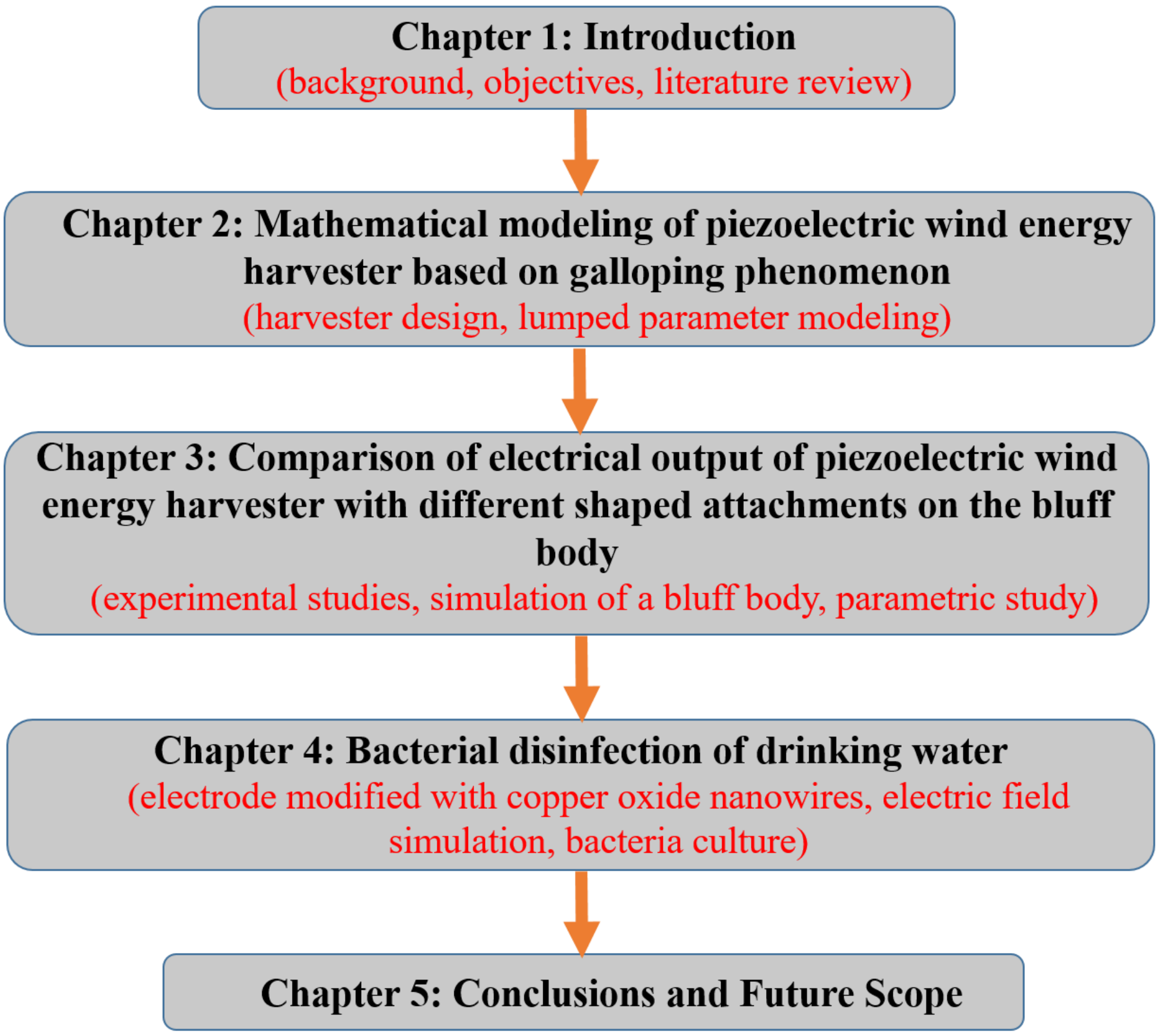

Figure 1.10. Thesis outline.

Note: Some chapters of this thesis are based on research findings previously published in the journals, *energies* and *Global Challenges*.

# CHAPTER 2: Mathematical modeling of a piezoelectric wind energy harvester based on galloping phenomenon

In this chapter, a lumped parameter mathematical model of a piezoelectric wind energy harvester is presented. An electromechanical modeling of the piezoelectric wind energy harvester is developed to predict the electrical power based on the galloping phenomenon. An aerodynamic modeling of the harvester is performed in order to model the wind force acting on the energy harvester. Finally, a set of coupled governing equations is derived, which will be used for the numerical studies conducted in the thesis.

## 2.1. Electromechanical Modeling

Piezoelectric wind energy harvester that operates on galloping phenomenon can be modeled using the lumped parameter model, Euler-Bernoulli distributed parameter model, and the finite element model. Zhou et al. [50] performed the modeling of a galloping based piezoelectric wind energy harvester with a lumped parameter model. A distributed parameter electromechanical model based on Euler-Bernoulli beam theory has been presented by Abdelkefi et al. [51]. Dash et al. [52] presented a finite element model of the galloping based piezoelectric energy harvester which is able to explain the complex nonlinear dynamics of the harvester. In this thesis, a lumped parameter model is considered to develop a mathematical model that eliminates the complexities associated with the fluid structure interaction and electromechanical coupling of the harvester.

Figure 2.1 (a) shows a piezoelectric wind energy harvester that consists of a piezoelectric beam fixed at one end, and a bluff body attached at the other end. A unimorph piezoelectric sheet is attached on the surface of the elastic beam near the fixed end, as shown in Figure 2.1 (a). The direction of the wind is perpendicular to the surface of the bluff body, as shown in the figure. When the air flows over the bluff body with an asymmetric cross section, a coupling between aerodynamic forces and the deflection of the bluff body occurs that cause the oscillating motion. The vibrational motion of the bluff body due to the galloping phenomenon is input to the system. Finally, this motion is converted into electrical energy by the piezoelectric sheets. Different shaped attachments, namely circular, square, triangular, Y-shaped, and curved-shaped are used in this research work, and are shown in Figure 2.1 (b). These different shaped attachments are used in order to compare the performance of different shaped harvesters and identify the appropriate energy harvester that produce high electrical voltage and power. For all shaped bluff bodies, the two attachments are fixed on a bluff body making an angle of $2\theta = 120^{\circ}$ with each other, as shown in Figure 2.1 (b).

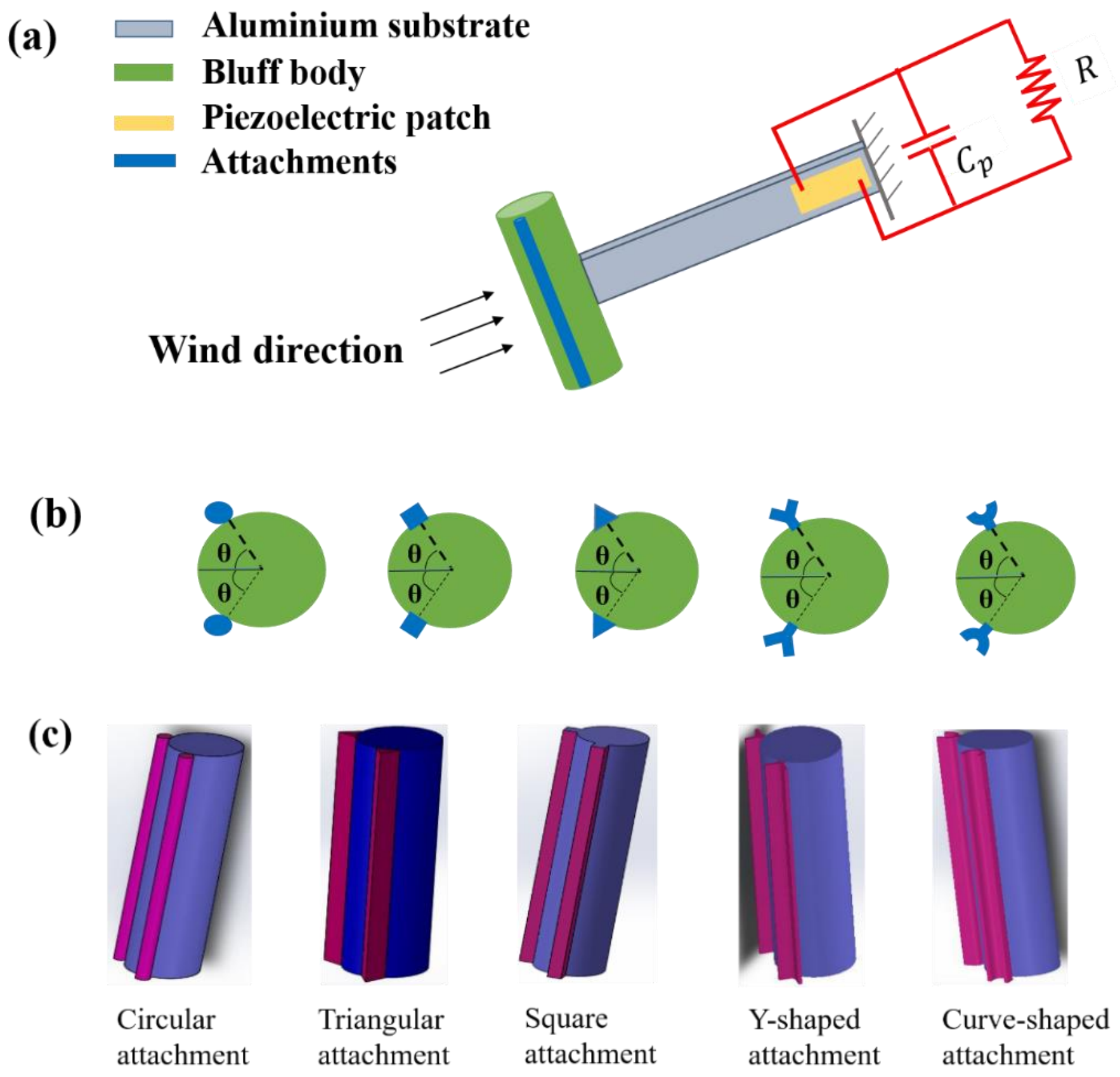

Figure 2.1. Schematic of the wind energy harvester: (a) structural diagram with circular attachments to a bluff body; (b) different shaped attachments (circular, square, triangular, Y-shaped and curve shaped); (c) 3D view of different shaped attachments to a bluff body.

In order to develop a mathematical model, following assumptions are assumed,

- Air flow is uniform and quasi-steady hypothesis is assumed.
- Energy harvester always faces the air flow.
- Unimorph piezoelectric layer is considered.
- The piezoelectric and substructure layers are assumed to be perfectly bonded to each other.
- Piezoelectric material is working under the $d_{31}$ mode.
- A resistive electrical load (R) is considered in the circuit along with the internal capacitances of the piezoelectric layer.
- Lumped parameter model is considered to model the beam.

The energy harvester consists of both the mechanical and electrical system, thus a complete system can be split into the mechanical and electrical system. Further, the mechanical system (beam) can be modeled using a lumped mass approach, as shown in Figure 2.2(a).

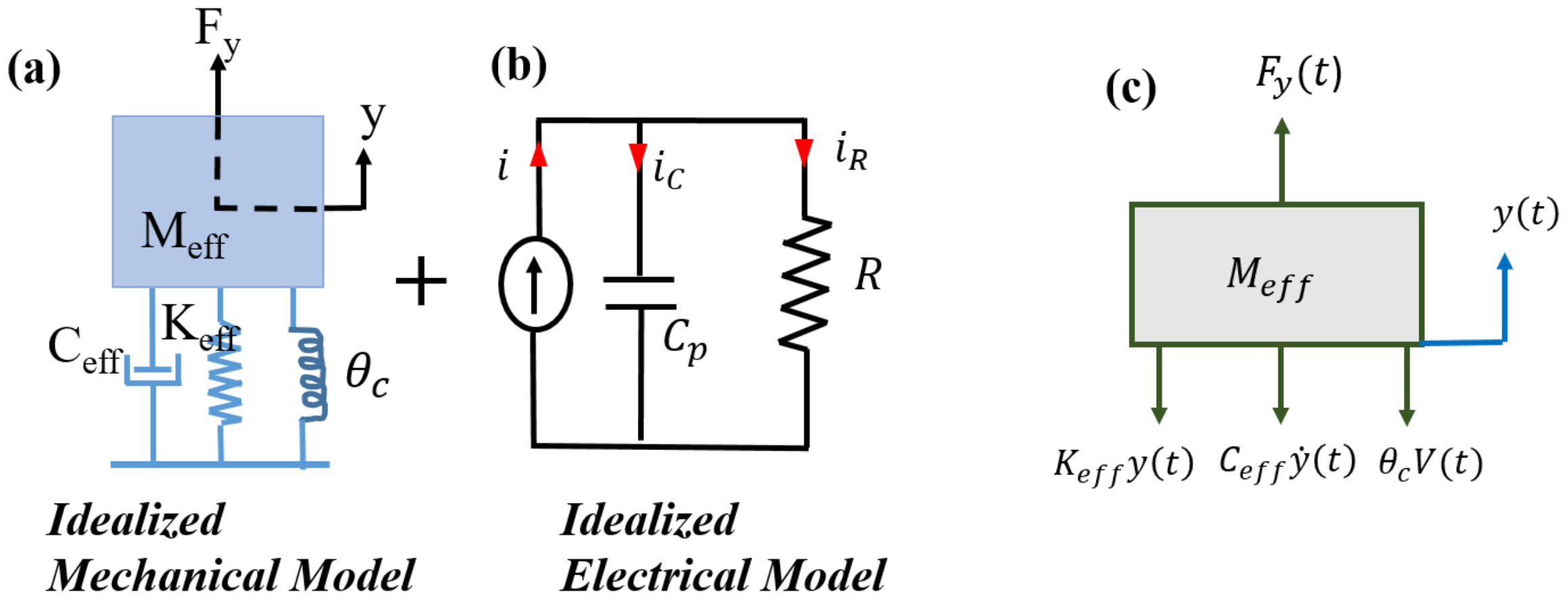

Figure 2.2. (a) Idealized mechanical model of a wind energy harvester; (b) idealized electrical model; (c) free body diagram of a mechanical model.

Using the force balance concept, which states the force applied on the energy harvester is equal to the resistive force offered by the system. The free body diagram of the forces involved in the energy harvester is as shown in Figure 2.2 (c). Thus, we can obtain,

$$M_{eff}\ddot{y}(t) + C_{eff}\dot{y}(t) + K_{eff}y(t) + \theta_c V(t) = F_y(t) \quad (1)$$

Thus, the equivalent mass, $M_{eff} = \left(\frac{33}{140}\right)m_1 + m_2 + m_3$, where, $m_1$, $m_2$ and $m_3$ are the masses representing the piezoelectric beam, bluff body and the two attachments fixed to the bluff body, respectively. The equivalent damping $C_{eff} = 2\xi w_n M_{eff}$ and equivalent stiffness $K_{eff} = w_n^2 M_{eff}$ are determined through experiments. A logarithmic decrement technique is employed to obtain the damping ratio, $\xi$. Similarly, a free decay test is used to calculate the natural frequency, $w_n$ of the system. Here, $y(t)$ is the linear displacement of the bluff body in the direction perpendicular to that of wind flow. The output parameter of the harvester is the voltage $V(t)$, which is measured across the load resistance, $R$. The parameter, $\theta_c$ is called as electromechanical coupling coefficient.

Now, applying Kirchhoff's current law in Figure 2.2 (b), we can obtain,

$$i = i_R + i_C$$

$$\theta_c \dot{y}(t) = C_p \dot{V}(t) + \frac{1}{R}V(t)$$

$$C_p\dot{V}(t) + \frac{1}{R}V(t) - \theta_c\dot{y}(t) = 0 \tag{2}$$

To visualize the coupling effect, the mechanical system can be converted into the electrical system using the force-voltage analogy, as shown in Figure 2.3.

**Force voltage analogy**

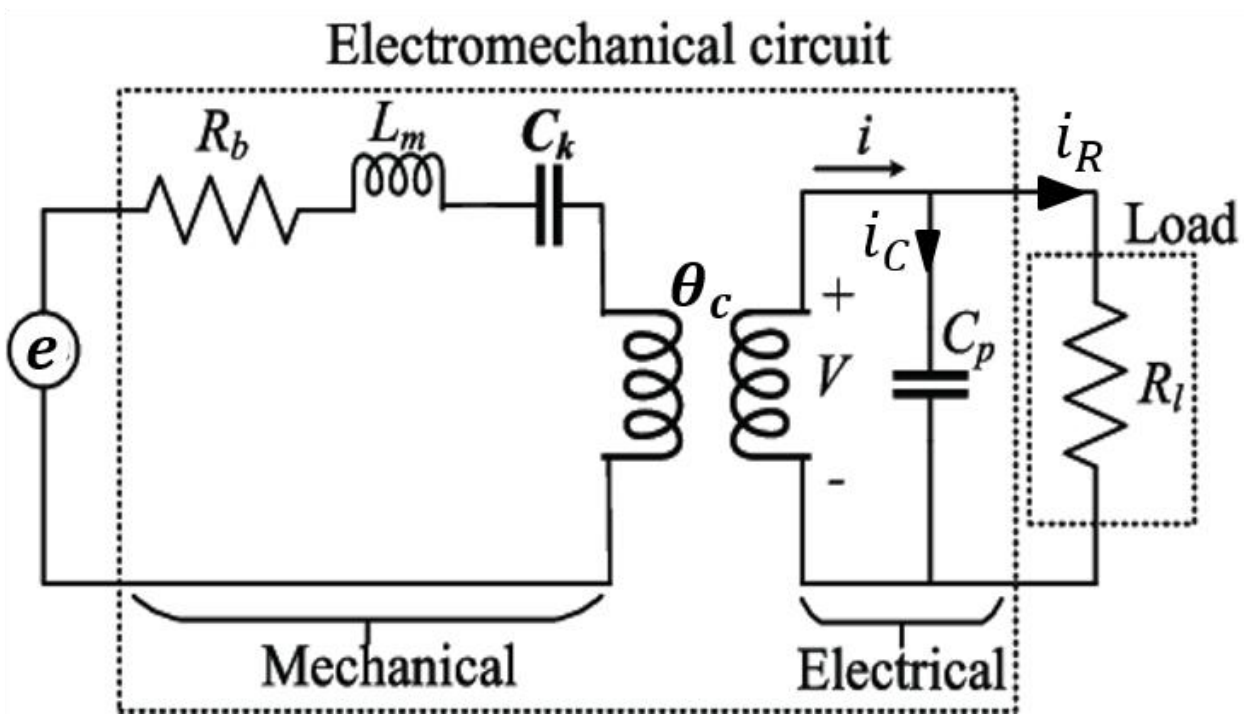

Figure 2.3. Equivalent electrical system the piezoelectric wind energy harvester.

The analogy between the electrical and mechanical parameters of the system can be defined as,

$$e = F_y, \quad L_m = M_{eff}, \quad R_b = C_{eff}, \quad C_k = \frac{1}{K_{eff}}, \quad q = y, \quad i = v = \dot{y}$$

Using the short circuit criteria, $V = 0$ and rearranging, equation (1) becomes,

$$\ddot{y}(t) + \frac{C_{eff}}{M_{eff}}\dot{y}(t) + \frac{K_{eff}}{M_{eff}}y(t) = \frac{F_y(t)}{M_{eff}} \tag{3}$$

The natural frequency representing the above equation (3) in short circuit condition is written as,

$$(w_n)_{sc} = \sqrt{\frac{K_{eff}}{M_{eff}}} \tag{4}$$

For the open circuit condition, the current, $I = \frac{V}{R} = 0$. Substituting this into equation (2), we get,

$$C_p\dot{V}_e(t) + 0 - \theta_c\dot{y}(t) = 0 \tag{5}$$

On integrating equation (5), we can obtain,

$$V = \frac{\theta_c}{C_p}y(t) + c_1 \tag{6}$$

Substituting the voltage value in equation (1), following expression can be derived as,

$$\ddot{y}(t) + \frac{C_{eff}}{M_{eff}}\dot{y}(t) + \frac{K_{eff}}{M_{eff}}y(t) + \frac{\theta_c}{M_{eff}}\left(\frac{\theta_c}{C_p}y(t)\right) = \frac{F_y(t)}{M_{eff}} \tag{7}$$

$$\ddot{y}(t) + \frac{C_{eff}}{M_{eff}}\dot{y}(t) + \left(\frac{K_{eff}}{M_{eff}} + \frac{{\theta_c}^2}{M_{eff}C_p}\right)y(t) = \frac{F_y(t)}{M_{eff}} \tag{8}$$

Equation (8) can be used to define the open circuit natural frequency as,

$$(w_n)_{oc} = \sqrt{\left(\frac{K_{eff}}{M_{eff}} + \frac{{\theta_c}^2}{M_{eff}C_p}\right)} \tag{9}$$

Rearranging equation (4) and (9), the expression for the electromechanical coupling coefficient can be defined as,

$$\theta_c = \sqrt{((w_n)_{oc}^2 - (w_n)_{sc}^2)M_{eff}C_p} \tag{10}$$

The natural frequency, $(w_n)_{oc}$ defined at the open circuit condition and the natural frequency, $(w_n)_{sc}$ defined at the short circuit condition can be calculated by experimental measurements. The capacitance of the piezoelectric patch, $C_p$ is obtained from the manufacturer's formula.

By performing experiments and using the modeling formulation discussed above, the model parameters of the energy harvester are obtained and tabulated in Table 2.1.

**Table 2.1**. Parameters of the piezoelectric wind energy harvester.

| **Properties** | **Value** |
|---|---|
| Mass of cantilever beam | 2.54 g |
| Mass of bluff body | 2.52 g |
| Mass of curve-shaped attachments | 3.45 g |
| Effective mass, $M$ | 7.5 g |
| Effective damping, $C$ | 0.0059 N/(m/s) |
| Effective stiffness, $K$ | 6.8359 N/m |
| Open circuit natural frequency, $f_{oc}$ | 4.8211 Hz |
| Short circuit natural frequency, $f_{sc}$ | 4.8141 Hz |
| Capacitance of piezoelectric patch, $C_p$ | 1.3574 x $10^{-8}$ F |
| Electromechanical coupling coefficient, $\theta_c$ | 2.24 x $10^{-5}$ N/V |
| Damping ratio, $\xi$ | 0.013 |

## 2.2. Aerodynamic Modeling

The aerodynamic force, $F_y(t)$ that acts on the bluff body due to galloping phenomenon is given as [53], [54],

$$F_y(t) = \frac{1}{2}\rho U^2 dh C_{F_y} \tag{11}$$

Where, $d$ and $h$ are the frontal dimensions of the bluff body facing the direction of wind, $\rho$ is the density of the air and $U$ represents the speed of the wind. $C_{F_y}$ denotes the coefficient of the aerodynamic force in y- direction. The aerodynamic force coefficient is an important parameter in design of the energy harvester and depends on the shape of a bluff body. The galloping phenomenon that is responsible for the motion of the bluff body can be explained by considering the force acting on a body, other than circular cross-section as shown in Figure 2.4.

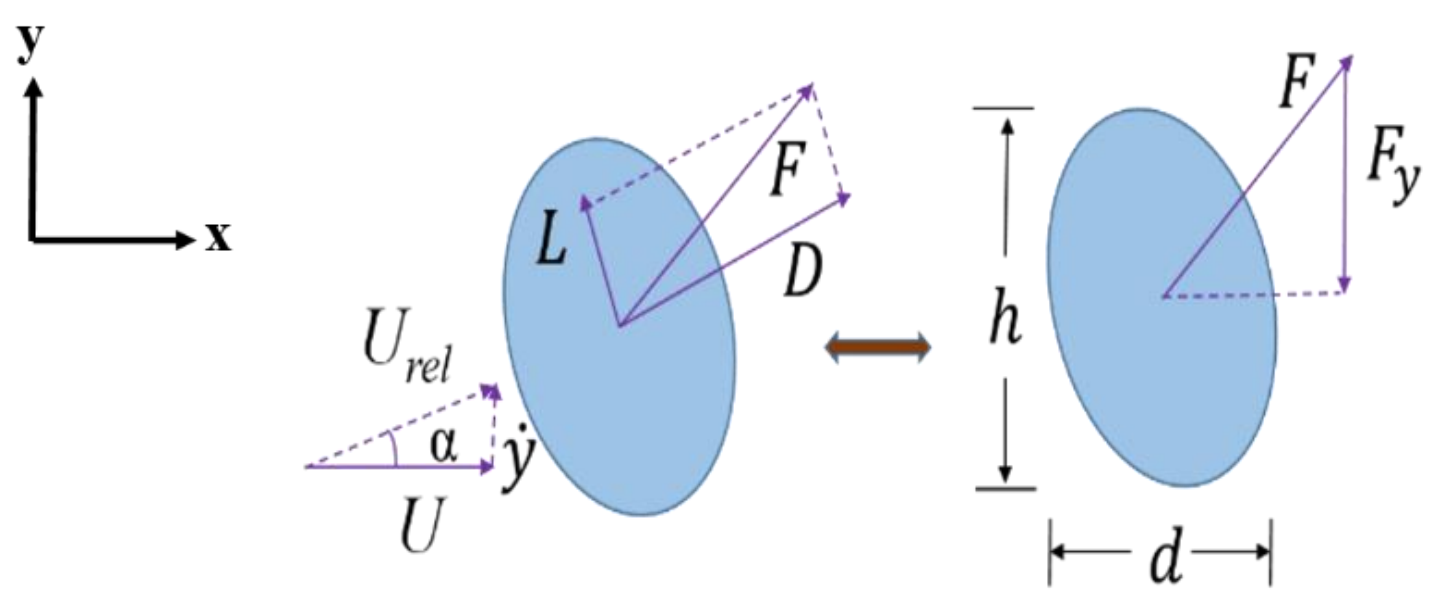

Figure 2.4. A bluff body subjected to galloping.

The force $F_y$ acting on a bluff body is expressed in terms of lift and drag forces [53] as,

$$F_y = -(Lcos\alpha + Dsin\alpha) \tag{12}$$

Assuming a quasi-static hypothesis, the aerodynamic force acting on the oscillating body is equivalent to the force acting on a steady body, measured at an equivalent angle of attack, considering low oscillation of the body. For a body undergoing only translational vibration motion, without any rotational motion, the angle of attack $\alpha$ is given as,

$$\alpha = \tan^{-1}\left(\frac{\dot{y}(t)}{U}\right) \tag{13}$$

The value of $C_{F_y}$ depends on the angle of attack and an approach to obtain its value is to express it in a cubic polynomial form as [53], [55],

$$C_{F_y} = \left(a_1 \frac{\dot{y}(t)}{U} + a_3 \left(\frac{\dot{y}(t)}{U}\right)^3\right) \tag{14}$$

The empirical coefficients $a_1$ and $a_3$ are achieved by curve fitting of $C_{F_y}$ versus $\alpha$ curve. The plot of $C_{F_y}$ versus $\alpha$ curve is obtained experimentally in a static test with varying angle of attack. Den Hartog [20] explained the instability of the galloping phenomenon, which can be expressed as,

$$\frac{\partial C_L}{\partial \alpha} + C_D < 0 \tag{15}$$

Where, $C_L = 2L/(\rho U^2 hd)$ is the lift coefficient, $C_D = 2D/(\rho U^2 hd)$ is the drag coefficient. $L$ and $D$ are the lift and drag forces acting on a body respectively. Equation (15) indicates that for a specified orientation of a bluff body with a small oscillation and a small change in angle of attack, the galloping instability of the body requires negative slope of the lift coefficient [56].

If we consider the rotation effect of the bluff body, the polynomial expansion of $C_{F_y}$ can be revised as,

$$C_{F_y} = \left(a_1\left(\frac{\dot{y}(t)}{U} + y'(t)\right) + a_3\left(\left(\frac{\dot{y}(t)}{U}\right) + y'(t)\right)^3\right) \tag{16}$$

Where, $y'(t) = \mu y(t)$ and, $\mu$ is the coefficient that relates the transverse displacement and rotation at the free end of the cantilever beam and is given as, $\mu = {}^{1.5}/_{l}$

The force acting on the wind energy harvester with different shaped bluff bodies is different, depending on the values of the aerodynamic force coefficient, $C_{F_y}$. The empirical coefficients, $a_1$ and $a_3$ related to the aerodynamic force coefficients are calculated experimentally and listed on Table 2.2. The complete theoretical model presented here predicts the mechanism of piezoelectric wind energy harvester under galloping phenomenon.

Table 2.2: Empirical coefficients $a_1$ and $a_3$ for different shaped bluff bodies.

| **Bluff body shapes** | Plain circular | Triangular attachments | Circular attachments | Square attachments | Y-attachments | Curve-attachments |
|---|---|---|---|---|---|---|
| $\boldsymbol{a_1}$ | 2.9 | 0.9 | 2.09 | 1.5 | 2.17 | 2.29 |
| $\boldsymbol{a_3}$ | −178 | −1628 | −1268 | −1552 | −913 | −726 |

The values of various parameters of the piezoelectric wind energy harvester are substituted in the coupled equations (1) and (2), which are solved using MATLAB Simulink to obtain the output voltage across different electrical load resistances. The average power output of the harvester is calculated using, $P = \frac{V_{rms}^2}{R}$ . The MATLAB Simulink model used to solve the governing equations is shown in Figure 2.5. In the Simulink model, the initial conditions are considered as:

$$At\ time\ (t = 0),\ displacement\ (y = 0),\ and\ velocity\ (\dot{y} = 0)$$

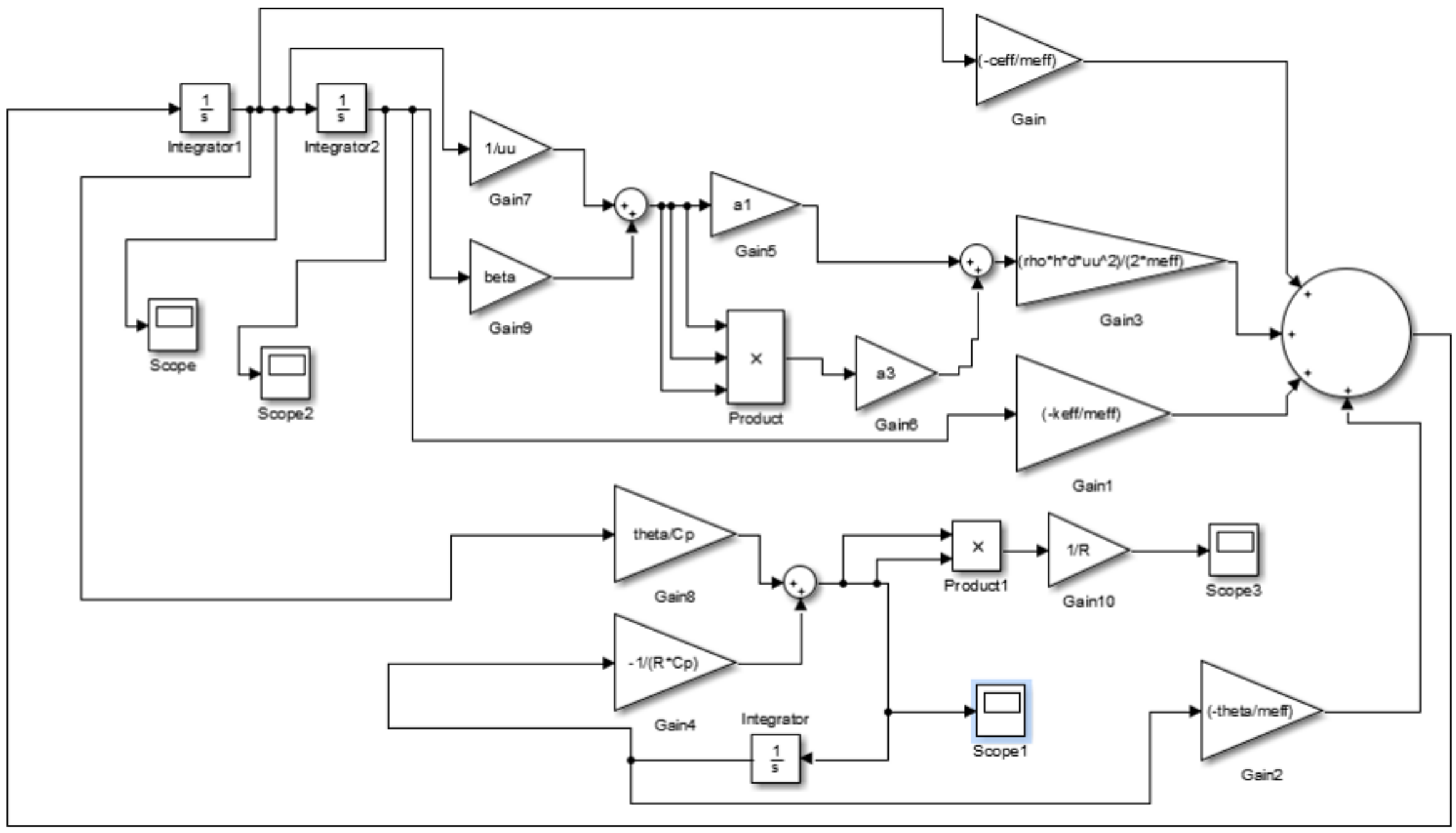

Figure 2.5. MATLAB Simulink model.

## 2.3. Conclusions

A set of coupled equations is developed in order to perform the numerical studies of the piezoelectric wind energy harvester. Various parameters, such as: effective mass, effective damping, effective stiffness, electromechanical coupling coefficient, natural frequency, etc. of the wind energy harvester are defined and also determined based on the formulation presented in this chapter. The empirical coefficients that define the force coefficients for different shaped bluff bodies under the study are determined. Based on the model developed in this chapter, the comparison of the output electrical voltage and power of different shaped harvesters will be carried out in the next chapter.

# CHAPTER 3: Comparison of an Output of a Piezoelectric Wind Energy Harvester with Different Shaped Attachments on the Bluff Body

In this chapter, a comparative analysis of different shaped attachments on a bluff body is carried out to illustrate their effect on the output voltage and output power of a piezoelectric wind energy harvester. The voltage and power produced by the harvester with several different shaped attachments is compared; these include circular, triangular, square and Y-shaped attachments.

This chapter is adapted from the author's publication [57].

## 3.1. Introduction

Harvesting wind energy from a piezoelectric cantilever beam with a buff body attached to the free end is an effective approach. A number of researchers have investigated the shape of the bluff body and proposed the optimum shapes to obtain a high output in simple wind harvesting systems. Hu et al. [58] investigated the efficiency of the harvester by attaching two small rods to the main circular cylinder and the study showed that the output voltage can be greatly influenced by adding attachments to the bluff body. A high-performance harvester with Y-shaped attachments was proposed by Wang et al. [59] and the transition of VIV to galloping phenomenon was studied. A comparison of the harvester with and without attachments revealed that a simple modification of the bluff body was sufficient to achieve optimum performance at low wind speeds. In order to achieve the optimum performance, a harvester with a bluff body that contains curve shaped attachments is presented in this chapter. A comparison of the harvester is undertaken with circular, triangular, square, and Y-shaped attachments. A theoretical model of the harvester is developed, and a variety of experiments are carried out to validate the model and undertake a parametric study of different design parameters.

## 3.2. Simulation Analysis of the Bluff Body

To demonstrate the mechanism of galloping-based wind energy harvesting using different shaped attachments, a two-dimensional model was developed. We performed simulations with the standard k-ε turbulence model in order to obtain better computational accuracy with high stability. The length and width of the computational domain used for the simulation are 100 mm and 80 mm respectively. A free triangular mesh is adopted, with three different meshing sizes, namely 65652 (coarse), 89532 (medium) and 102568 (fine). When the coarse mesh is replaced by a medium mesh, the lift and drag coefficients changes by nearly 12%. Similarly, when the resolution of the medium mesh is adjusted to fine, the coefficients of lift and drag forces is changed by less than 2%. Thus, the medium mesh resolution is chosen for our simulation analysis.

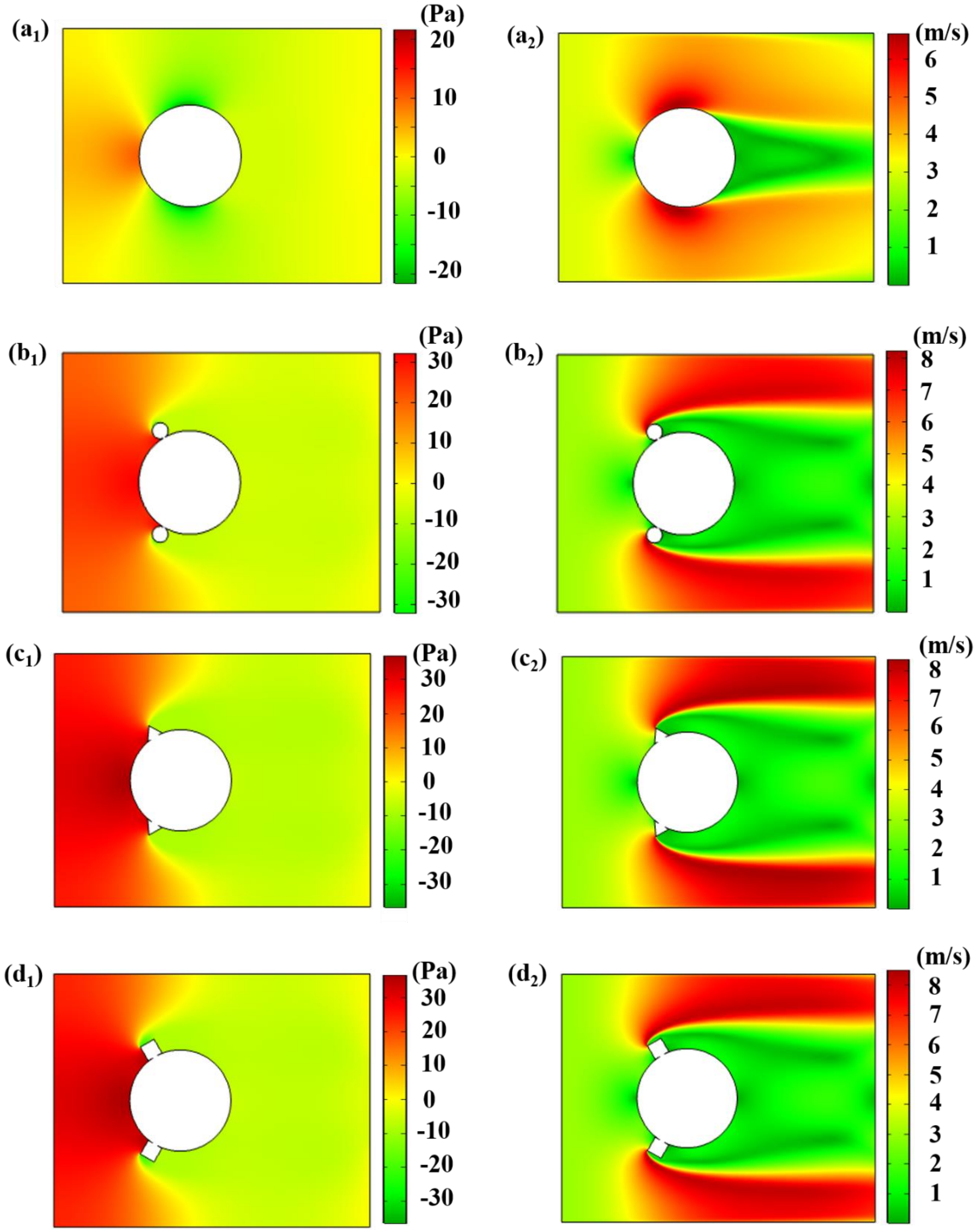

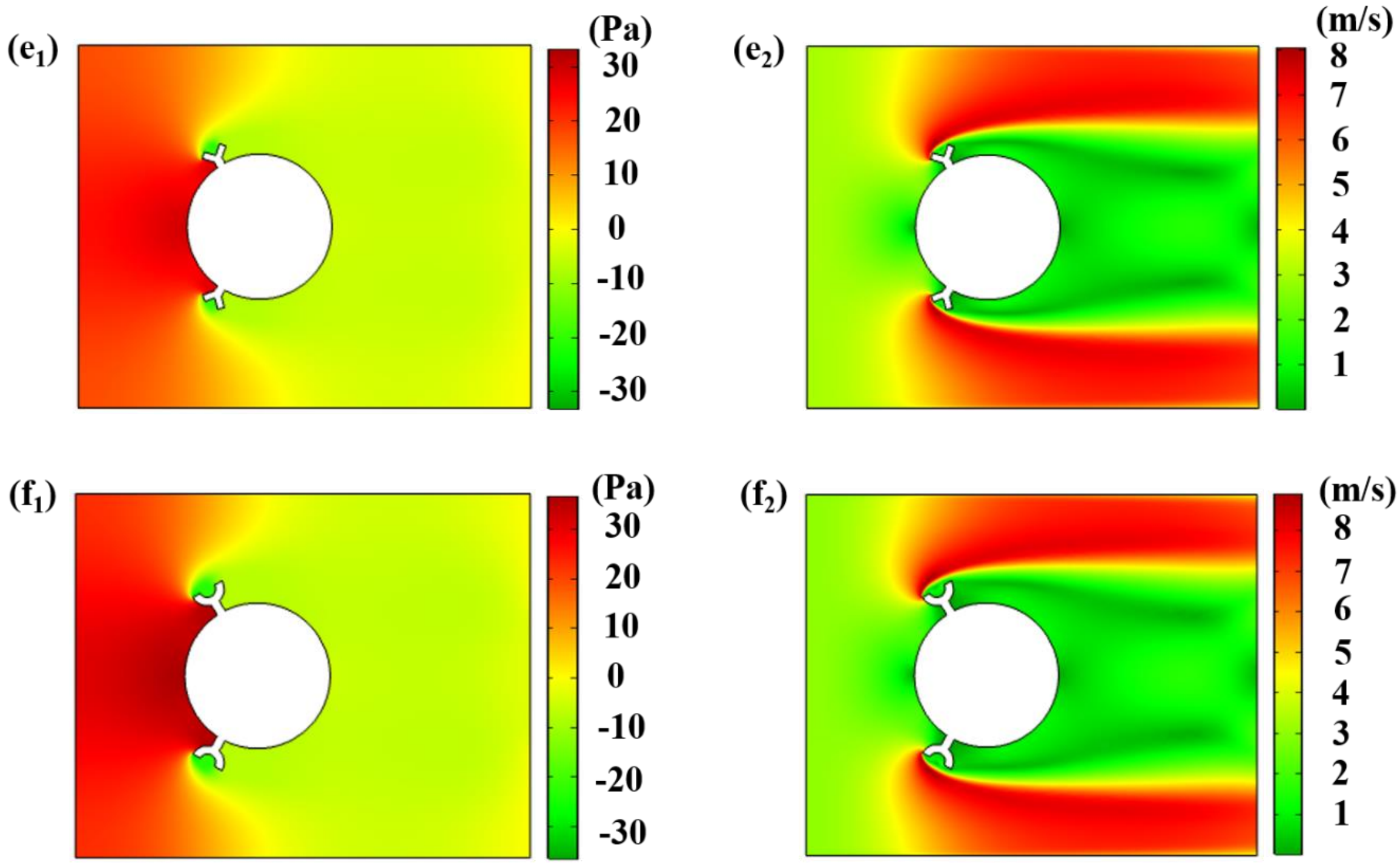

Figure 3.1. 2-D flow field simulation: ($a_1$)-($f_1$) Pressure field; ($a_2$)-($f_2$) Velocity field of different shaped attachments in a bluff body; (a) plain cylinder; (b) circular attachments; (c) triangular attachments; (d) square attachments; (e) Y-shaped attachments; (f) curve-shaped attachments.

The incoming air velocity at the inlet boundary is taken as 3 m/s and the air is supposed to flow in the direction perpendicular to the inlet domain. The pressure is considered zero at the outlet boundary of the domain and the top and bottom boundaries are considered to be fixed. The variation of the pressure field and the velocity field around the bluff body with different shaped attachements is shown in Figure 3.1. Figure 3.1 (a) shows the pressure and velocity variation when the air blows past the plain cylindrical bluff body. The pressure difference between the upstream and the downstream side of the bluff body reveals that there exists a lift force on the bluff body, which creates the transverse force component that is necessary for the galloping mechanism. Recently, Liu et.al [60] performed a CFD analysis based on COMSOL on a circular bluff body with double flat plates placed ahead of the bluff body for utilizing the wake flow produced by such plates for improved harvesting performance. Figure 3.1 ($b_1$-$f_1$) shows the pressure field variation around the bluff body with different shaped attachments, and the velocity field variation can be observed in Figure 3.1 ($b_2$-$f_2$). If we observe the pressure variation in bluff bodies with different shaped attachments, it is found that the minimum pressure occurs at the downstream of the body. The occurence of negative pressure and high lift force will make the system aerodynamically unstable, and this instability will eventually increase the amplitude of vibration of the body. The variation of velocity with maximum value at the upper and lower sides of the body, whereas the minimum value just behind the bluff body that is observed for different shaped attachments signifies that there is vibration in the bluff body.

## 3.3. Experimental Studies

In order to verify the results achieved from the theoretical model, an experimental setup was developed, and a range of experiments were performed in an open atmosphere. Figure 3.2 shows the experimental setup used for energy harvesting using different shaped attachments on the bluff bodies. A centrifugal air blower was used to supply the wind required to drive the harvester. The speed of the wind was measured using digital anemometer, with an accuracy of approximately ± 0.5 m/s. The required speed of wind is maintained by making appropriate adjustment of the blower. A PZT-5A (SP-5A, India) lead zirconate titanate piezoelectric sheet with dimensions of 50 × 20 × 0.4 $mm^3$ was placed at one end of pure aluminium beam with dimensions of 200 × 25× 0.6 $mm^3$; this a is relatively soft ferroelectric materials that has high piezoelectric activity, making it suitable for sensing and harvesting applications. A circular bluff body with a height of 120 mm and a diameter of 32 mm was made of expanded polystyrene (EPS) material with low density and the different shaped attachments required were fabricated using a 3D printer. The mass of the attachments considered in this thesis are same and the material used is polylactic acid (PLA). The length was made equal to that of the bluff body with diameter of 5 mm for circular attachments. The dimensions for triangular, square, Y-shaped, and curve-shaped attachments were chosen accordingly from the equal mass consideration. The output voltage obtained across the electric load resistance was measured using a digital oscilloscope (InfiniiVision DSO-X 3034A), with an input impendence of 10 MΩ.

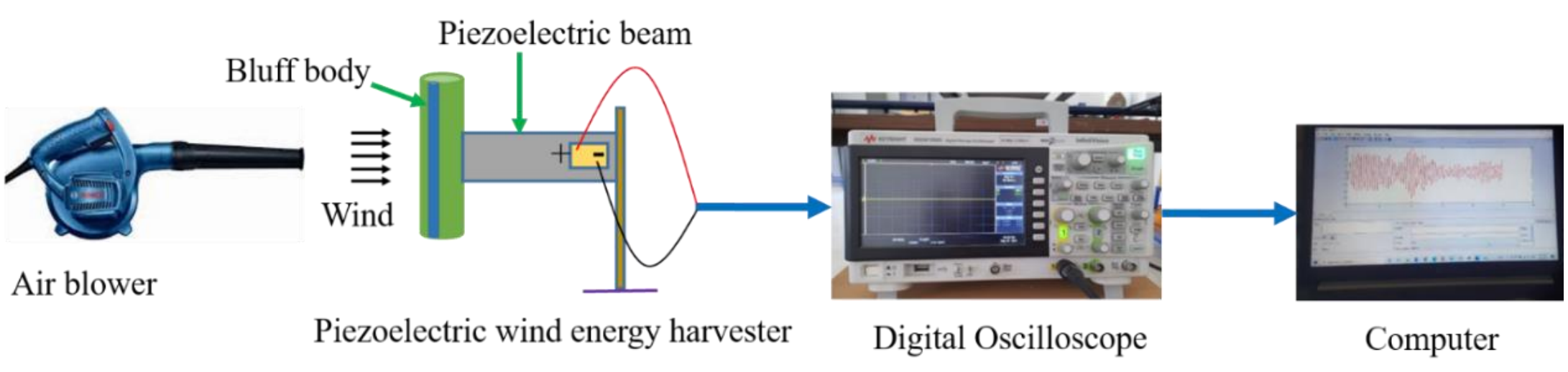

Figure 3.2. Experimental setup of piezoelectric wind energy harvester.

The experimental results obtained are analysed and compared with the simulation results. The experimental output voltage of the harvester with curved shaped attachments is found to be approximately 25 V and the simulated output voltage is about 29 V when the speed of wind is kept at 4 m/s, as shown in Figure 3.3 (a). The difference in the simulated and experimental value can be due to the assumptions used in modeling of the harvester. The modeling is based on the lumped parameter model and a quasi-steady hypothesis is assumed in the derivation of transverse force component with small angle of attack. In addition, the experiments are performed in an open environment, where it is difficult to predict the wind behaviour accurately. Figure 3.3 (b) compares the experimental and simulated output voltage of the harvester with Y-shaped attachments. Similarly, Figure 3.3 (c), 3.3 (d) and 3.3 (e) illustrate the comparisons for square, circular and triangular attachments respectively. If we compare the output voltage produced by different shaped harvesters, we can conclude that the harvester with curved shaped attachments provides the best performance, while the harvester with triangular attachments leads to the lowest output voltage and power. It can be seen that it requires a few seconds for the harvesters to produce stable output voltage for both simulation and experiment, as seen in Figure 3.3.

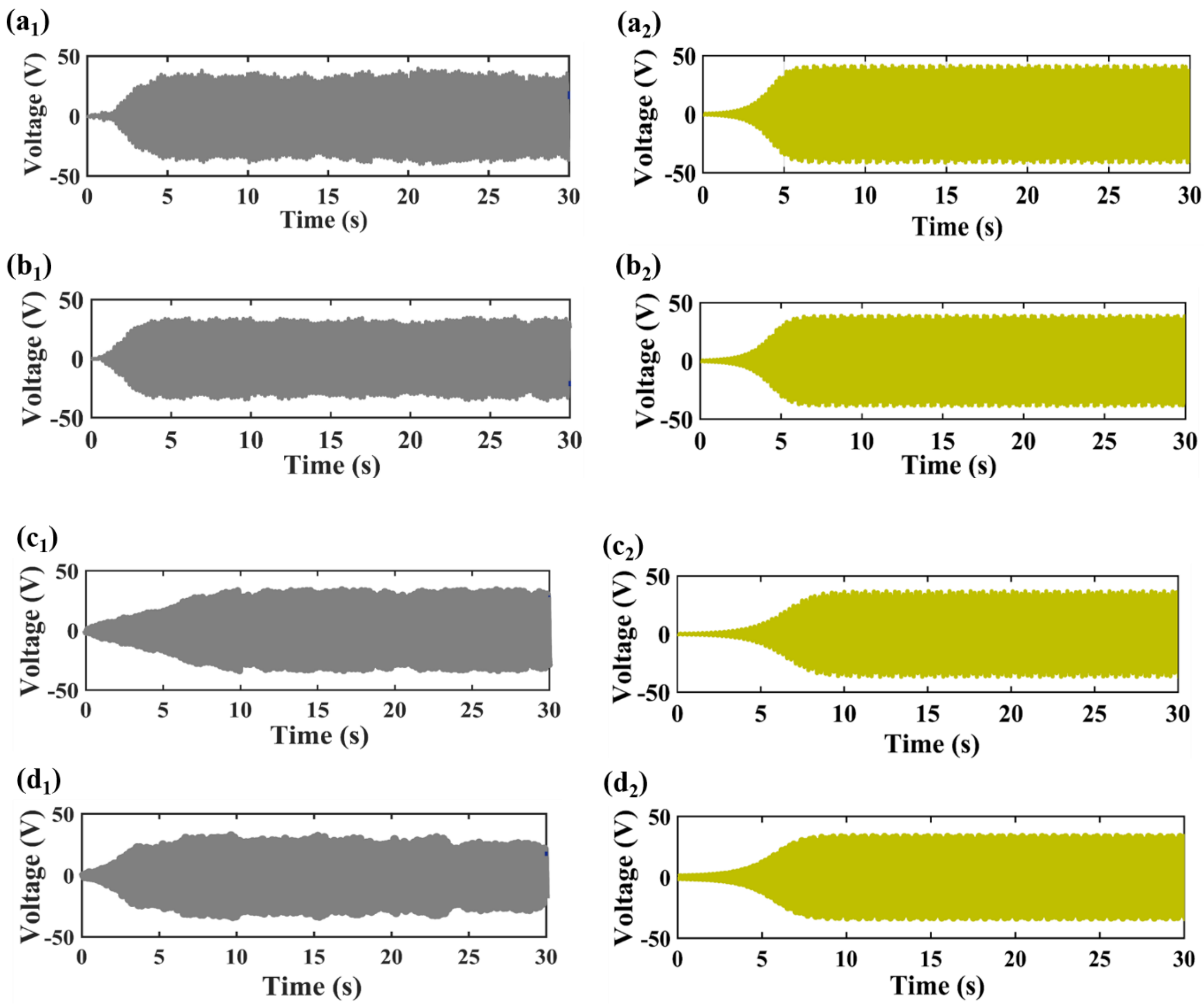

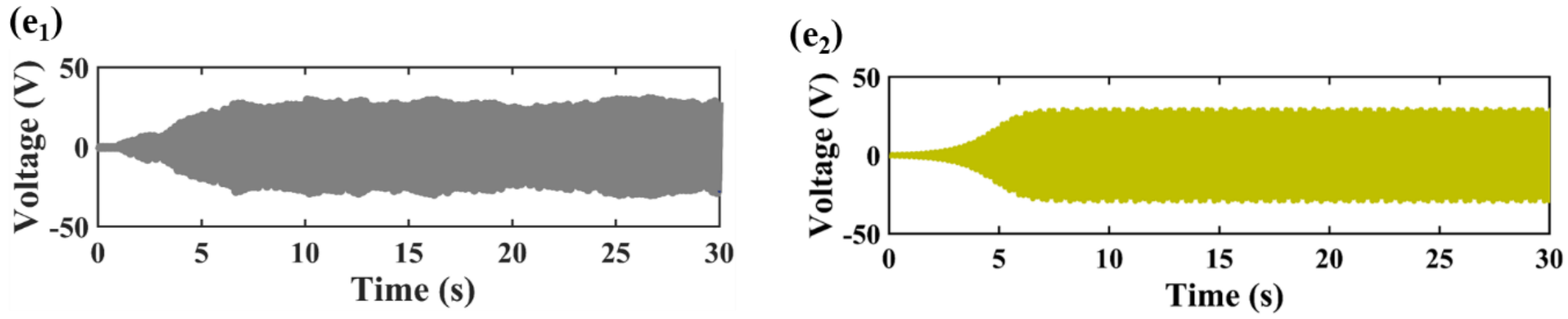

Figure 3.3. Experimental and simulation output voltage at 4 m/s wind speed: ($a_1$)-($e_1$) Experiment; ($a_2$)-($e_2$) Simulation of different shaped harvester; (a) curve-shaped attachments; (b) Y-shaped attachments; (c) square attachments; (d) circular attachments; (e) triangular attachments.

Figure 3.4 (a) illustrates the variation of output voltage with wind speed produced by the harvesters with different attachments. It can be seen that the output voltage is increased when the speed of wind is increased for the harvesters with attachments subject to galloping phenomenon. Galloping occurs when the wind speed is greater than the threshold value required for galloping and the experimentally measured threshold value is approximately 1.65 m/s. However, for the harvester without any attachments, there is vortex-induced vibration (VIV) phenomena. The maximum output voltage for the harvester with a plain circular bluff body occurs at a velocity where a lock-in region exists. In this lock-in region, the oscillating frequency is locked to the natural frequency of the harvester. Figure 3.4 (a) demonstrates that the lock-in region exists at approximately 1.2-1.5 m/s, where the harvester has maximum voltage of 6 V. It can also be observed that a post-synchronization stage exists after the lock-in region, where the output voltage starts decreasing with increasing velocity. It is important to understand that the harvester operating under vortex-induced vibration will enter the lock-in region at a velocity less than that required for galloping of the harvesters with different shaped harvesters. However, the harvesters operating under galloping will perform at a higher velocity as compared to the harvester with only a plain circular harvester. The variation of output voltage with different load resistance is shown in Figure 3.4 (b). Experiments were carried out for five different load resistances (0.1 MΩ, 0.5 MΩ, 1 MΩ, 2.5 MΩ and 5 MΩ) and the results show that the output voltage increases when we shift from 0.1 MΩ to 1 MΩ resistance, but it remains almost constant while resistance is increased from 1 MΩ to 5 MΩ. The output voltage of the harvester with curve-shaped attachments provides highest output voltage with different load resistances.

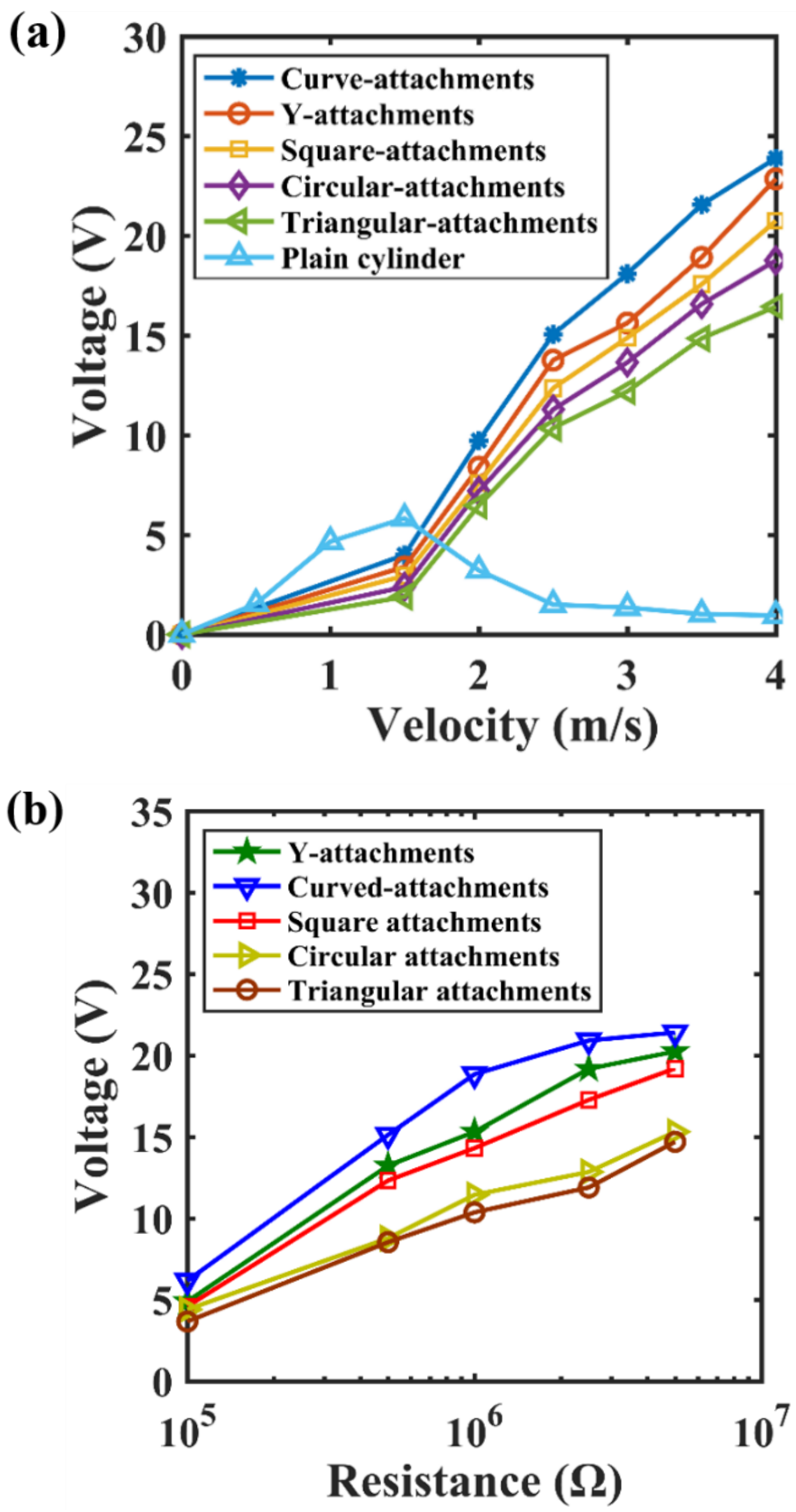

Figure 3.4. Experimental comparison of the output voltage with different attachments with: (a) Wind velocity; (b) Load resistance in log scale at 4 m/s wind speed.

The variation of the power output with wind speed and load resistance is illustrated in Figure 3.5 (a) and 3.5 (b) respectively. The power output provided by the harvesters operating at wind speeds lower than 1.5 m/s is low and beyond it, the power increases with increasing the wind speed. The power output of the harvester with curve-shaped attachments provides the maximum value of approximately 0.105 mW while operating at a wind speed of 4 m/s. Similarly, a lower power output is generated by harvester with triangular attachments and its value is 0.048 mW. By observing the variation of power output with electric load resistance, it is found that the harvester with curve-shaped attachments performs best under different load resistances. The maximum power output is obtained with curve-shaped attachments operating at 4 m/s wind speed with an electrical load resistance of 0.5 MΩ and its value is 0.46 mW as shown in Figure 3.5 (b).

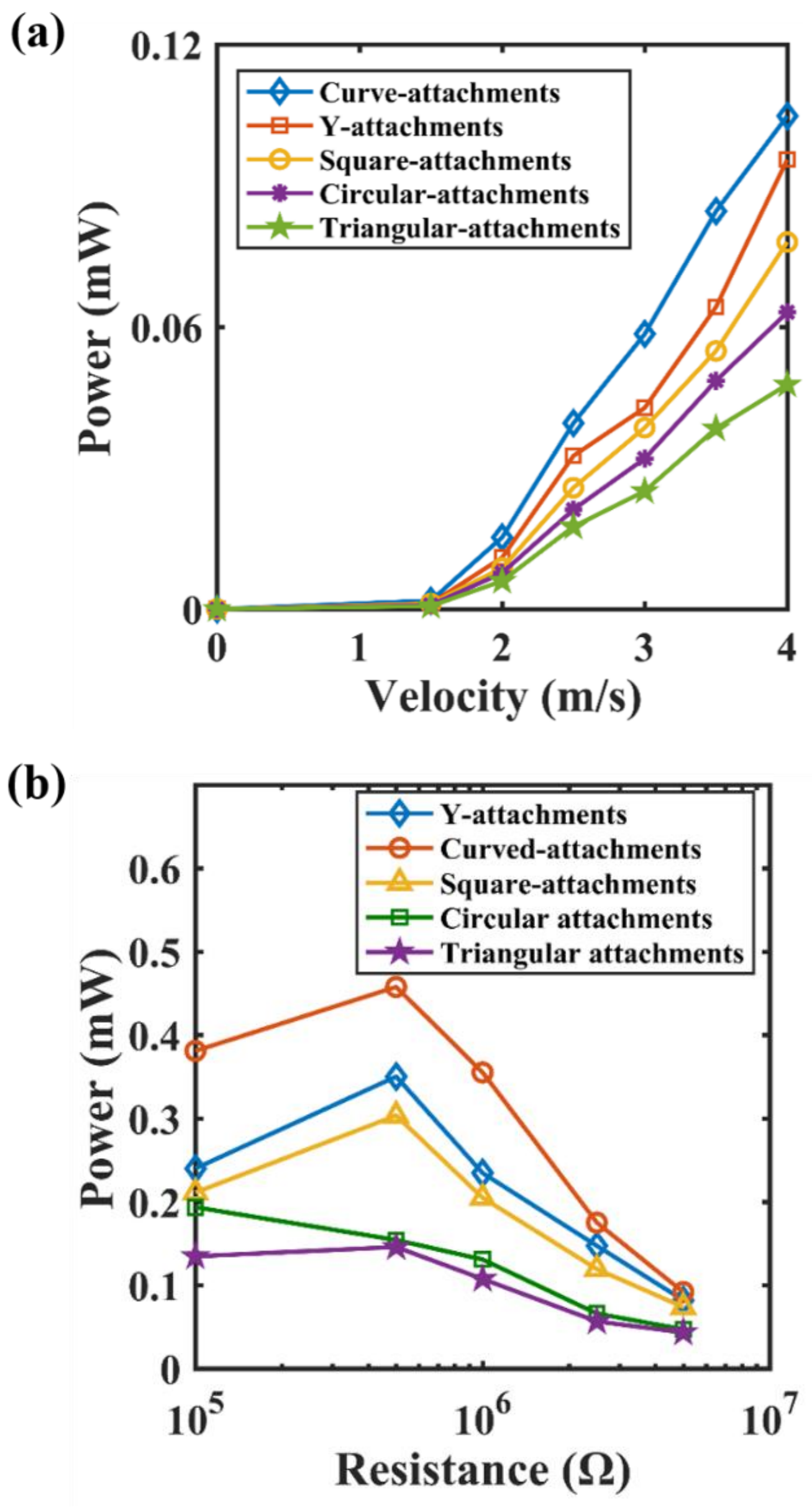

Figure 3.5. Experimental comparison of the power output of the harvester with different attachments with: (a) Wind velocity at 5 MΩ resistance; (b) Load resistance in log scale at 4 m/s wind speed.

The frequency domain diagram is plotted in Figure 3.6. The frequencies of vibration of the wind energy harvesters at different velocities are obtained using the Fast Fourier Transform (FFT) method. Figure 3.6 (a) and 3.6 (b) show the frequency of vibration of the harvester with a curve-shaped attachments, at the wind speeds of 1.5 m/s and 3.5 m/s and are found to be 4.05 Hz and 4.75 Hz, respectively. Similarly, the frequency of vibration of the harvester with a cylindrical shaped bluff body, at the wind speeds of 1.5 m/s and 3.5 m/s and are found to be 4.72 Hz and 6.06 Hz, respectively, as shown in Figure 3.6 (c) and 3.6 (d). The plot of frequencies of the two different harvesters consisting of a plain cylindrical shaped bluff body and curve-shaped attachments to the bluff body, at different wind speeds is shown in Figure 3.6 (e). The natural frequency of vibration for both shaped harvesters is 4.8 Hz. It can be seen that for the harvester with curve-shaped attachments, the frequency of vibration is less than the natural frequency of the system at low wind speeds. As the wind speed increases beyond 2 m/s, the frequency of vibration is close to the natural frequency of the harvester, thus providing the oscillations with higher amplitude. In case of the harvester without attachments, the frequency of vibration is close to the natural frequency at the wind speed of range 1m/s to 1.5 m/s. The frequency of the harvester is locked to the natural frequency of the system in this wind speed range, thus providing large voltage output from large amplitude oscillations. As the wind speed is increased beyond 1.5 m/s, the oscillating frequency of the harvester also increases, as shown in Figure 3.6 (e).

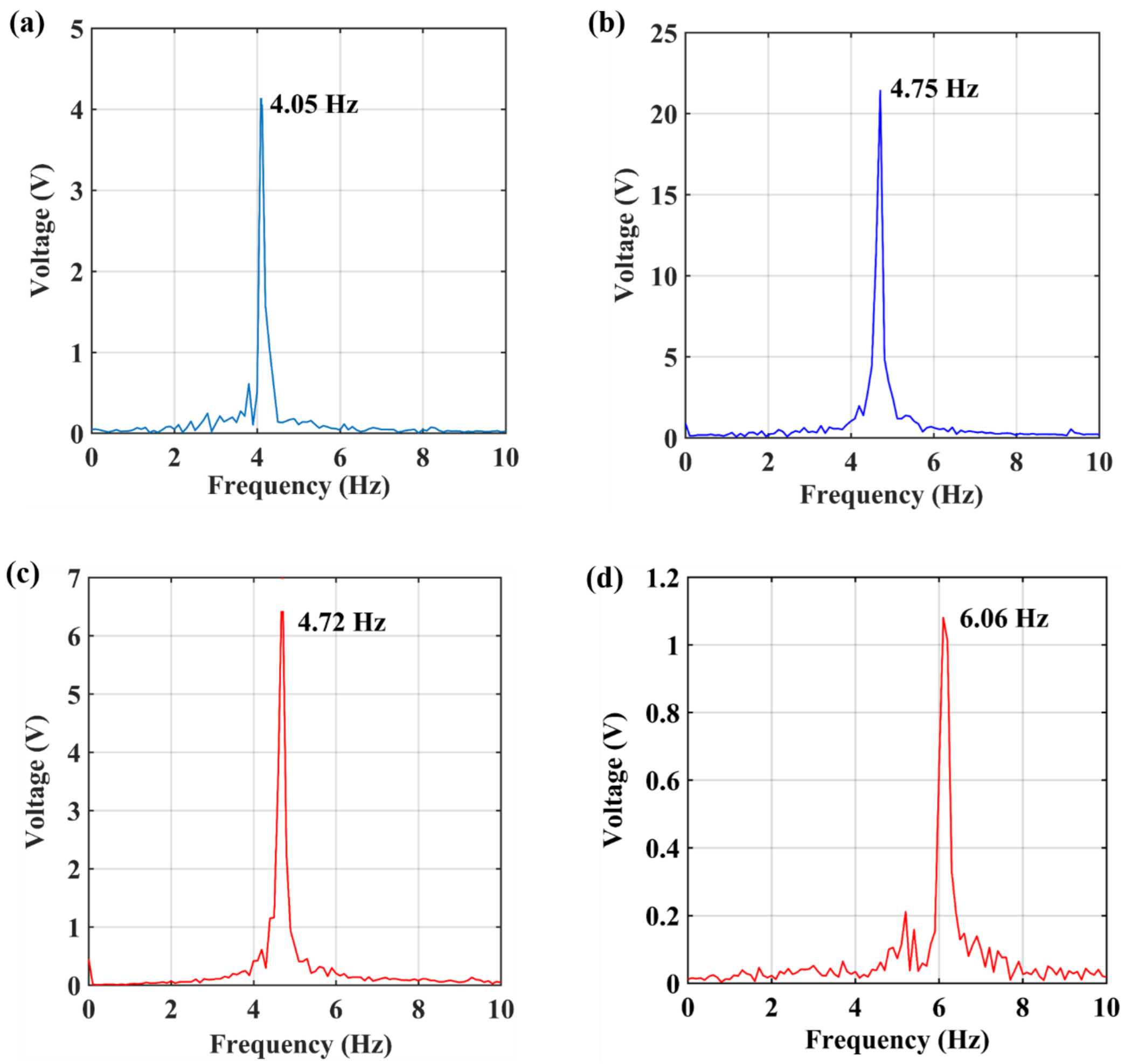

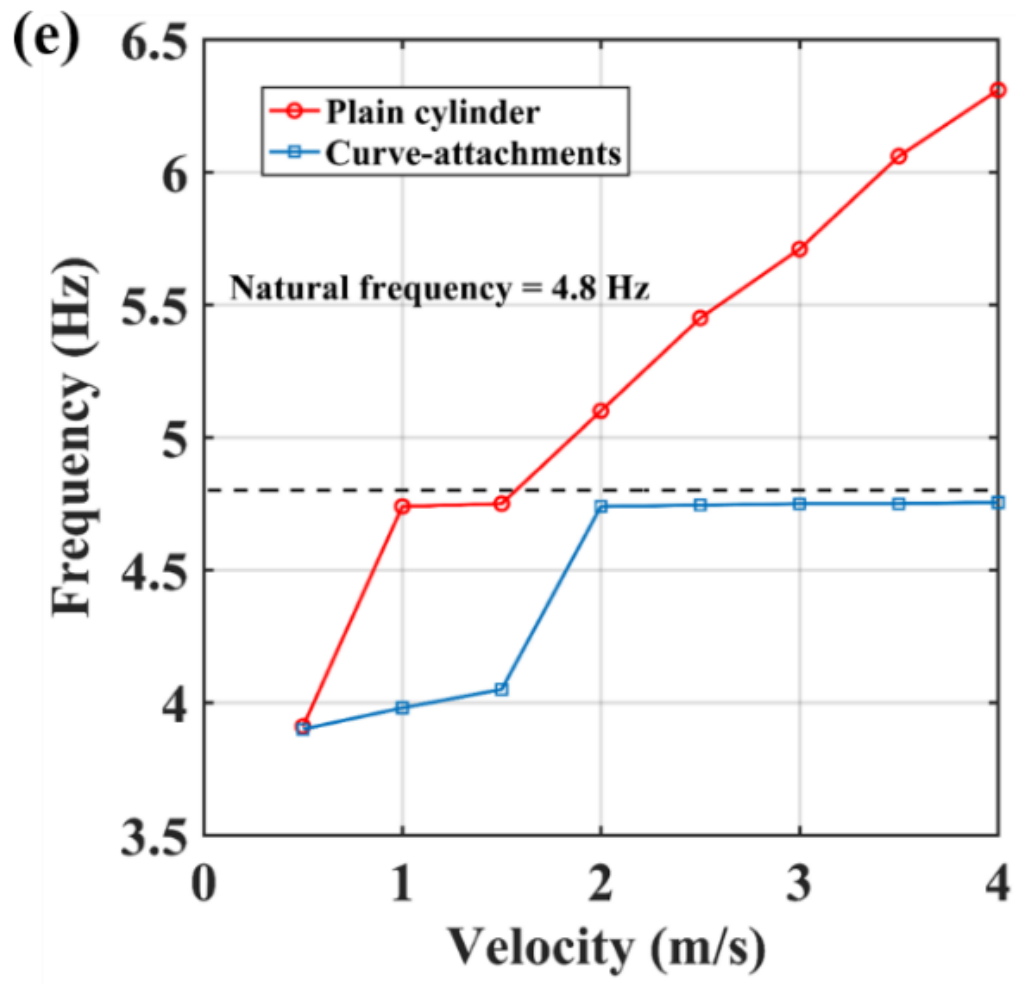

Figure 3.6. Frequency domain diagrams: (a) plain cylinder at 1.5 m/s; (b) plain cylinder at 3.5 m/s; (c) curve-attachments at 1.5 m/s; (d) curve-attachments at 3.5 m/s; (e) comparison of frequencies of plain cylinder and curve-attachments at different wind speeds.

Figure 3.7 represents the bar diagram showing the comparison of the output voltage produced by the harvesters with different shaped attachments under study. The output voltage and the output power of the piezoelectric wind energy harvester depends on the amplitude of oscillation of the bluff body attached to the beam. In case of curve-shaped attachments, the force coefficient acting on the bluff body is higher because the lift force that drives the harvester is more in compared to other shaped harvester. It can be observed that the curve-shaped attachments provide a high output voltage and power whereas, the triangular attachments provide a comparatively low output voltage and power.

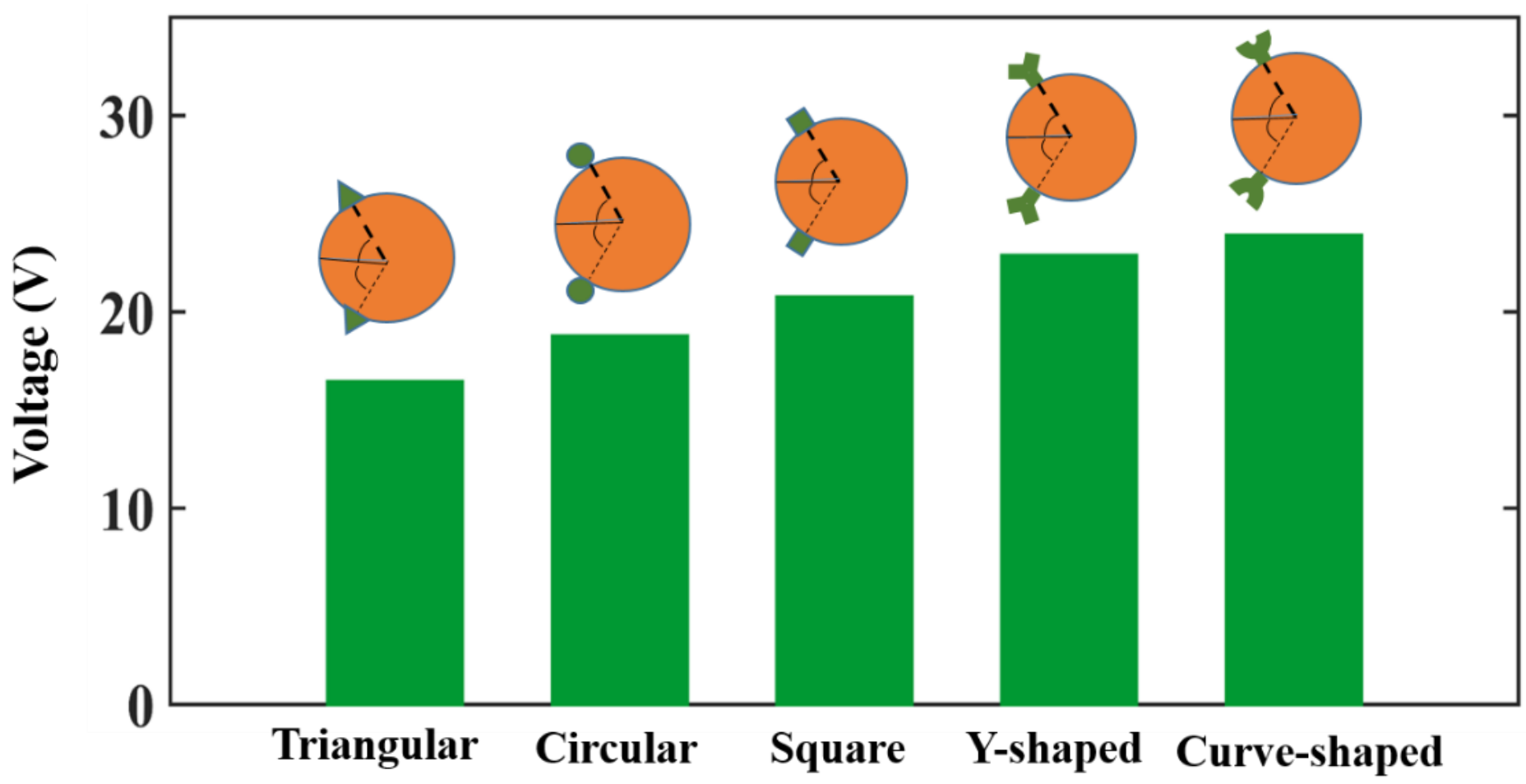

Figure 3.7. Maximum output voltage of harvesters with different shaped attachments measured at 4 m/s wind speed.

## 3.4. Conclusions

A new concept for improving the performance of the piezoelectric wind energy harvesters with different attachments on the bluff body is presented in this chapter. Simulation analysis performed on the bluff bodies with different attachments leads to a high-pressure difference on a bluff body with curved attachments compared to other attachments producing more vibration required for improved performance. The harvester with curve-shaped attachments provides an output voltage of 25 V and a power output of 0.105 mW at a wind speed of 4 m/s, which is higher than the output produced by harvesters with other shaped attachments under consideration. Similarly, the overall output produced by the harvester consisting of triangular attachments on a bluff body is the lowest, when compared with other shaped attachments. The proposed curved attachments on a bluff body lead to improved aerodynamic efficiency with design flexibility. The experimental analysis is performed in an open environment which allows good potential for different applications of harvested power as it is not always possible to harvest wind power in a wind tunnel.

In the next chapter, a bacterial disinfection of the drinking water will be presented utilizing the electrical output produced from the piezoelectric wind energy harvester. A piezoelectric wind energy harvester with the curve-shaped attachments on the bluff body is used for disinfection purpose since it produces high electrical output voltage as compared to the other shaped harvesters.

# CHAPTER 4: Bacterial Disinfection of Drinking Water

In this chapter, a water disinfection system based on the direct piezoelectric effect that is produced by beam vibration driven by wind is presented. The voltage produced is then supplied to the copper tube filled with bacteria-infected water. Nanowires grown on the central electrode enhances the electric field to a magnitude that is sufficient to disinfect the bacteria present in the water. The successful treatment of the infected water in a short period of time is demonstrated and the bacterial treatment achieved is found to be efficient.

This chapter is adapted from the author's publication [61].

## 4.1. Introduction

Pathogenic infections caused due to impure drinking water have become a major challenge for undeveloped as well as developing countries [62], [63], [64], [65]. Nearly 1 million people die yearly around the globe from waterborne diseases because of a lack of proper sanitation facilities [66]. Highly efficient disinfection methods with low energy consumption are required to safeguard the life of people from diseases caused by pathogenic infections. Disinfection techniques in centralized units may involve chlorination and ozonation for killing bacteria in water treatment [67]. Chlorination is adopted as an important water disinfection technique because of its high efficiency, low cost and reliable performance. However, chlorination causes the occurrence of carcinogenic disinfection by-products that pose a high risk to human health [68], [69], [70]. This problem of the formation of by-products can be resolved by the use of non-chlorine-based water disinfection techniques. Membrane filtration [71], [72] and UV disinfection [73], [74] are the alternative methods but are limited by high energy consumption, high cost and inability to produce residual antimicrobial power. The energy consumed during the water treatment process with membrane filtration technique is around 500- 5000JL$^{-1}$, whereas UV disinfection technique involves energy consumption of about 20- 60JL$^{-1}$ [75].

There is a need for water treatment systems offering low cost, high efficiency and less consumption of energy. Many studies have been performed on the water disinfection technique, that can be achieved by supplying a high strength electric field, both AC and DC, to the water [76], [77], [78]. Locally enhanced electric field treatment (LEEFT) is the phenomenon that implements the electroporation mechanism [79], [80]. When the bacteria cells are exposed to high electric field intensity, the permeability of the cell membrane increases [81]. When the supplied electric field reaches a high value, there occurs an irreversible damage to the cell membrane, causing cell death [82]. Implementing this phenomenon in the water disinfection process results in an efficient treatment without the formation of disinfection by-products. The requirement of high strength electric field will involve high voltage (>1 KV) that consumes high electrical energy with operational threats. The electrodes used in LEEFT can be modified with nanowires which can amplify the electric field intensity with many folds near the tips of nanowires [83], [84], [85]. Though a low voltage is supplied externally, there exists a strong electric field that is sufficient for the electroporation process. The energy consumed in this water treatment is very low compared to other existing disinfection methods such as: conventional methods, UV disinfection, and membrane filtration method. Zhou et al. [86] performed the chemically free water disinfection analysis in pipes using the center electrode modified with copper oxide nanowires and providing very high disinfection rate with the electric voltage of 1 V. The stability of copper oxide nanowire was improved using a protective polydopamine (PDA) coating, as illustrated by Huo et al. [87]. The enhancement of electrical field at the tips of the nanowires due to the lightening rod effect was described by Liu et al. [88]. Various methods of supplying the voltage for electric field enhancement at the tips of the nanowires have been employed for water disinfection techniques. The voltage in the range of few volts produced from the energy harvesting techniques can be suitable for supplying to disinfect water. Energy harvesting techniques based on the piezoelectric effect, triboelectric nanogenerators, and photoelectric effect are preferred as they have high power density with low frequency.

Kumar et al. [89] demonstrated the bacterial degradation using piezo-photocatalysis approach and the beam vibration. Bacterial disinfection obtained using $BaTiO_3$ ceramic at low frequency (8 Hz) was able to produce the required external voltage within 30 min of exposure. A tribopump with low cost and self-driven water disinfection system was proposed by Ding et al. [90] in which the water consisting of bacteria is pumped to the triboelectric nanogenerators system where disinfection occurs by the electric field produced in the system and the outlet is the water free from bacteria. Huo et al. [91] developed a localized electric field air disinfection method to destroy the outer membrane of the bacteria, driven by a triboelectric nanogenerators. The air disinfection system provides high performance in terms of microbial inactivation in a very short time period.

The previously reported bacterial disinfection systems either used the DC voltage generators or electric pulse generators to supply the electrical energy required for bacterial disinfection. However, in this thesis, an AC voltage that is generated form the piezoelectric wind energy harvester has been utilized in order to kill the bacteria present in the water. The disinfection technique present here is portable as it harvests the energy from the ambient wind.

## 4.2. Experimental Setup and Bacterial Disinfection

Figure 4.1 shows a tank consisting of bacterial water that is to be disinfected, and several cantilever-based wind energy harvesters are attached to the water tank in order to supply the electrical output power required for disinfection process. One of the piezoelectric wind energy harvesters as shown in Figure 4.1 is used to supply the electrical energy in order to disinfect the bacteria present in the water.

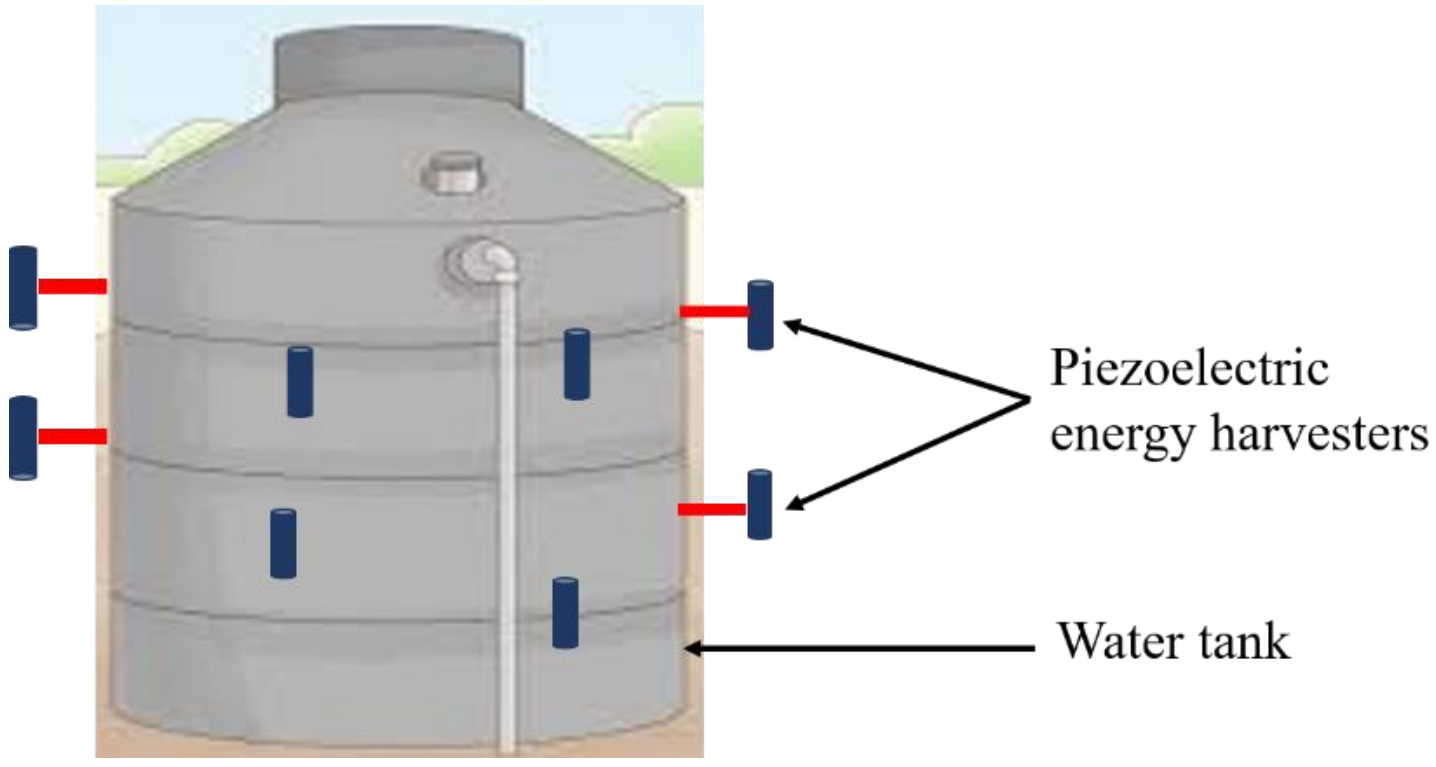

Figure 4.1. Bacterial disinfection in a water tank.

The experimental analysis in this study was performed in two phases. In the first phase, various experiments were conducted in order to produce the electric voltage by flowing the air through the bluff body in an open environment. The wind required to drive the harvester was supplied using an air blower and a digital anemometer was used to measure the wind speed. The required speed was maintained by making proper adjustment to a blower and the energy harvester. A commercially purchased piezoelectric patch, PZT-5A (SP-5A, India) with dimensions of 50 × 20 × 0.4 $mm^3$ was placed at one end of an aluminium beam, with the dimensions of 200 × 25× 0.6 $mm^3$. The piezoelectric patch was attached on the beam using a strong epoxy adhesive. The piezoelectric energy harvester was operated in $d_{31}$ working mode and the piezoelectric coefficient of the patch was -190 Coul/N × $10^{-12}$. A circular bluff body with a height of 120 mm and 32 mm diameter was attached at the other end of the beam, and the curve-shaped attachments required were 3D printed. The length of the attachments was made equal to the height of the bluff body with frontal dimension of 5 mm. Air was supplied to the harvester at different speeds, and comparisons were made to obtain an optimum value of the electrical voltage. The output voltage across the electric load resistance of the harvester was measured using a digital oscilloscope (InfiniiVision DSO-X 3034A) with an input impendence of 10 MΩ.

After the successful generation of the output voltage from the piezoelectric wind energy harvester, the second phase of the experiment was performed, which involved the disinfection of the bacteria present in the water. The experimental setup of the piezoelectric wind energy harvester based on galloping and its application on the degradation of bacteria is shown in Figure 4.2.

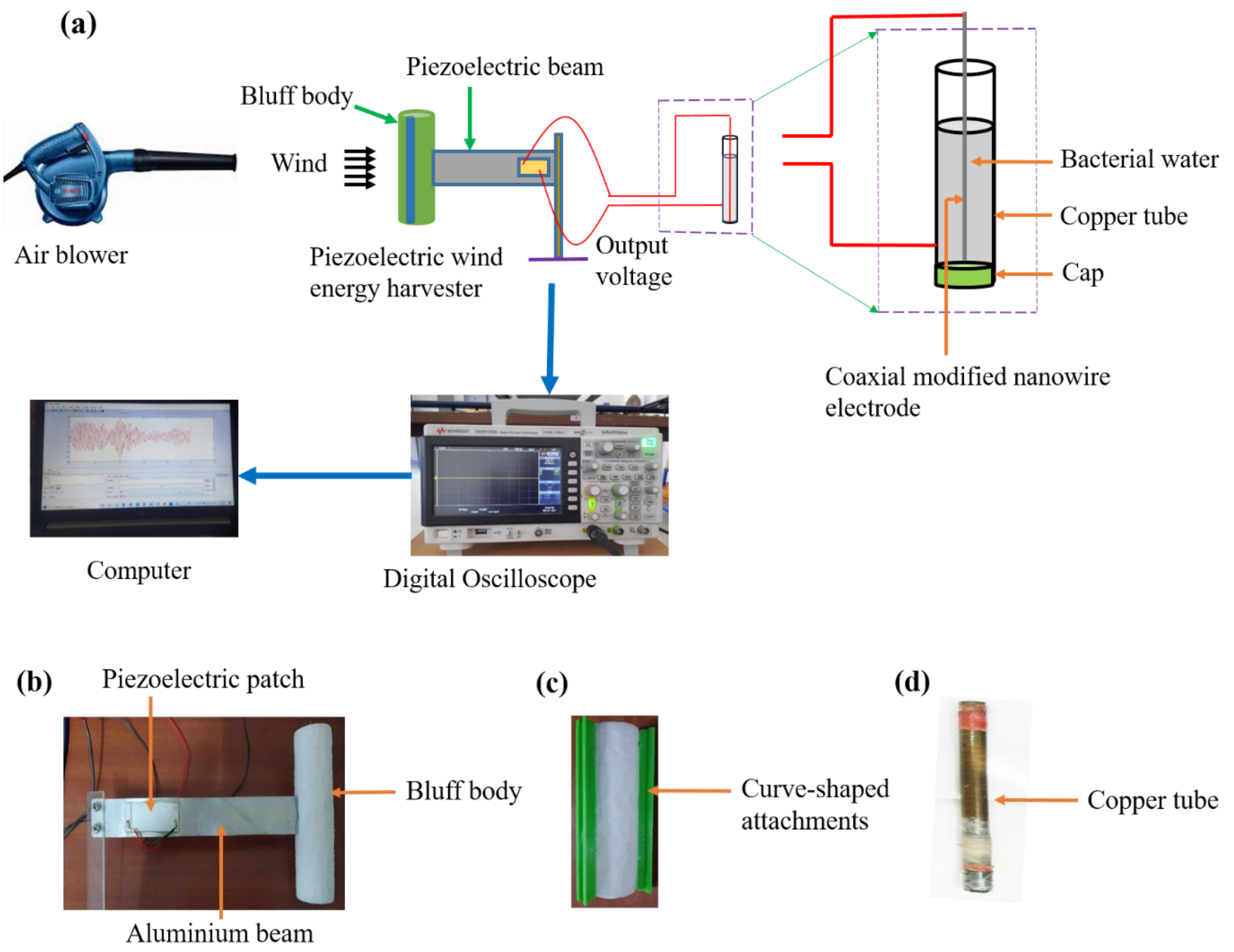

Figure 4.2. Experimental setup: (a) Bacterial disinfection using galloping piezoelectric wind energy harvester; (b) Piezoelectric wind energy harvester; (c) Bluff body with curve-shaped attachments; (d) Copper tube.

The electrical voltage produced from the harvester was supplied to the water consisting of Escherichia coli (E.coli) bacteria on the copper tube, as illustrated in the Figure 4.2 (a). As the AC electrical voltage generated was not enough to disinfect the bacteria, the concept of center copper oxide modified nanowires was applied, which enhances the electrical field by several times at the tips of the nanowires that is adequate to disinfect the bacteria [81]. The tube that contained bacterial water was made up of copper, with a diameter of 94 mm and length of 300 mm, and was closed from one side, as shown in Figure 4.2 (d). A fine copper wire (coaxial electrode) with a diameter of 0.35 mm was heated at around 400°C for 2 hours that is followed by cooling down to a room temperature in an open atmosphere in order to produce the copper oxide modified nanowires (CuONWs). It was observed that the nanowires grew perpendicular to the surface of the copper wire. A two-dimensional model of the copper tube with a center copper electrode modified with nanowires was developed to illustrate the electric field around the tip of the nanowire. An electrostatic model is used which is defined by,

$$E = -\nabla V, \tag{19}$$

where, $E$ is the electric field developed and $V$ is the electric potential. E.coli bacteria required for the disinfection analysis was cultured in broth at 37°C and 200 rpm for 12-18 hours. The cultured bacteria solution was then diluted with autoclaved (20 PSI, 15 minute) distilled water to a dilution factor of 1/10. The concentration of the E.coli bacteria after dilution was $2\times10^7$, $3\times10^7$, $5\times10^7$ colony forming unit per mL (CFU/mL) for triplet experiments. 20 ml of diluted bacteria solution was filled in a disinfection device. Coaxial electrode and copper tube were connected to the output voltage field of the harvester. Air was supplied from the air blower at a speed of 4 m/s and the oscillating behavior of the bluff body was observed. After every 5 minutes of the test, 100µml of the bacterial sample was removed from the tube and spread on the agar plate. It was stored at 37°C temperature for 12 hours for the final count. Three complete disinfection experiments were performed to observe the repeatability of the study. CFU's were counted for every test sample, and log inactivation efficiency was calculated using equation (11),

$$Log\ inactivation\ efficeiency = -log_{10}\left(\frac{C_{eff}}{C_{in}}\right) \tag{20}$$

Where, $C_{eff}$ and $C_{in}$ are the concentrations of the effluent and influent, respectively.

## 4.3. Results and Discussion

### 4.3.1 Simulation and experimental results of the piezoelectric wind energy harvester

A two-dimensional model is developed for performing the CFD simulation of the bluff body with curved shaped attachments. The standard k-ε turbulence model is used for simulation in order to obtain better computational accuracy with great stability. The computational domain considered is of size 100 mm long and 80 mm wide. Figure 4.3 shows the pressure and velocity field developed on the vicinity of the bluff body when air flows through the body at a speed of 4 m/s. The direction of air flow is supposed to be perpendicular to the inlet boundary. At the outlet boundary, the pressure is set as zero and the top and bottom boundaries are considered to be fixed.

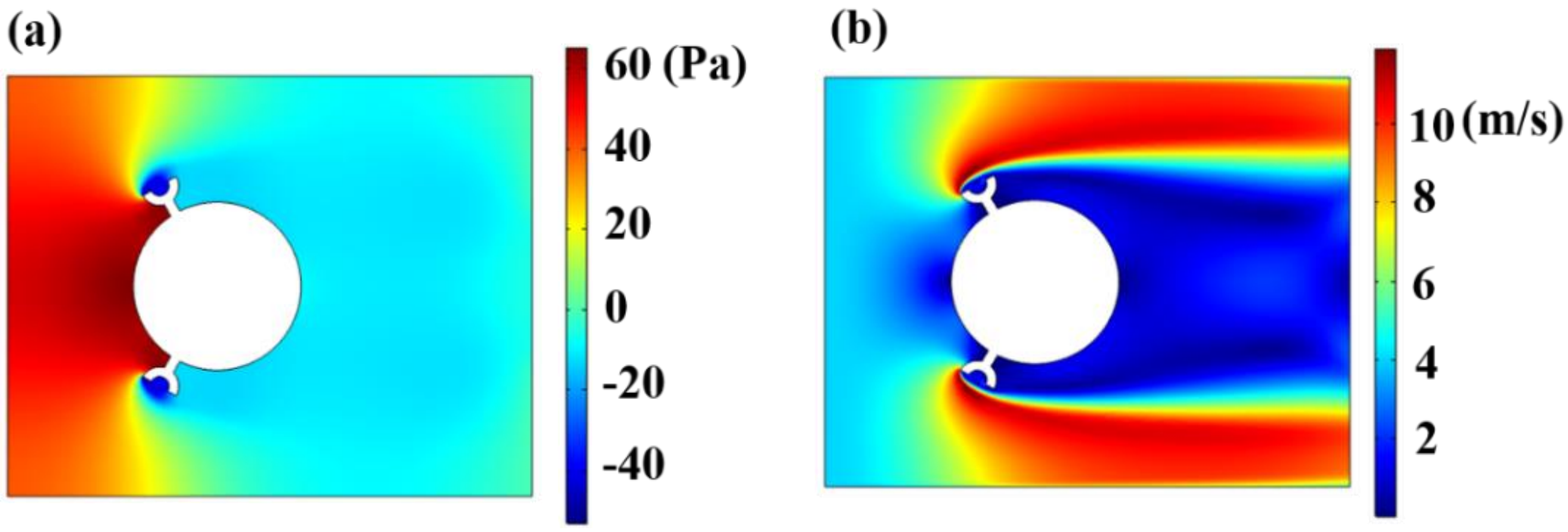

Figure 4.3. 2-D flow field simulation of a bluff body with curved-shaped attachments: (a) Pressure field; (b) Velocity field.

The pressure difference existing between the upstream and the downstream side of the bluff body reveals that there exists the lift force on the bluff body, which eventually creates the transverse force component required for galloping mechanism as observed in Figure 4.3 (a). Pressure difference ranging from 60 Pa just ahead of the bluff body to −40 Pa at the vicinity of the curve-shaped attachments gives large magnitude force component which eventully produce the required electric voltage for disinfection. There is variation of velocity with maximum value at the upper and lower side of the bluff body, where, minimum value just behind the bluff body observed in different shaped attachments signifies the occurrence of vibrations on the bluff body. It is important to simulate the bluff body to observe the distrubance produced in the air flow.

Furthermore, the experimental electrical output voltage measured at different wind speeds is compared with the electrical voltage obtained from the simulation. The experimental output voltage of the harvester with curved shaped attachments is found to be about 25 V and the simulated output voltage is about 29 V when the speed of wind is kept as 4 m/s, as shown in Figure 4.3 (a) and Figure 4.3 (b), respectively. The modeling of the harvester is based on the lumped parameter model and the quasi-steady hypothesis is assumed in the derivation of the transverse force component with small angle of attack. These assumptions are responsible for the difference in the simulated and experimental output values.  It requires some period of seconds for the harvesters to produce stable output voltage for both the simulation and experiment as observed in the Figure 4.3.

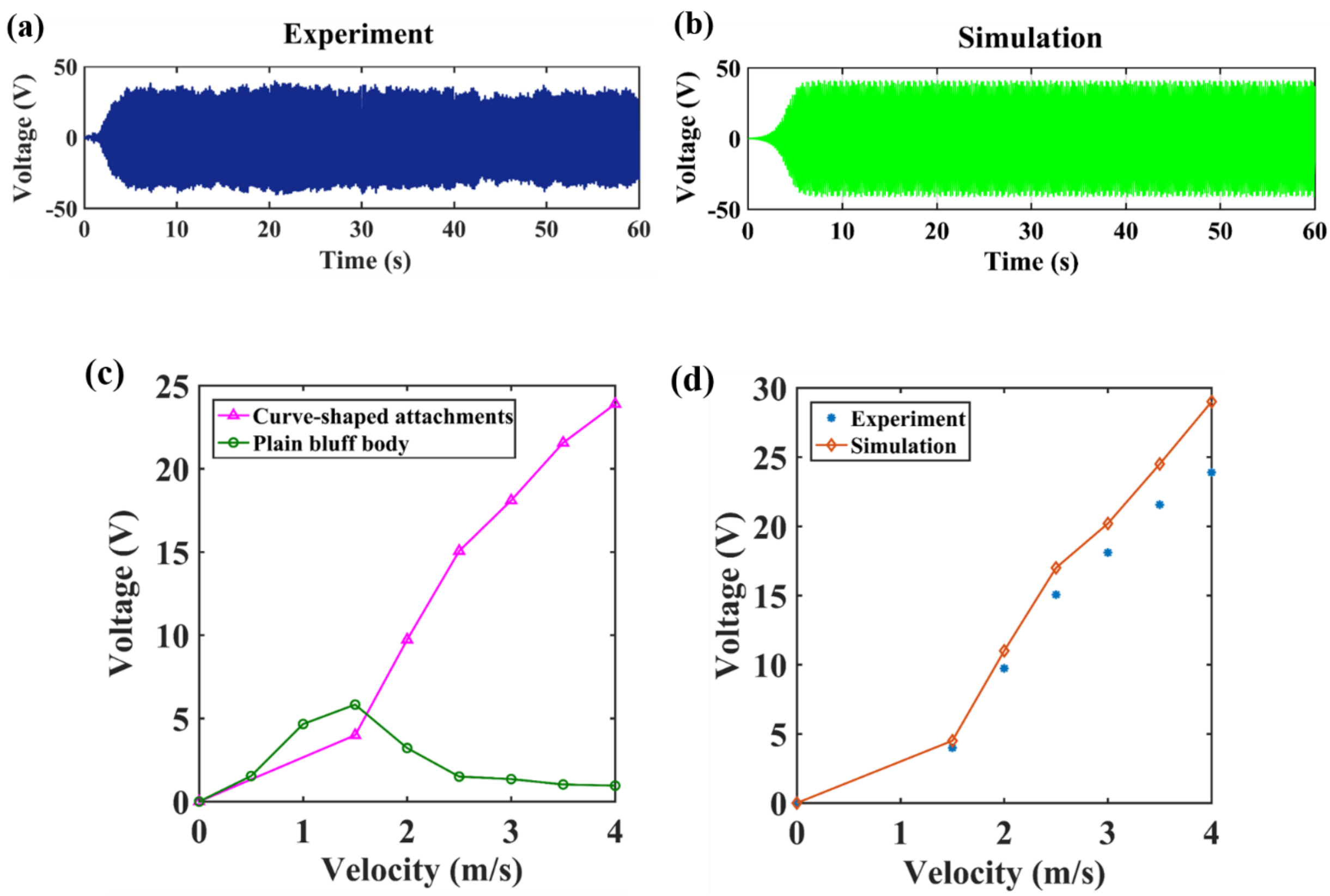

Figure 4.4. Experimental and simulation output voltage: (a) Experiment at 4 m/s wind speed; (b) Simulation at 4 m/s wind speed; (c) Experimental output voltage with and without attachments (d) Comparison at different wind speeds.

The voltage output of the energy harvester without curve-shaped attachment is low as compared to the output with curve-shaped attachments, as shown in Figure 4.4 (c). The maximum voltage obtained with the harvester without any attachments is around 6 V at a wind speed of 1.5 m/s. For the harvester, without any attachments, it follows vortex-induced vibration phenomena, due to the circular cross-section of the bluff body. Figure 4.4 (d) compares the simulation and experimental electrical voltage produced by the harvester at different wind speeds. It is observed that for galloping phenomena, the electrical voltage is increased when we increase the wind speed. Since the harvester, when operating at 4 m/s, produces the maximum voltage in our study, we use this speed for the disinfection of bacteria. The comparison of the output voltage of several typical piezoelectric wind energy harvester based on galloping phenomenon is listed on Table 4.1. It can be observed that using the curve-shaped attachment, an RMS open circuit voltage of 25 V can be generated, which is not possible with other shapes.

**Table 4.1.** Comparison results of various piezoelectric galloping wind energy harvesters.

| No. | Reference | Bluff body | | | Piezoelectric beam | | Wind velocity (m/s) | Output RMS voltage (V) |
|---|---|---|---|---|---|---|---|---|
| | | Shape | Windward width (mm) | Height (mm) | Material | Dimensions in mm (length x width x thickness) | | |
| 1. | Zhao et al. [92] | Square | 40 | 150 | Aluminum | 150 × 30 × 0.6 | 8 | 30 |
| 2. | Zhou et al. [93] | Curved plate | 35 | 100 | Tin bronze | 180 × 10 × 0.8 | 5 | 5 |
| 3. | Gang et al. [58] | Cylindrical body with circular attachments | 48 | 240 | Stainless Steel | 200 × 26 × 0.95 | 8 | 19 |
| 4. | Ding et al. [94] | Cylindrical body with fin-shaped attachments | 30 | 150 | Stainless Steel | 120 × 15 × 0.5 | 7 | 14 |
| 5. | Present work | Cylindrical body with curve-shaped attachments | 32 | 120 | Aluminum | 200 × 25× 0.6 | 4 | 25 |

### 4.3.2 Electric field enhancement

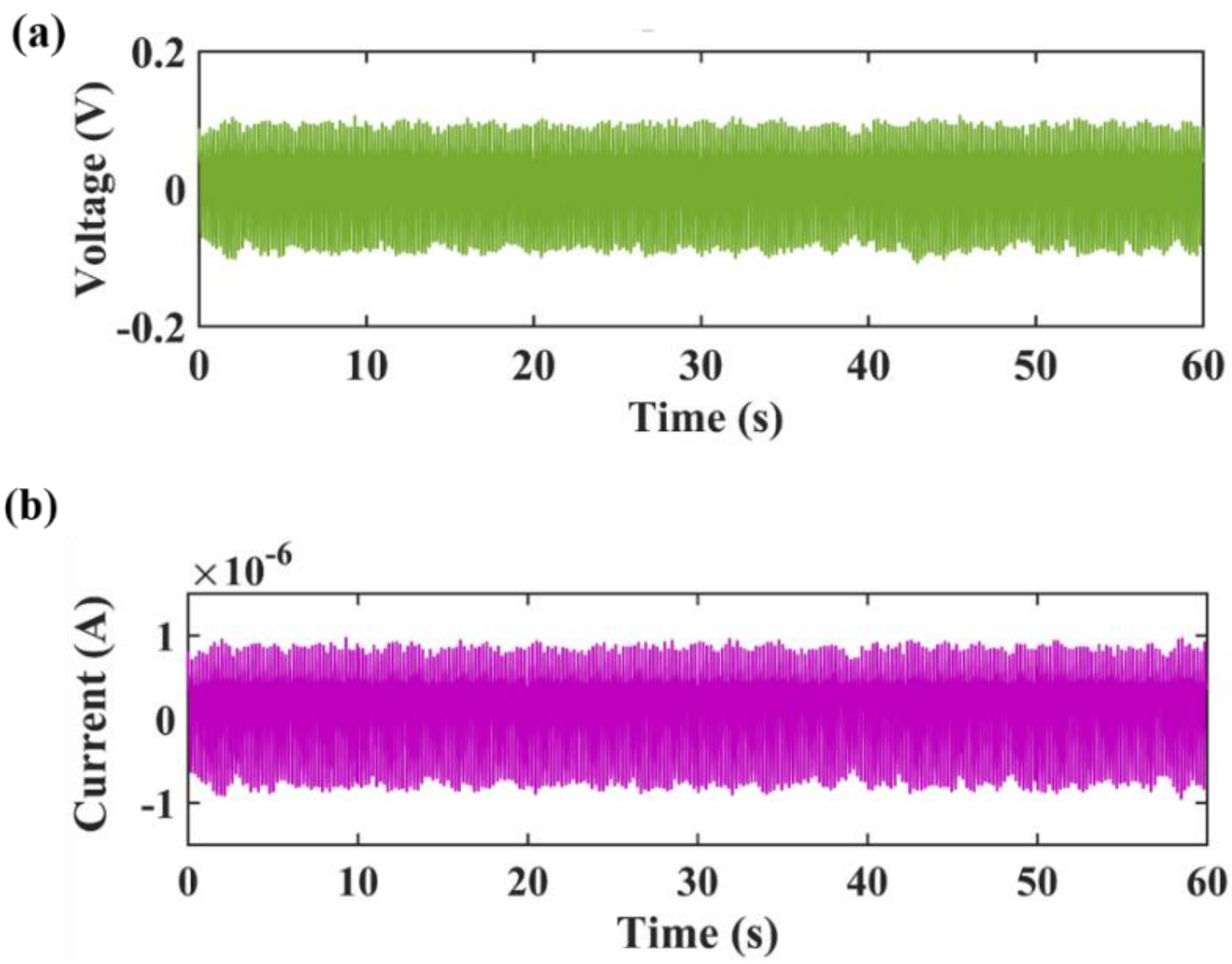

Figure 4.5. Variation of electric supply during disinfection: (a) Output voltage vs. time plot; (b) Current vs. time plot.

When the output of the galloping piezoelectric wind energy harvester is connected with the center coaxial electrode and copper tube, the voltage that is supplied to the bacterial water decreases. It is because the bacterial water offers some resistance as the electric potential is supplied. The resistance offered by the bacterial water, which is the mixture of deionized water and E.coli bacteria, is around 85 KΩ and the voltage measured across this resistance is 0.1V, which can be seen in Figure 4.5 (a). Similarly, the electric current measured during the disinfection experiment is around 0.8 μA, as shown in Figure 4.5 (b).

As a high electric field, in the range of (1 to 10 KV/cm) [95] is required for the degradation of bacteria, there is a need to enhance the electric field by the use of electrode with modified nanowires. The modification of fine copper wire was achieved by an oxidation process in which oxygen was combined with copper and deposited on the surface of the wire. Figure 4.6 (a) shows the X-ray diffraction (XRD) image of the coaxial centre copper electrode, and the data was matched with JCPDS database. Visualizing the image, we can clarify that CuO (01-089-2529) is present with (200), (210) and (222) planes, $Cu_2O$ (01-078-2076) is present with (111), (200) and (211) planes and Cu (03-065-9743) is also present with (211), (220) and (400) planes.

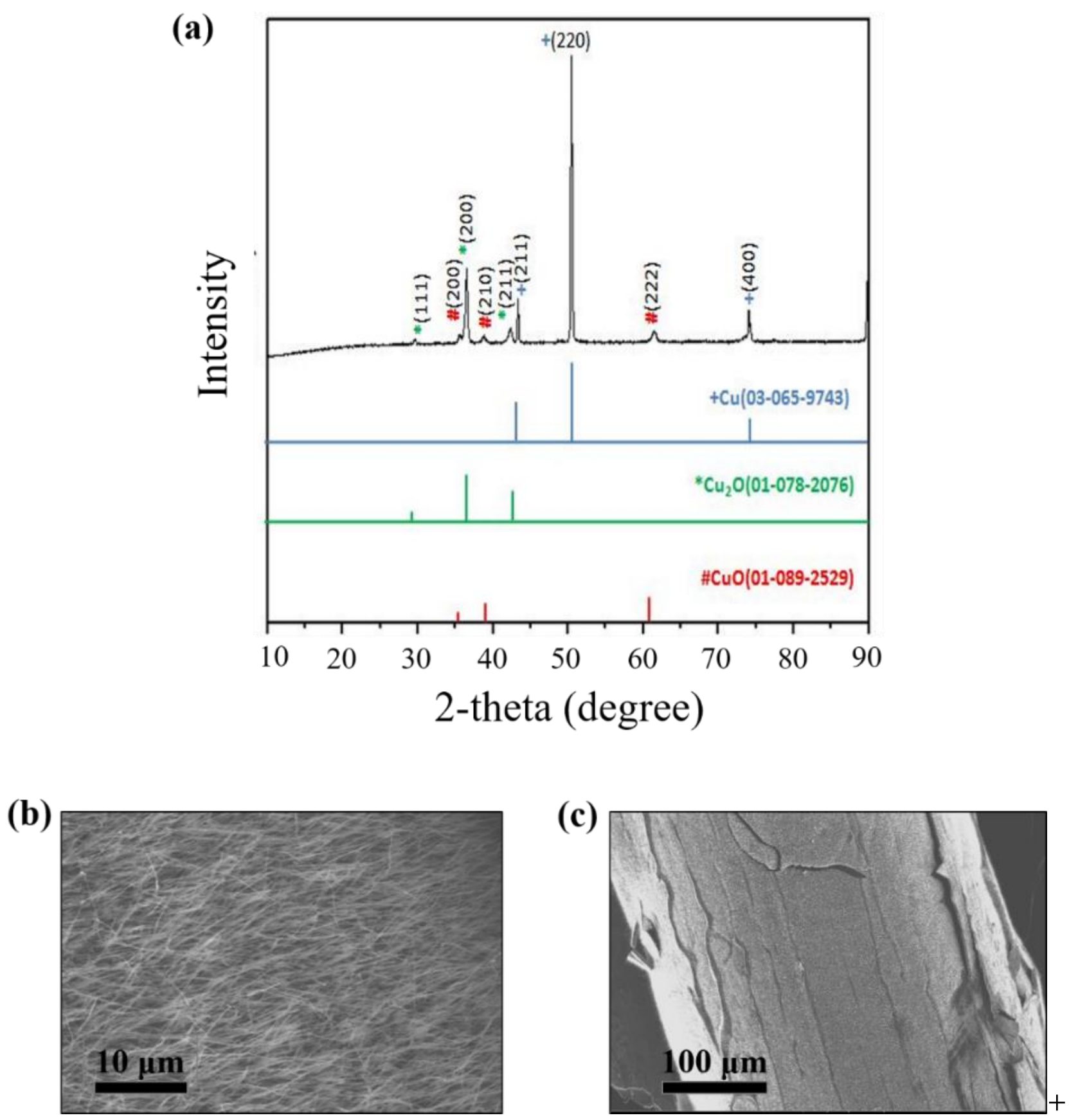

Figure 4.6. (a) XRD spectra of coaxial center copper electrode; (b) SEM image of nanowires generated on the surface of the coaxial copper electrode; (c) SEM image of coaxial center copper electrode.

Scanning electron microscopy (SEM) image of coaxial Cu wire as shown in Figure 4.6 (b) reveal that nanowires with 2-2.5 μm length and 65-80nm diameter that are perpendicular to the surface are grown. When coaxial Cu wire was heated $Cu^+$ is generated, which reacts with $O_2$ present in the atmosphere and generates $Cu_2O$. Later $Cu_2O$ is converted to CuO. The generation of $Cu_2O$ is in the form of a thin layer, whereas CuO is generated in the form of nanowires. According to Wagner's theory [96], at low temperature (~0.3-0.5 melting point) a metal is oxidized through short-circuit diffusion (diffusion along the sub-boundaries, dislocations) of atoms or ions during the reaction in the $Cu_2O$ layer. The melting temperature of Cu, $Cu_2O$, and CuO are 1083, 1235, and 1446°C, respectively. At 400°C, short-circuit diffusion (diffusion along dislocations) is dominated, and CuONWs generation is achieved [97]. Due to the domination of CuO, which propagate in the form of wire-like structure, nanowires are generated, which can be verified by SEM images.

To observe the effect of the copper oxide modified nanowires on the electric field enhancement, a two-dimensional simulation of the disinfection device is performed. As seen in Figure 4.7, the outer circle represents the circumferential outline of the copper tube consisting of bacterial water. The inner circle represents the coaxial copper wire modified with nanowire. For the sake of simplicity, we consider a single nanowire to perform the simulation analysis. The output electric potential obtained from the piezoelectric wind energy harvester is connected to the center coaxial copper wire as well as to the outer copper tube. An electric potential of 0.1 V is supplied for the enhancement purpose. It can be observed that there is high strength of electric field produced at the tip of the nanowire. Electric field as high as 14000 V/m is generated by the use of a nanowire as illustrated in the Figure 4.7. Several other nanowires are generated on the surface of the copper wire, which can produce enough electric field for bacterial degradation.

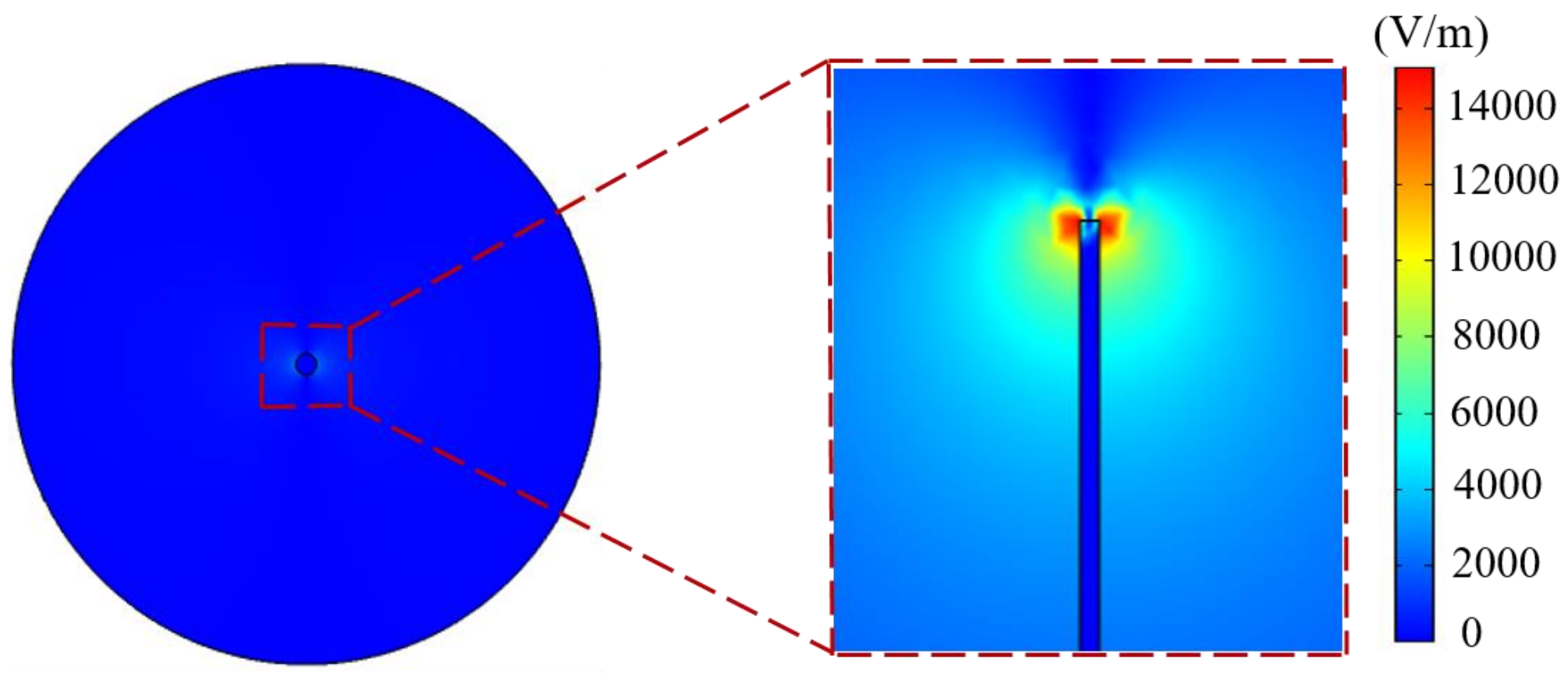

Figure 4.7. Electric field simulation performed on the cross-section of the copper tube using copper oxide modified nanowire.

### 4.3.3 Bacterial disinfection

Figure 4.8 shows the spread plate technique implemented on the agar plate for different periods of operation. Three different tests were performed to confirm the repeatability of the experiment. Here, the bacterial disinfection analysis performed for a single test is illustrated. The initial sample of the bacterial water was removed and spread on the plate showing bacterial CFUs. After supplying the electrical potential for 5 minutes, another sample was removed and spread on the agar plate. This process continued for a time period of 25 minutes, and the results confirmed that for each sample, there is a decrease in the bacterial CFUs as shown in the Figure 4.7. It can be observed that the initial samples on a petri dish consists of a large number of CFUs of bacteria, while no viable bacterial cell is seen after 25 minutes of treatment. The variation of log inactivation efficiency with the treatment time is plotted with standard deviation as shown in Figure 4.8. Bacterial log removal efficiency is found to be 2.33.

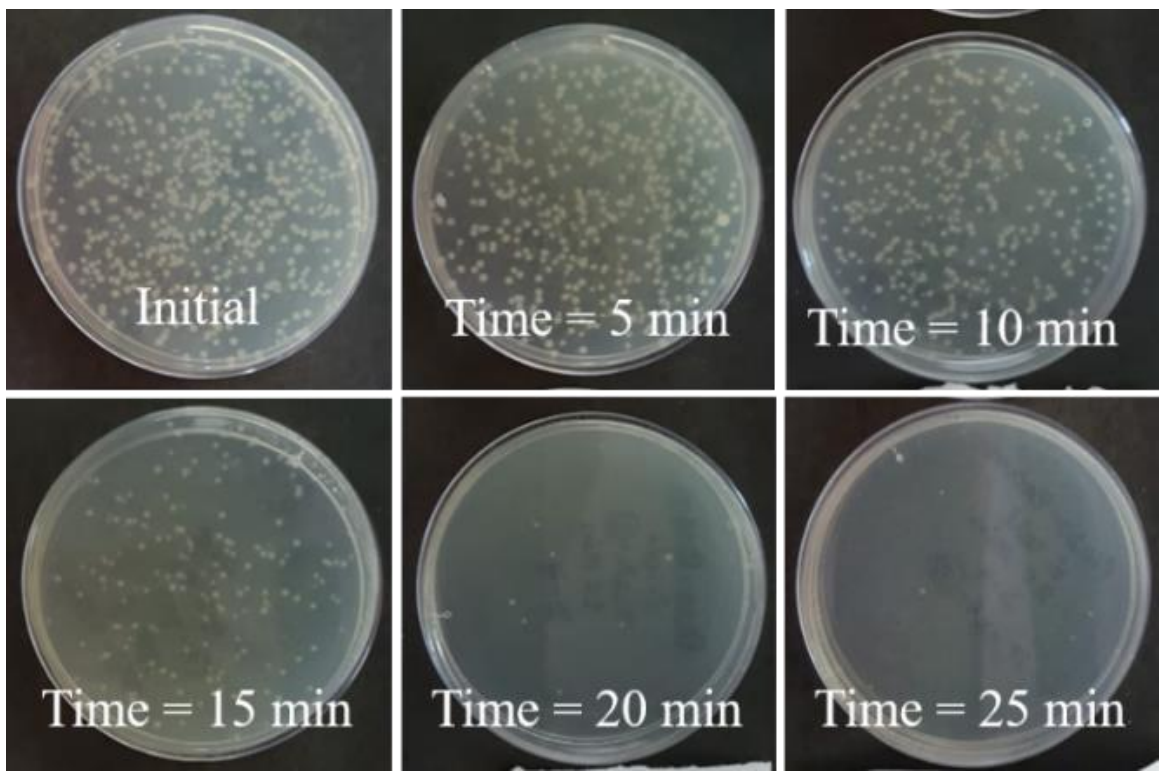

Figure 4.8. Agar plates showing bacterial concentration for different periods of treatment.

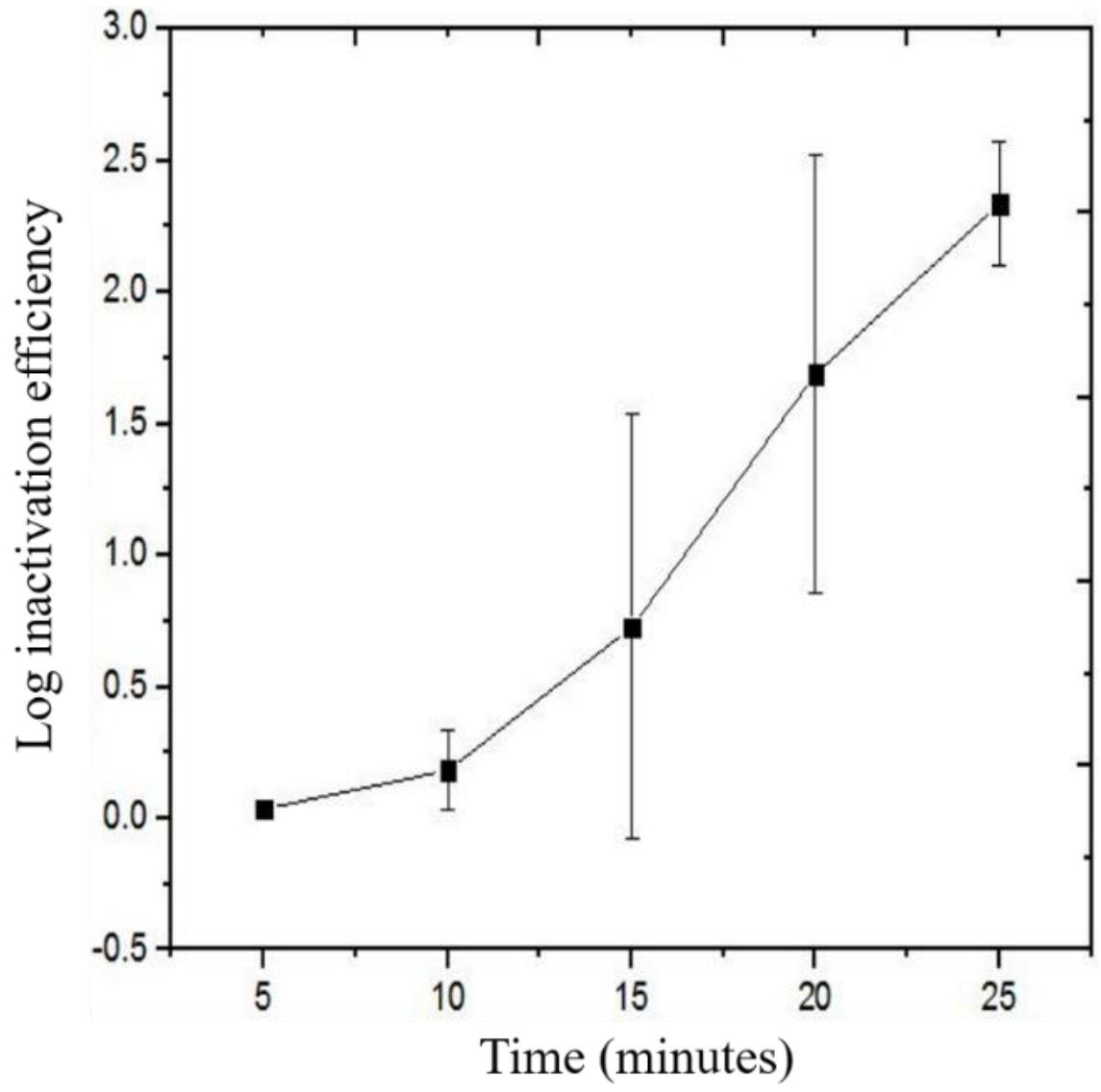

Figure 4.9. Bacterial removal efficiency of E.coli bacteria.

SEM images of the coaxial CuONWs as illustrated in Figure 4.10 (a) and 4.10 (b), reveal that CuONWs break during the bacterial inactivation experiment. The disinfection device was turned upside down for taking the samples out for plating. At that moment, turbulence was generated, and CuONWs are not strong enough to withstand the turbulence. CuONWs can be coated with polydopamine (PDA) [98] for improving their stability which is not done in our study. Figure 4.10 (c) and 4.10 (d) show the SEM images of E.coli bacteria before and after the treatment, respectively.  During the disinfection, bacterial cells get dragged towards the modified coaxial center electrode by various forces such as electrophoresis and dielectrophoresis forces where high electric field is present [99]. In the electric field, bacteria behave like a capacitor. Charged irons inside and outside of the cell move towards the cell membrane and generate transmembrane potential [95]. Under this potential, water molecule tends to penetrate the membrane which leads to the formation of permanent pores and cracks. The comparison of previously reported LEEFT water disinfection system with our study is presented on Table 4.2.

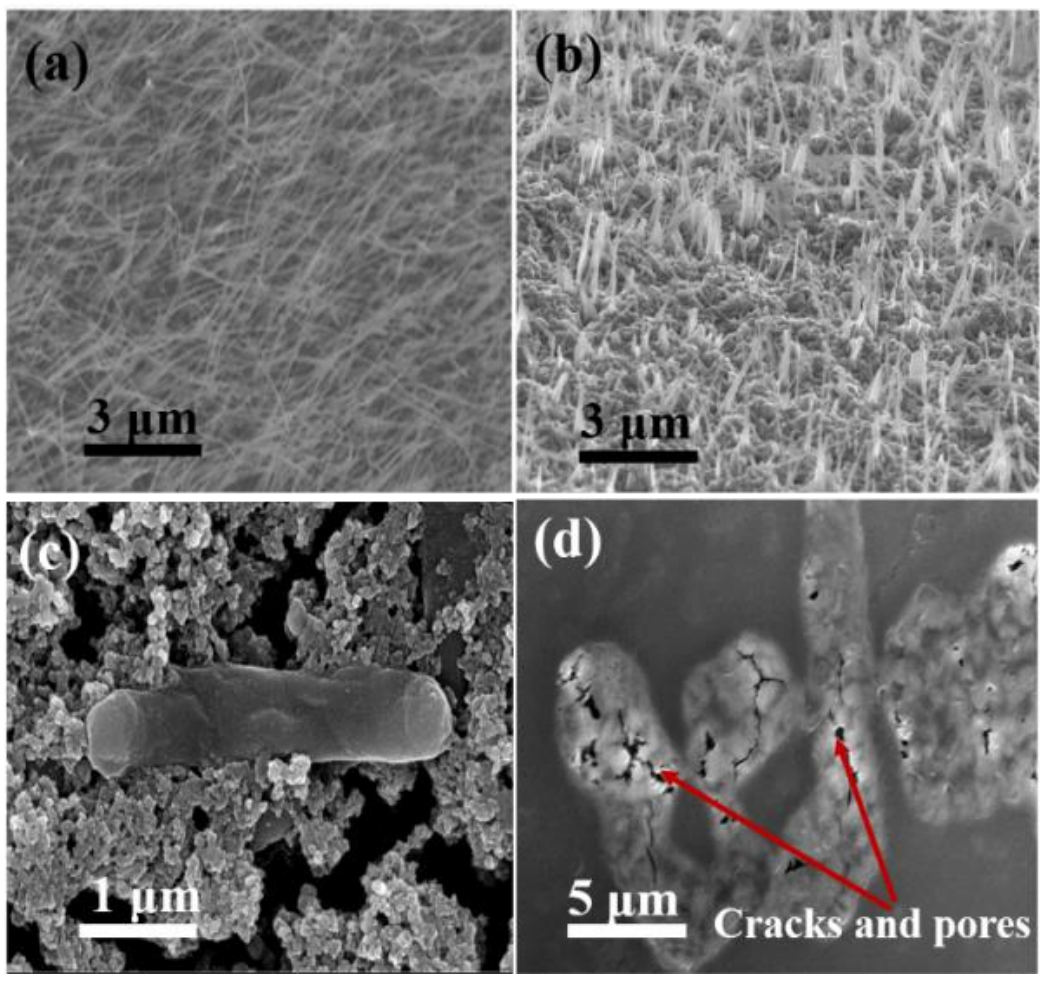

Figure 4.10. SEM images: (a) Copper oxide nanowires before experiment; (b) Copper oxide nanowires after experiment (broken); (c) Live E.coli bacteria (before treatment); (d) E.coli bacteria (after treatment).

**Table 4.2**. Comparison of previously reported LEEFT water disinfection.

| No. | Reference | Study | Applied Voltage (electrical parameters) | Portable | Self-powered | Flow/steady mode | Capacity | Log inactivation efficiency |
|---|---|---|---|---|---|---|---|---|
| 1. | Xie et al. [100] | LEEFT with Ozonation | 1.2V DC | × | × | Flow | 4mL/min, 2.2HRT(min) | 6-log |
| 2. | Zhou et al. [101] | Smart phone powered LEEFT | 2VDC | Yes | × | Flow | 10mL/min, 0.88HRT(min) | 3-log |
| | | | ~1.8VDC | | | | 5mL/min, 1.77HRT(min) | 6-log |
| | | | ~ 1.4 | | | | 2mL/min, 4.43HRT(min) | |
| 3. | Ding et al. [90] | Tribo-pump | $I_{RMS}$=80μm R=several kΩ | Yes | Yes | Flow | 2-2.5HRT(min) | 6-log |
| 4. | Zhou et al. [86] | LEEFT in pipes | 0.5V DC | × | × | Flow | >1mL/min >8.86HRT(min) | 1-log |
| | | | 1V DC | | | | 1mL/min, 8.86HRT(min) | 6-log |
| | | | 2V DC | | | | 1mL/min, 8.86HRT(min) | |
| 5. | Present work | Piezoelectric wind energy harvester for LEEFT | 0.1V | Yes | Yes | Steady | 0.8 mL/min | 2.33-log |

## 4.4. Conclusions

This chapter presents an effective method for bacterial disinfection of water by utilizing the energy produced from a piezoelectric wind energy harvester. The electrical output of the harvester when connected to the bacterial water is about 0.1 V which is further enhanced by generating copper oxide nanowires (CuONWs) at the center copper wire. A high electric field around 14000 V/m is produced due to the local enhancement of the electric field that is demonstrated in the simulation analysis performed on a single nanowire. Complete bacterial disinfection is achieved by supplying electric potential for 25 minutes, and the log efficiency of the bacterial removal is calculated for each time interval of 5 minutes. The bacterial removal mechanism is confirmed by counting bacterial colonies using the spread plate technique. Piezoelectric harvesting technique based on wind flow can be an important concept for disinfection of bacteria in the water distribution systems. It is suitable to implement such approaches in the treatment of drinking water in areas where there is no availability of power supply. This water disinfection technique is efficient, low cost, by-products free, and shows strong potential in water treatment in the storage systems.

Based on the studies presented so far, the next chapter presents the overall conclusions of the thesis. Moreover, the challenges and recommendations for the future work in this direction are presented as the future scope in the next chapter.

# CHAPTER 5: CONCLUSIONS AND FUTURE SCOPE

## 5.1. Conclusions

This thesis consists of theoretical, numerical, experimental and the application in bacterial disinfection of a galloping piezoelectric wind energy harvester. The objectives of the thesis were successfully achieved through the research work. The enhanced performance of the wind energy harvester is achieved with modification in the shape of the bluff body.

A detailed modeling of a piezoelectric wind energy harvester based on galloping phenomenon has been carried out and experimental studies are conducted to validate the numerical results. The parameters of the energy harvester, such as damping ratio, natural frequency, lift and drag forces acting on a bluff body were determined experimentally in a static test. The key conclusions drawn from the studies presented in the thesis are summarized as follows,

- The electrical output voltage and output power of the harvester with curved shaped attachment to the bluff body is higher than the harvesters with other shaped attachments; namely, triangular, circular, square, and Y-shaped attachments to the bluff body.
- At a wind speed of 4 m/s, the output voltage and output power of the harvester consisting of curve-shaped attachment are 25 V and 0.105 mW, respectively.
- Within the wind speed range of 1 m/s to 1.5 m/s, the harvester with plain cylindrical bluff body provides high amplitude oscillations with the oscillating frequency of 4.72 Hz, which is close to the natural frequency of the system.
- Beyond the wind speed of 2 m/s, the frequency of vibration of the curve-shaped wind energy harvester is 4.75 Hz, which is close to natural frequency of the system.
- The resistance offered by the bacterial water, which is the mixture of deionized water and E.coli bacteria, is around 85 KΩ. Thus, the actual electrical voltage supplied for the disinfection process is 0.1 V that is measured across the resistance offered by the bacterial water.
- Copper oxide modified nanowires enhances the electric field up to 14000 V/m at a supply input voltage of 0.1 V.
- For complete disinfection, the bacterial log inactivation efficiency for the time interval of 5 minutes is found to be 2.33.

- The proposed disinfection method based on the piezoelectric wind energy harvester can be suitable for water treatment in areas, where there is an unavailability of electrical power supply.

- The water disinfection technique based on energy supplied from wind energy harvester is efficient, low cost, disinfection by-products free, and shows strong potential in water treatment in storage systems.

Overall, this thesis presents the improved performance of a piezoelectric wind energy harvester, based on galloping phenomenon and its potential application in inactivation of bacterial cells that may be present in drinking water. This concept can be very useful for providing clean drinking water in remote places, where there is unavailability of electrical power and other disinfection means.

## 5.2. Future Scope

The results presented in this thesis provide understandings into the small-scale wind energy harvester based on piezoelectricity, and its practical application in bacterial disinfection. At the same time, the implications of this thesis open up new avenues for future research. The possible future directions for this research work are as follows,

- Lumped parameter modeling of the piezoelectric wind energy harvester was considered in this thesis with some assumptions. However, in order to obtain more accurate results, finite element modeling of the energy harvester can be performed.
- In this thesis, a constant wind velocity is considered for the numerical as well as experimental analysis. However, in real situation, the wind velocity is random. The wind energy harvester that operates in random velocity field can be studied in future work.
- The wind energy harvester is always supposed to face the direction of air flow. But in reality, it is not possible for a harvester to always face the air flow, as the direction of air flow is random in nature. Wind energy harvester with a yaw mechanism can be developed, thus enabling the harvester to continuously face the flowing air, as in case of wind turbines.
- In this thesis, wind energy harvester provides electrical output based on the piezoelectric effect only. Hybrid wind energy harvester that incorporates both piezoelectric and triboelectric effect can be developed for obtaining even enhanced performance.

- This thesis presents the practical application of piezoelectric wind energy harvester in bacterial disinfection in water storage systems. This thesis work can be extended to the bacterial disinfection of drinking water in distribution systems in pipelines. Water flow parameters must be included in detail in the analysis.
- The centre coaxial electrode used to supply electrical power during disinfection is modified with copper oxide nanowires without any coatings. Thus, these grown nanowires are susceptible to breaking during experimental study due to the turbulence created in the copper tube. Copper oxide nanowires with coatings, such as polydopamine coating, will enhance the electrode stability. These coatings could be included in future work for better performance.

# References

[1] S. P. Beeby, M. J. Tudor, and N. M. White, "Energy harvesting vibration sources for microsystems applications," *Meas Sci Technol*, vol. 17, no. 12, 2006, doi: 10.1088/0957-0233/17/12/R01.

[2] F. K. Shaikh and S. Zeadally, "Energy harvesting in wireless sensor networks: A comprehensive review," *Renewable and Sustainable Energy Reviews*, vol. 55, pp. 1041–1054, 2016, doi: 10.1016/j.rser.2015.11.010.

[3] C. Wei and X. Jing, "A comprehensive review on vibration energy harvesting: Modelling and realization," *Renewable and Sustainable Energy Reviews*, vol. 74, no. December 2016, pp. 1–18, 2017, doi: 10.1016/j.rser.2017.01.073.

[4] S. Nabavi and L. Zhang, "Portable wind energy harvesters for low-power applications: A survey," *Sensors (Switzerland)*, vol. 16, no. 7, 2016, doi: 10.3390/s16071101.

[5] C. R. Farrar, G. Park, T. Rosing, M. D. Todd, and W. Hodgkiss, "Energy harvesting for structural health monitoring sensor networks," *Structural Health Monitoring 2007: Quantification, Validation, and Implementation - Proceedings of the 6th International Workshop on Structural Health Monitoring, IWSHM 2007*, vol. 2, no. 1, pp. 1773–1780, 2007, doi: 10.1061/(asce)1076-0342(2008)14:1(64).

[6] S. Saadon and O. Sidek, "A review of vibration-based MEMS piezoelectric energy harvesters," *Energy Convers Manag*, vol. 52, no. 1, pp. 500–504, 2011, doi: 10.1016/j.enconman.2010.07.024.

[7] C. Klinger, T. Michael, and D. Bettge, "Fatigue cracks in railway bridge hangers due to wind induced vibrations - Failure analysis, measures and remaining service life estimation," *Eng Fail Anal*, vol. 43, pp. 232–252, 2014, doi: 10.1016/j.engfailanal.2014.02.019.

[8] L. E. Cross, R. E. Newnham, and W. A. Schulze, "Perforated Pzt-Polymer Composites For Piezoelectric Transducer Applications," *Ferroelectrics*, vol. 41, no. 1, pp. 197–205, 1982, doi: 10.1080/00150198208210624.

[9] R. M. Martin, "PHYSIC AL REVIEW B VOLUME 5, NUMB ER 4 Piezoelectricity," *Phys. Rev.*, vol. 5, no. February, p. 1607, 1972.

[10] C. Wang, S. Wang, Q. J. Li, X. Wang, Z. Gao, and L. Zhang, "Fabrication and performance of a power generation device based on stacked piezoelectric energy-harvesting units for pavements," *Energy Convers Manag*, vol. 163, no. February, pp. 196–207, 2018, doi: 10.1016/j.enconman.2018.02.045.

[11] Y. B. Jeon, R. Sood, J. H. Jeong, and S. G. Kim, "MEMS power generator with transverse mode thin film PZT," *Sens Actuators A Phys*, vol. 122, no. 1 SPEC. ISS., pp. 16–22, 2005, doi: 10.1016/j.sna.2004.12.032.

[12] L. Zhou *et al.*, "A model for the energy harvesting performance of shear mode piezoelectric cantilever," *Sens Actuators A Phys*, vol. 179, pp. 185–192, 2012, doi: 10.1016/j.sna.2012.02.041.

[13] C. H. K. Williamson, “Vortex dynamics in the cylinder wake,” *Annu Rev Fluid Mech*, vol. 28, pp. 477–539, 1996, doi: 10.1146/annurev.fl.28.010196.002401.

[14] M. L. Facchinetti, E. de Langre, and F. Biolley, “Coupling of structure and wake oscillators in vortex-induced vibrations,” *J Fluids Struct*, vol. 19, no. 2, pp. 123–140, 2004, doi: 10.1016/j.jfluidstructs.2003.12.004.

[15] R. Nakayama, Y. Nakamura, and O. Shigehira, “A numerical study of vortex shedding from flat plates with square leading and trailing edges,” *J Fluid Mech*, vol. 236, no. 445, pp. 445–460, 1992, doi: 10.1017/S0022112092001472.

[16] G. Alonso, J. Meseguer, and I. Pérez-Grande, “Galloping stability of triangular cross-sectional bodies: A systematic approach,” *Journal of Wind Engineering and Industrial Aerodynamics*, vol. 95, no. 9–11, pp. 928–940, 2007, doi: 10.1016/j.jweia.2007.01.012.

[17] G. Alonso, E. Valero, and J. Meseguer, “An analysis on the dependence on cross section geometry of galloping stability of two-dimensional bodies having either biconvex or rhomboidal cross sections,” *European Journal of Mechanics, B/Fluids*, vol. 28, no. 2, pp. 328–334, 2009, doi: 10.1016/j.euromechflu.2008.09.004.

[18] M. A. Baenziger, W. D. James, B. Wouters, and L. Li, “Dynamic loads on transmission line structures due to galloping conductors,” *IEEE Transactions on Power Delivery*, vol. 9, no. 1, pp. 40–49, 1994, doi: 10.1109/61.277678.

[19] M. Cai, X. Yang, H. Huang, and L. Zhou, “Investigation on Galloping of D-Shape Iced 6-Bundle Conductors in Transmission Tower Line,” *KSCE Journal of Civil Engineering*, vol. 24, no. 6, pp. 1799–1809, 2020, doi: 10.1007/s12205-020-0595-z.

[20] J. P. DEN HARTOG, “Transmission Line Vibration Due to Sleet,” *Transactions of the American Institute of Electrical Engineers*, vol. 51, no. 4, pp. 1074–1076, 1932, doi: 10.1109/T-AIEE.1932.5056223.

[21] C. De Marqui, A. Erturk, and D. J. Inman, “Piezoaeroelastic modeling and analysis of a generator wing with continuous and segmented electrodes,” *J Intell Mater Syst Struct*, vol. 21, no. 10, pp. 983–993, 2010, doi: 10.1177/1045389X10372261.

[22] D. M. Tang, H. Yamamoto, and E. H. Dowell, “Flutter and limit cycle oscillations of two-dimensional panels in three-dimensional axial flow,” *J Fluids Struct*, vol. 17, no. 2, pp. 225–242, 2003, doi: 10.1016/S0889-9746(02)00121-4.

[23] P. Hémon, E. de Langre, and P. Schmid, “Experimental evidence of transient growth of energy before airfoil flutter,” *J Fluids Struct*, vol. 22, no. 3, pp. 391–400, 2006, doi: 10.1016/j.jfluidstructs.2005.11.005.

[24] S. Orrego *et al.*, “Harvesting ambient wind energy with an inverted piezoelectric flag,” *Appl Energy*, vol. 194, pp. 212–222, 2017, doi: 10.1016/j.apenergy.2017.03.016.

[25] S. Li, J. Yuan, and H. Lipson, “Ambient wind energy harvesting using cross-flow fluttering,” *J Appl Phys*, vol. 109, no. 2, pp. 1–4, 2011, doi: 10.1063/1.3525045.

[26] M. Eugeni *et al.*, “Numerical and experimental investigation of piezoelectric energy harvester based on flag-flutter,” *Aerosp Sci Technol*, vol. 97, p. 105634, 2020, doi: 10.1016/j.ast.2019.105634.

[27] S. Tokoro, H. Komatsu, M. Nakasu, K. Mizuguchi, and A. Kasuga, “Study on wake-galloping employing full aeroelastic twin cable model,” *Journal of Wind Engineering and Industrial Aerodynamics*, vol. 88, no. 2–3, pp. 247–261, 2000, doi: 10.1016/S0167-6105(00)00052-0.

[28] M. Usman, A. Hanif, I. H. Kim, and H. J. Jung, “Experimental validation of a novel piezoelectric energy harvesting system employing wake galloping phenomenon for a broad wind spectrum,” *Energy*, vol. 153, pp. 882–889, 2018, doi: 10.1016/j.energy.2018.04.109.

[29] L. B. Zhang, H. L. Dai, A. Abdelkefi, and L. Wang, “Experimental investigation of aerodynamic energy harvester with different interference cylinder cross-sections,” *Energy*, vol. 167, pp. 970–981, 2019, doi: 10.1016/j.energy.2018.11.059.

[30] J. Wang, L. Tang, L. Zhao, and Z. Zhang, “Efficiency investigation on energy harvesting from airflows in HVAC system based on galloping of isosceles triangle sectioned bluff bodies,” *Energy*, vol. 172, pp. 1066–1078, 2019, doi: 10.1016/j.energy.2019.02.002.

[31] D. G. J. Blatt, P. A. Schroeder, C. L. Foiles, “Thermopower of metals,” *Thermopower of metals*, p. 5, 1976.

[32] N. Asim *et al.*, “A review on the role of materials science in solar cells,” *Renewable and Sustainable Energy Reviews*, vol. 16, no. 8, pp. 5834–5847, 2012, doi: 10.1016/j.rser.2012.06.004.

[33] W. P. Mason, “Piezoelectricity, its history and applications,” *Journal of the Acoustical Society of America*, vol. 70, no. 6, pp. 1561–1566, 1981, doi: 10.1121/1.387221.

[34] F. R. Fan, Z. Q. Tian, and Z. Lin Wang, “Flexible triboelectric generator,” *Nano Energy*, vol. 1, no. 2, pp. 328–334, 2012, doi: 10.1016/j.nanoen.2012.01.004.

[35] C. R. Bowen, J. Taylor, E. Leboulbar, D. Zabek, A. Chauhan, and R. Vaish, “Pyroelectric materials and devices for energy harvesting applications,” *Energy Environ Sci*, vol. 7, no. 12, pp. 3836–3856, 2014, doi: 10.1039/c4ee01759e.

[36] D. A. Howey, A. Bansal, and A. S. Holmes, “Design and performance of a centimetre-scale shrouded wind turbine for energy harvesting,” *Smart Mater Struct*, vol. 20, no. 8, 2011, doi: 10.1088/0964-1726/20/8/085021.

[37] M. Y. Zakaria, D. A. Pereira, and M. R. Hajj, “Experimental investigation and performance modeling of centimeter-scale micro-wind turbine energy harvesters,” *Journal of Wind Engineering and Industrial Aerodynamics*, vol. 147, pp. 58–65, 2015, doi: 10.1016/j.jweia.2015.09.009.

[38] S. Priya, “Modeling of electric energy harvesting using piezoelectric windmill,” *Appl Phys Lett*, vol. 87, no. 18, pp. 1–3, 2005, doi: 10.1063/1.2119410.

[39] Y. Yang, Q. Shen, J. Jin, Y. Wang, W. Qian, and D. Yuan, “Rotational piezoelectric wind energy harvesting using impact-induced resonance,” *Appl Phys Lett*, vol. 105, no. 5, pp. 0–4, 2014, doi: 10.1063/1.4887481.

[40] J. M. McCarthy, S. Watkins, A. Deivasigamani, and S. J. John, “Fluttering energy harvesters in the wind: A review,” *J Sound Vib*, vol. 361, pp. 355–377, 2016, doi: 10.1016/j.jsv.2015.09.043.

[41] D. Zhu, S. Beeby, J. Tudor, N. White, and N. Harris, “A novel miniature wind generator for wireless sensing applications,” *Proceedings of IEEE Sensors*, vol. 1, pp. 1415–1418, 2010, doi: 10.1109/ICSENS.2010.5690505.

[42] M. Zhang and J. Wang, “Experimental study on piezoelectric energy harvesting from vortex-induced vibrations and wake-induced vibrations,” *J Sens*, vol. 2016, 2016, doi: 10.1155/2016/2673292.

[43] H. L. Dai, A. Abdelkefi, and L. Wang, “Theoretical modeling and nonlinear analysis of piezoelectric energy harvesting from vortex-induced vibrations,” *J Intell Mater Syst Struct*, vol. 25, no. 14, pp. 1861–1874, 2014, doi: 10.1177/1045389X14538329.

[44] W. Sun and J. Seok, “A novel self-tuning wind energy harvester with a slidable bluff body using vortex-induced vibration,” *Energy Convers Manag*, vol. 205, no. January, p. 112472, 2020, doi: 10.1016/j.enconman.2020.112472.

[45] L. B. Zhang, A. Abdelkefi, H. L. Dai, R. Naseer, and L. Wang, “Design and experimental analysis of broadband energy harvesting from vortex-induced vibrations,” *J Sound Vib*, vol. 408, pp. 210–219, 2017, doi: 10.1016/j.jsv.2017.07.029.

[46] M. Argentina and L. Mahadevan, “Fluid-flow-induced flutter of a flag,” *Proc Natl Acad Sci U S A*, vol. 102, no. 6, pp. 1829–1834, 2005, doi: 10.1073/pnas.0408383102.

[47] A. Abdelkefi, M. R. Hajj, and A. H. Nayfeh, “Piezoelectric energy harvesting from transverse galloping of bluff bodies,” *Smart Mater Struct*, vol. 22, no. 1, 2013, doi: 10.1088/0964-1726/22/1/015014.

[48] J. Sirohi and R. Mahadik, “Harvesting wind energy using a galloping piezoelectric beam,” *Journal of Vibration and Acoustics, Transactions of the ASME*, vol. 134, no. 1, pp. 1–8, 2012, doi: 10.1115/1.4004674.

[49] L. Tang, L. Zhao, Y. Yang, and E. Lefeuvre, “Equivalent circuit representation and analysis of galloping-based wind energy harvesting,” *IEEE/ASME Transactions on Mechatronics*, vol. 20, no. 2, pp. 834–844, 2015, doi: 10.1109/TMECH.2014.2308182.

[50] C. F. Zhou, H. X. Zou, K. X. Wei, and J. G. Liu, “Enhanced performance of piezoelectric wind energy harvester by a curved plate,” *Smart Mater Struct*, vol. 28, no. 12, 2019, doi: 10.1088/1361-665X/ab525a.

[51] A. Abdelkefi, Z. Yan, and M. R. Hajj, “Modeling and nonlinear analysis of piezoelectric energy harvesting from transverse galloping,” *Smart Mater Struct*, vol. 22, no. 2, 2013, doi: 10.1088/0964-1726/22/2/025016.

[52] R. C. Dash, D. K. Maiti, and B. N. Singh, “A finite element model to analyze the dynamic characteristics of galloping based piezoelectric energy harvester,” *Mechanics of Advanced Materials and Structures*, vol. 0, no. 0, pp. 1–14, 2021, doi: 10.1080/15376494.2021.1921316.

[53] G. Parkinson, “Phenomena and modelling of flow-induced vibrations of bluff bodies,” *Progress in Aerospace Sciences*, vol. 26, no. 2, pp. 169–224, 1989, doi: 10.1016/0376-0421(89)90008-0.

[54] A. Barrero-Gil, A. Sanz-Andrés, and M. Roura, “Transverse galloping at low Reynolds numbers,” *J Fluids Struct*, vol. 25, no. 7, pp. 1236–1242, 2009, doi: 10.1016/j.jfluidstructs.2009.07.001.

[55] C. Scruton and D. E. J. Walshe, “On the aeroelastic instability of bluff cylinders,” *Journal of Applied Mechanics, Transactions ASME*, vol. 29, no. 1, pp. 215–216, 1960, doi: 10.1115/1.3636467.

[56] G. Alonso and J. Meseguer, “A parametric study of the galloping stability of two-dimensional triangular cross-section bodies,” *Journal of Wind Engineering and Industrial Aerodynamics*, vol. 94, no. 4, pp. 241–253, 2006, doi: 10.1016/j.jweia.2006.01.009.

[57] P. Poudel *et al.*, “Enhancing the Performance of Piezoelectric Wind Energy Harvester Using Curve-Shaped Attachments on the Bluff Body,” *Global Challenges*, vol. 7, no. 4, Apr. 2023, doi: 10.1002/gch2.202100140.

[58] G. Hu, K. T. Tse, M. Wei, R. Naseer, A. Abdelkefi, and K. C. S. Kwok, “Experimental investigation on the efficiency of circular cylinder-based wind energy harvester with different rod-shaped attachments,” *Appl Energy*, vol. 226, no. May, pp. 682–689, 2018, doi: 10.1016/j.apenergy.2018.06.056.

[59] J. Wang, S. Zhou, Z. Zhang, and D. Yurchenko, “High-performance piezoelectric wind energy harvester with Y-shaped attachments,” *Energy Convers Manag*, vol. 181, 2019, doi: 10.1016/j.enconman.2018.12.034.

[60] F. R. Liu *et al.*, “Performance enhancement of wind energy harvester utilizing wake flow induced by double upstream flat-plates,” *Appl Energy*, vol. 257, 2020, doi: 10.1016/j.apenergy.2019.114034.

[61] P. Poudel *et al.*, “The Bacterial Disinfection of Water Using a Galloping Piezoelectric Wind Energy Harvester,” *Energies (Basel)*, vol. 15, no. 17, Sep. 2022, doi: 10.3390/en15176133.

[62] M. Allaire, H. Wu, and U. Lall, “National trends in drinking water quality violations,” *Proc Natl Acad Sci U S A*, vol. 115, no. 9, pp. 2078–2083, 2018, doi: 10.1073/pnas.1719805115.

[63] P. W. Bohn, M. Elimelech, J. G. Georgiadis, B. J. Mariñas, A. M. Mayes, and A. M. Mayes, “Science and technology for water purification in the coming decades,” *Nanoscience and Technology: A Collection of Reviews from Nature Journals*, vol. 452, no. March, pp. 337–346, 2009, doi: 10.1142/9789814287005_0035.

[64] P. H. Gleick, “Dirty Water : Estimated Deaths from Water-Related Diseases 2000-2020,” *Pacific Institute Research Report*, pp. 1–12, 2002.

[65] A. Prüss-Üstün, R. Bos, F. Gore, and J. Bartram, “Safer water, better health,” *World Health Organization*, p. 53, 2008.

[66] UNICEF & WHO, “Progress on household drinking water, sanitation and hygiene, 2000-2017,” *Unicef/Who*, p. 140, 2019.

[67] M. Peter-Varbanets, C. Zurbrügg, C. Swartz, and W. Pronk, “Decentralized systems for potable water and the potential of membrane technology,” *Water Res*, vol. 43, no. 2, pp. 245–265, 2009, doi: 10.1016/j.watres.2008.10.030.

[68] T. C. Chalmers and I. F. Angelillo, “Chlorination, Chlorination By-products, and,” *Am J Public Health*, vol. 82, no. 7, pp. 955–963, 1992.

[69] C. Tian, R. Liu, H. Liu, and J. Qu, “Disinfection by-products formation and precursors transformation during chlorination and chloramination of highly-polluted source water: Significance of ammonia,” *Water Res*, vol. 47, no. 15, pp. 5901–5910, 2013, doi: 10.1016/j.watres.2013.07.013.

[70] M. Deborde and U. von Gunten, “Reactions of chlorine with inorganic and organic compounds during water treatment-Kinetics and mechanisms: A critical review,” *Water Res*, vol. 42, no. 1–2, pp. 13–51, 2008, doi: 10.1016/j.watres.2007.07.025.

[71] K. P. Goswami and G. Pugazhenthi, “Credibility of polymeric and ceramic membrane filtration in the removal of bacteria and virus from water: A review,” *J Environ Manage*, vol. 268, no. March, p. 110583, 2020, doi: 10.1016/j.jenvman.2020.110583.

[72] L. Deng, H. H. Ngo, W. Guo, and H. Zhang, “Pre-coagulation coupled with sponge-membrane filtration for organic matter removal and membrane fouling control during drinking water treatment,” *Water Res*, vol. 157, pp. 155–166, 2019, doi: 10.1016/j.watres.2019.03.052.

[73] I. Kim and H. Tanaka, “Photodegradation characteristics of PPCPs in water with UV treatment,” *Environ Int*, vol. 35, no. 5, pp. 793–802, 2009, doi: 10.1016/j.envint.2009.01.003.

[74] T. Zhang *et al.*, “Removal of antibiotic resistance genes and control of horizontal transfer risk by UV, chlorination and UV/chlorination treatments of drinking water,” *Chemical Engineering Journal*, vol. 358, pp. 589–597, 2019, doi: 10.1016/j.cej.2018.09.218.

[75] Y. Chang, P. Kwan, and G. Pierson, *Dynamic energy consumption of advanced water and wastewater treatment technologies*. 2006.

[76] K. H. Schoenbach, R. P. Joshi, R. H. Stark, F. C. Dobbs, and S. J. Beebe, “Bacterial decontamination of liquids with pulsed electric fields,” *IEEE Transactions on Dielectrics and Electrical Insulation*, vol. 7, no. 5, pp. 637–645, 2000, doi: 10.1109/94.879359.

[77] P. Peng *et al.*, “Bacterial inactivation of liquid food and water using high-intensity alternate electric field,” *J Food Process Eng*, vol. 40, no. 4, 2017, doi: 10.1111/jfpe.12504.

[78] R. K. Singh, V. Babu, L. Philip, and S. Ramanujam, “Disinfection of water using pulsed power technique: Effect of system parameters and kinetic study,” *Chemical Engineering Journal*, vol. 284, pp. 1184–1195, 2016, doi: 10.1016/j.cej.2015.09.019.

[79] J. Zhou, T. Wang, C. Yu, and X. Xie, “Locally enhanced electric field treatment (LEEFT) for water disinfection,” *Front Environ Sci Eng*, vol. 14, no. 5, 2020, doi: 10.1007/s11783-020-1253-x.

[80] S. Pudasaini, A. T. K. Perera, S. S. U. Ahmed, Y. B. Chong, S. H. Ng, and C. Yang, “An electroporation device with microbead-enhanced electric field for bacterial inactivation,” *Inventions*, vol. 5, no. 1, pp. 1–11, 2020, doi: 10.3390/inventions5010002.

[81] Z. Y. Huo *et al.*, “Cell Transport Prompts the Performance of Low-Voltage Electroporation for Cell Inactivation,” *Sci Rep*, vol. 8, no. 1, pp. 1–10, 2018, doi: 10.1038/s41598-018-34027-0.

[82] S. Pudasaini, A. T. K. Perera, D. Das, S. H. Ng, and C. Yang, “Continuous flow microfluidic cell inactivation with the use of insulating micropillars for multiple electroporation zones,” *Electrophoresis*, vol. 40, no. 18–19, pp. 2522–2529, 2019, doi: 10.1002/elps.201900150.

[83] J. Zhou, C. Yu, T. Wang, and X. Xie, “Development of nanowire-modified electrodes applied in the locally enhanced electric field treatment (LEEFT) for water disinfection,” *J Mater Chem A Mater*, vol. 8, no. 25, pp. 12262–12277, 2020, doi: 10.1039/d0ta03750h.

[84] J. Zhou, T. Wang, and X. Xie, “Rationally designed tubular coaxial-electrode copper ionization cells (CECICs) harnessing non-uniform electric field for efficient water disinfection,” *Environ Int*, vol. 128, no. March, pp. 30–36, 2019, doi: 10.1016/j.envint.2019.03.072.

[85] Z. Yu, H. Zeng, X. Min, and X. Zhu, “High-performance composite photocatalytic membrane based on titanium dioxide nanowire/graphene oxide for water treatment,” *J Appl Polym Sci*, vol. 137, no. 12, pp. 2–14, 2020, doi: 10.1002/app.48488.

[86] J. Zhou, T. Wang, W. Chen, B. Lin, and X. Xie, “Emerging investigator series: Locally enhanced electric field treatment (LEEFT) with nanowire-modified electrodes for water disinfection in pipes,” *Environ Sci Nano*, vol. 7, no. 2, pp. 397–403, 2020, doi: 10.1039/c9en00875f.

[87] Z. Y. Huo, H. Liu, C. Yu, Y. H. Wu, H. Y. Hu, and X. Xie, “Elevating the stability of nanowire electrodes by thin polydopamine coating for low-voltage electroporation-disinfection of pathogens in water,” *Chemical Engineering Journal*, vol. 369, no. March, pp. 1005–1013, 2019, doi: 10.1016/j.cej.2019.03.146.

[88] C. Liu *et al.*, “Static electricity powered copper oxide nanowire microbicidal electroporation for water disinfection,” *Nano Lett*, vol. 14, no. 10, pp. 5603–5608, 2014, doi: 10.1021/nl5020958.

[89] S. Kumar, M. Sharma, A. Kumar, S. Powar, and R. Vaish, “Rapid bacterial disinfection using low frequency piezocatalysis effect,” *Journal of Industrial and Engineering Chemistry*, vol. 77, pp. 355–364, 2019, doi: 10.1016/j.jiec.2019.04.058.

[90] W. Ding *et al.*, “TriboPump: A Low-Cost, Hand-Powered Water Disinfection System,” *Adv Energy Mater*, vol. 9, no. 27, pp. 1–8, 2019, doi: 10.1002/aenm.201901320.

[91] Z. Y. Huo *et al.*, “Triboelectrification induced self-powered microbial disinfection using nanowire-enhanced localized electric field,” *Nat Commun*, vol. 12, no. 1, pp. 1–11, 2021, doi: 10.1038/s41467-021-24028-5.

[92] Y. Yang, L. Zhao, and L. Tang, “Comparative study of tip cross-sections for efficient galloping energy harvesting,” *Appl Phys Lett*, vol. 102, no. 6, 2013, doi: 10.1063/1.4792737.

[93] C. F. Zhou, H. X. Zou, K. X. Wei, and J. G. Liu, “Enhanced performance of piezoelectric wind energy harvester by a curved plate,” *Smart Mater Struct*, vol. 28, no. 12, 2019, doi: 10.1088/1361-665X/ab525a.

[94] L. Ding, L. Yang, Z. Yang, L. Zhang, C. Wu, and B. Yan, “Performance improvement of aeroelastic energy harvesters with two symmetrical fin-shaped rods,” *Journal of Wind Engineering and Industrial Aerodynamics*, vol. 196, no. September 2019, p. 104051, 2020, doi: 10.1016/j.jweia.2019.104051.

[95] T. Wang, H. Chen, C. Yu, and X. Xie, “Rapid determination of the electroporation threshold for bacteria inactivation using a lab-on-a-chip platform,” *Environ Int*, vol. 132, no. April, p. 105040, 2019, doi: 10.1016/j.envint.2019.105040.

[96] C. Wagner, “Investigations on silver sulfide,” *J Chem Phys*, vol. 21, no. 10, pp. 1819–1827, 1953, doi: 10.1063/1.1698670.

[97] C. H. Xu, C. H. Woo, and S. Q. Shi, “Formation of CuO nanowires on Cu foil,” *Chem Phys Lett*, vol. 399, no. 1–3, pp. 62–66, 2004, doi: 10.1016/j.cplett.2004.10.005.

[98] Z. Y. Huo *et al.*, “Low-voltage alternating current powered polydopamine-protected copper phosphide nanowire for electroporation-disinfection in water,” *J Mater Chem A Mater*, vol. 7, no. 13, pp. 7347–7354, 2019, doi: 10.1039/C8TA10942G.

[99] T. Kotnik, W. Frey, M. Sack, S. Haberl Meglič, M. Peterka, and D. Miklavčič, “Electroporation-based applications in biotechnology,” *Trends Biotechnol*, vol. 33, no. 8, pp. 480–488, 2015, doi: 10.1016/j.tibtech.2015.06.002.

[100] X. Xie, J. Zhou, and T. Wang, “Locally Enhanced Electric Field Treatment (LEEFT) promotes the performance of ozonation for bacteria inactivation by disrupting the cell membrane,” *Environ Sci Technol*, vol. 54, no. 21, pp. 14017–14025, 2020, doi: 10.1021/acs.est.0c03968.

[101] J. Zhou, F. Yang, Y. Huang, W. Ding, and X. Xie, “Smartphone-powered efficient water disinfection at the point of use,” *NPJ Clean Water*, vol. 3, no. 1, 2020, doi: 10.1038/s41545-020-00089-9.

# List of Publications

1. Poudel P., Kumar R., Narain V., Jain S.C. (2022) Power Optimization of a Wind Turbine Using Genetic Algorithm. Machines, Mechanism and Robotics, Lecture Notes in Mechanical Engineering. Springer, Singapore. https://doi.org/10.1007/978-981-16-0550-5_171.

2. Poudel, P.; Sharma, S.; Ansari, M.N.M.; Kumar, P.; Ibrahim, S.M.; Vaish, R.; Kumar, R.; Thomas, P. The Bacterial Disinfection of Water Using a Galloping Piezoelectric Wind Energy Harvester. *Energies 2022*.

3. Poudel, P.; Sharma, S.; Ansari, M.N.M.; Bowen, C.; Ibrahim, S.M.; Vaish, R.; Kumar, R.; Thomas, P. Enhancing the performance of a piezoelectric wind energy harvester using curved-shaped attachments on the bluff body. *Global Challenges 2022*.